\documentclass{agujournal2019}
\journalname{Journal of Advances in Modeling Earth Systems (JAMES)}

\usepackage{url} 
\usepackage{lineno}
\usepackage{soul}

\usepackage[utf8]{inputenc}
\usepackage{upgreek,adjustbox,float}

\begin{document}

\title{Immersion freezing in particle-based aerosol-cloud microphysics: a~probabilistic perspective on singular and time-dependent models}

\authors{Sylwester~Arabas\affil{1}\thanks{Research carried out in part while at the University of Illinois and at the Jagiellonian University.}, Jeffrey~H.~Curtis\affil{2}, Israel~Silber\affil{3}\thanks{Now at Atmospheric, Climate, and Earth Sciences Division, Pacific Northwest National Laboratory.}, 
Ann~M.~Fridlind\affil{4}, Daniel~A.~Knopf\affil{5},\\~\\ Matthew~West\affil{6} and Nicole~Riemer\affil{2}
}

\affiliation{1}{Faculty of Physics and Applied Computer Science, AGH University of Krakow, Kraków,~Poland\\~}
\affiliation{2}{Department of Climate, Meteorology \& Atmospheric Sciences, University of Illinois at Urbana-Champaign, Urbana,~IL~61801,~USA\\~}
\affiliation{3}{Department of Meteorology and Atmospheric Science, Pennsylvania State University, University~Park,~PA~16801,~USA\\~}
\affiliation{4}{Goddard Institute for Space Studies, National Aeronautics and Space Administration, New~York,~NY~10025,~USA\\~}
\affiliation{5}{School of Marine and Atmospheric Sciences, Stony Brook University, Stony Brook,~NY~11794,~USA\\~}
\affiliation{6}{Department of Mechanical Science and Engineering, University of Illinois at Urbana-Champaign, Urbana,~IL~61801,~USA}

\correspondingauthor{Sylwester Arabas}{sylwester.arabas@agh.edu.pl}

\begin{keypoints}
    \item Discussion of origins, congruence and limitations of the singular and the time-dependent immersion freezing modeling approaches.
    \item Zero- and two-dimensional particle-based microphysics simulations illustrate similarities and differences between those two approaches.
    \item A singular active-sites parmeterization leads to a laboratory-cooling-rate signature that is absent in a time-dependent model. 
\end{keypoints}

\linespread{1.1}
\begin{abstract}
  Cloud droplets containing ice-nucleating particles (INPs) may freeze at temperatures above the homogeneous freezing threshold temperature. This process, referred to as immersion freezing, is one of the modulators of aerosol-cloud interactions in the Earth’s atmosphere. In modeling studies, immersion freezing is often described using either so-called “singular” or “time-dependent” parameterizations. Here, we juxtapose both approaches and discuss them in the context of probabilistic particle-based cloud microphysics modeling. First, using a box model, we contrast how both parameterizations respond to different idealized ambient cooling rate proﬁles and quantify the impact of the polydispersity of the immersed surface spectrum on the frozen fraction evolution. Second, using a prescribed-flow two-dimensional cloud model, we illustrate the implications of applying the singular model in simulations with ﬂow regimes relevant to ambient cloud conditions rather than to the cloud-chamber experiments on which these parameterizations are built upon. We discuss the critical role of the attribute-space sampling strategy for particle-based model simulations in modeling heterogeneous ice nucleation which is contingent on the presence of relatively sparse immersed INPs. The key takeaways include: (i) The singular approach, constituting a time-integrated form of a more general time-dependent approach, is only applicable under a limited range of ambient cooling rates. (ii) The time-dependent approach, especially when based on water-activity, is suitable for integration with particle-based model components of detailed aerosol composition and collisional growth/breakup. (iii) A ﬂow-coupled aerosol-budget-resolving simulation shows the benefits and challenges of modeling cloud condensation nuclei activation and immersion freezing on insoluble ice nuclei with super-particle methods.
\end{abstract}

%\section*{Plain Language Summary}
%Clouds are composed of water droplets and/or ice particles.
%One of the ways ice forms in clouds is called immersion freezing.
%It happens when an ice nucleus is immersed in a liquid water droplet at subzero temperature.
%Among examples of immersion freezing nuclei, there are grains of minerals, proteins or organic layers.
%Without presence of the nuclei, freezing requires a lower temperature. 
%Consequently, constraining immersion freezing helps to understand at what temperature and, hence, when 
%  clouds may transition from the liquid to solid (ice) phase of water. 
%Here, we focus on the ways of including the immersion freezing process in simulations of clouds.
%Furthermore, we discuss the recurrent question in this field of research, namely the role of time in the freezing process. 
%The tool we use (a computer program) is a so-called particle-based simulation which relies on tracking the movement of notional particles, each representing 
%  a large number of droplets or ice particles.
%This study confirms that treating freezing as characterized by a chance of freezing (rather than the contrasting approach involving a fixed freezing temperature for each droplet) 
%  is technically feasible.
%It makes the simulations costlier (longer to run) but results in higher fidelity of the simulation results.

\newpage
\section{Introduction}

Mixed-phase clouds consist of supercooled droplets, ice particles and the moist air surrounding them.
Nucleation of both water drops and ice crystals predominantly relies on the presence
  of aerosol particles serving as nuclei facilitating the onset of phase transition.
Ice nucleating particles (INPs) include, e.g., dust and sea spray
 \cite<see>[for reviews]{Kanji_et_al_2017,Knopf_and_Alpert_2023}.
Cloud heights overlap with the topmost layer of the biosphere, and inorganic, biogenic, and organic matter, while less abundant, can also serve as INPs
  \cite<see, e.g.,>[and references therein]{Despres_et_al_2012,Froechlich_Nowoisky_et_al_2016,Knopf_et_al_2018}.

Clouds affect 
  the Earth's energy budget and climate \cite<e.g.,>{Pincus_and_Chepfer_2020}.
Particles of both natural and anthropogenic origin contribute to cloud condensation nuclei (CCN) and INP populations.
Clouds thus play a key role in the indirect effects of anthropogenic particle emissions
  on the radiative properties of the atmosphere \cite<see>[for a review]{Bellouin_et_al_2020}.
There are numerous open questions linked with the mixed-phase
  aspects of cloud evolution.
Examples range from the impact of aerosol on the persistence of shallow Arctic clouds
  \cite{Morrison_et_al_2012,Fridlind_and_Ackerman_2018} to the role of aerosol in determining the dynamics of deep
  convection \cite{Marinescu_et_al_2021}. 

Tackling research questions (and climate-change policy questions alike) related to mixed-phase clouds 
  continues to pose significant challenges originating from knowledge gaps in both observational and modeling domains
  \cite<see, e.g.>{Burrows_et_al_2022}.
Among the modeling challenges that are particularly relevant to aerosol-cloud interactions in mixed-phase clouds are:
  (i)~the representation of interconnected aerosol, droplet and ice particle concentration budgets 
    and the aerosol-activation and ice-nucleation processes dependent on aerosol and INP concentrations \cite<e.g.,>{Stevens_et_al_2018};
  (ii)~the diversity of both size and composition of atmospheric aerosol serving as CCN and INPs
 \cite<e.g.,>{Andreae_and_Rosenfeld_2008,Knopf_et_al_2021};
  (iii)~the two-way nature of aerosol-cloud interactions in which ambient aerosol particles shape cloud microstructure, and
    cloud processes shape ambient aerosol through scavenging \cite{Wood_et_al_2012} 
    as well as resuspension \cite{Solomon_et_al_2015} 
    of chemically or physically processed nuclei \cite{Kilchhofer_et_al_2021}.

Improving our ability to accurately model mixed-phase cloud processes in high-resolution models, which are the focus of this work, 
  can enhance our understanding of these clouds and their feedback within climate systems.
This knowledge is crucial for improving global circulation models, which currently have uncertainties related to the representation of mixed-phase clouds
  \cite<see, e.g.>[for a review]{Ceppi_et_al_2017}.
The modeling technique which, by-design, can consider all of the above-listed challenges is referred to as
  super-particle or particle-based cloud microphysics, a probabilistic approach gaining significant momentum
  in high-resolution atmospheric cloud modeling \cite{Morrison_et_al_2020}.
In this work, we focus on representation in particle-based models of the immersion freezing phenomenon
  upon which the presence of INPs within supercooled droplets allows freezing to occur at temperatures higher than the homogeneous freezing threshold.
Our contribution builds upon earlier developments detailed in \citeA{Alpert_and_Knopf_2016} and \citeA{Shima_et_al_2020} and
  compares how the so-called ``singular'' and the alternative ``time-dependent'' models of immersion freezing can be used
  in Monte-Carlo particle-resolved simulations of mixed-phase clouds.

Particle-based cloud microphysics modeling is based on the concept of
  splitting the simulation of the continuous and dispersed phases of a particle-laden flow
  into an Eulerian and a Lagrangian bi-directionally coupled simulation components \cite<for an overview, see>{Grabowski_et_al_2019,Morrison_et_al_2020}.
The Eulerian component is a computational fluid dynamics solver handling the motion of moist air, as well as
  mass and heat budget corresponding to phase changes of water. 
The Eulerian component employs gridded representation of scalar fields such as water vapor density.
The Lagrangian component solves for the location in both physical and 
  attribute space of computational particles, each representing a large multiplicity
  of real aerosol or cloud particles (hence the terms simulation particles, super-particles and super-droplets).
The Lagrangian component employs discrete particle representation with each particle being assigned a number
  of attributed (e.g., mass of water, phase of water, surface of immersed aerosol).
The key advantages of particle-resolved microphysics models, as opposed to continuous-field
  formulations also termed bulk or bin models, include \cite<see, e.g.,>{Grabowski_et_al_2019}: 
  (i)~maintaining the identity of particles throughout the simulation,
  (ii)~favorable scaling with the attribute space dimensionality and 
  (iii)~by-design, strict preservation of the positivity of derived density fields, while
  their numerical diffusion can be either zero (trajectory integration) or adapted to match
  the Eulerian component numerical diffusivity \cite{Curtis_et_al_2024}.
Noteworthily, both particle-based and bin-resolved models of the evolution of aerosol and cloud droplet
  size spectrum are bound by spectral resolution limitations stemming from the limited number 
  of super-particles or bins, both of which represent large numbers of real particles.

The identity preservation is particularly useful for research on aerosol-cloud interactions.
Unlike bin or bulk cloud microphysics models, particle-based techniques
  do not differentiate, at the level of simulation, if a given super-particle represents an aerosol particle, a cloud droplet or a rain drop -- such 
  categorization is done only at the level of simulation output analysis based on 
  values of particle attributes (i.e. the wet size in the case of differentiation
  between aerosol or droplet).
For instance, all liquid particles, regardless of their size, are subject to the same set of basic
  processes such as sedimentation, transport by the flow, condensational growth or evaporation and
  collisions.
Phenomena such as CCN activation, aerosol washout or resuspension are effectively simulated through the
  combination of the above basic processes.
The aerosol reservoir dynamics are thus inherently resolved,
  which enables studies on aerosol processing by clouds \cite<understood as a combination of both chemical processing and physical processing 
  caused by particle collisions, for a recent discussion, see e.g.>{Hoffmann_and_Feingold_2023}.
This capability has been exemplified in particle-resolved studies of in-cloud chemistry followed 
  by aerosol re-suspension on cloud droplet evaporation \cite{Jaruga_and_Pawlowska_2018,Yao_et_al_2021}.
Furthermore, what is also relevant in the context of triggering of freezing, particle-based models offer 
  flexibility in terms of handling particle response to ambient temperature and supersaturation
  fluctuations \cite{Abade_et_al_2018,Hoffmann_et_al_2019,Abade_and_Albuquerque_2024} or even tracking the heat content
  of each particle and resolving the heat accommodation inertia \cite{Richter_et_al_2021}.
In general however, particle-based microphysics models capturing aerosol-cloud interactions impose
  demanding constraints on the temporal and spatial resolution of the host flow solver,
  and typically are used in Large Eddy Simulation (LES) frameworks. 
From the point of view of resolving aerosol composition diversity, particle-resolved models can
  track the aerosol mixing state \cite[Sect.~6.3.2]{Riemer_et_al_2019} which is 
  also relevant for determining the ice formation potential of atmospheric aerosol \cite{Knopf_et_al_2018,Lata_et_al_2021,Knopf_et_al_2022,Burrows_et_al_2022,Knopf_et_al_2023_ACP}.
An objective of this work is to inform the future development of climate model aerosol physics schemes such as MATRIX \cite{Bauer_et_al_2008}. 
The MATRIX mixing state scheme has been evaluated against single-particle mass spectrometry measurements of composition \cite{Bauer_et_al_2013}, 
  but the results of such an evaluation for ice nucleation potential would depend on the selection between parameterization that may differ as investigated in this work.

Several authors have used particle-based concepts in studies on ice-phase and mixed-phase cloud microphysics modeling
  \cite<e.g.,>{Jensen_and_Pfister_2004,Paoli_et_al_2004,Soelch_and_Kaercher_2012,Shirgaonkar_and_Lele_2012,Brdar_and_Seifert_2018,Seifert_et_al_2019,Shima_et_al_2020,Welss_et_al_2022}.
This work builds upon the work of \citeA{Shima_et_al_2020} where an aerosol-budget coupled representation of immersion freezing was proposed.
The model formulated in \citeA{Shima_et_al_2020} features a probabilistic formulation of the singular representation of immersion freezing
  which embodies the assumption that ice nucleation only depends on INP-characteristic freezing temperatures
  (which are sampled taking into account the distribution of immersed insoluble surface area)
  and thus is time-independent.
The singular (time-independent) approximation of the immersion freezing process has long been challenged by evidence supporting the stochastic 
  (and thus time-dependent) nature of the process \cite<see>{Knopf_et_al_2020,Knopf_and_Alpert_2023}. 
Moreover, from an implementation point of view, the singular models are found to be grossly inadequate for diagnostically representing the INP reservoir 
  dynamics owing to neglected losses whereas prognostic implementations are found to be extremely expensive owing to the number of 
  additional variables required to properly track such losses
 \cite{Fridlind_et_al_2012,Kaercher_and_Marcolli_2021,Burrows_et_al_2022,Knopf_et_al_2023}.
Such findings motivate this work to inform large-scale models that currently require a balance of physical accuracy and numerical efficiency.

In this work, both the time-dependent as well as the singular freezing models are cast in 
  a probabilistic, nuclei-reservoir-resolving and super-particle-number-conserving form.
We highlight how a simple Poissonian model of the rate of heterogeneous freezing in time constitutes a common base for both approaches,
  and how it entails limitations in robustness of singular schemes to varying flow regimes.
We also comment on the importance of representing the diversity of immersed surface areas (as opposed to monodispersity)
  of which both approaches are capable of representing.

\section{Models of immersion freezing}\label{sec:freezing}

\subsection{Origins of models and nomenclature}

Among the seminal works for the discussion presented herein, there is Bernard \citeauthor{Vonnegut_1948}'s \citeyear{Vonnegut_1948}
  paper reporting on a series of quasi-isothermal experiments exploring freezing of droplets suspended in oil,
  carried out at different temperatures.
\citeauthor{Vonnegut_1948}'s work depicted (Fig.~1 therein) both strong temperature dependence of the overall 
  nucleation rate, as well as persistent time-dependent nature of the process.
Quoting \citeA[]{Vonnegut_1949}, {``\it time required for these drops to freeze could be best 
  explained on the basis of the chance formation of stable nuclei on the foreign surfaces''}.
In \citeyear{Levine_1950}, \citeauthor{Levine_1950} reported a seminal statistical theory of 
  heterogeneous freezing, a work referred to as the origin of the ``singular'' hypothesis.
It veils the role of time and puts forward the ansatz of each singularity (mote) causing
  freezing to be associated with a characteristic freezing temperature.
For further references and a detailed recount on the earlier works in both the meteorological as well as other domains, see e.g. \citeA{Vali_1971} and
  \citeA[Appendix~A therein]{Vali_2014}.

In an attempt to revisit Levine's theoretical considerations with inclusion of dependence on time, 
  in \citeauthor{Bigg_1953a} and \citeauthor{Bigg_1953b}, a unifying probabilistic description of the process was presented 
  which admits both the time-dependent and the singular characteristics.
Albeit Bigg's theory was developed ''{\it without appealing to the action of foreign ice-forming nuclei}'' (and in general
  without involving a description of any microscopic mechanism),
  its reinterpretation and applicability to heterogeneous freezing was highlighted 
  already in \citeA{Mossop_1955,Langham_and_Mason_1958} and henceforth had been referred to as Mason-Biggs theory \cite<e.g.,>{Michel_1967}.
An alternative, simpler derivation of the results obtained by Bigg was subsequently 
  presented by \citeA{Carte_1959}.
The ``stochastic'' label to describe these developments was first used by \citeA{Stansbury_1961} and \citeA{Marshall_1961}.
A review of these developments in the meteorological context is presented in \citeA[section 9.2.5]{Pruppacher_and_Klett_2010}.

The early works in the development of the theory and nomenclature of heterogeneous freezing
  outlined above do not feature the later-defined terminology \cite<e.g.,>{Isaac_and_Douglas_1972} which in particular
  differentiates: (i)~condensation-, (ii)~immersion- and (iii)~contact modes of freezing as three different pathways
  of triggering heterogeneous nucleation of supercooled droplets \cite<see, e.g.>[for a recent overview]{Laaksonen_and_Malila_2022}.
Herein, we focus on the immersion freezing mode, and contrast the two antipodal approaches of time-dependent and singular
  conceptualization of the process, neglecting hybrid models that bridge both descriptions 
  \cite<e.g.,>{Wright_and_Petters_2013, Vali_and_Snider_2015}.

\subsection{Poissonian counting and active sites}\label{sec:poisson}

The starting point for description of heterogeneous freezing in the treatments akin to \citeA{Bigg_1953a} and \citeA{Carte_1959}
  is the Poisson counting process implying the following form of cumulative probability 
  of $k$ freezing-triggering events occurring in time $t$ in a population of INPs \cite<see, e.g. sect.~7.2.3>[presented in the context of homogeneous freezing]{Pruppacher_and_Klett_2010}:
\begin{equation}
  P^{*}(\textrm{$k$ events in time $t$})=\frac{(rt)^k \exp(-rt)}{k!}
\end{equation}
where $r$ is a process rate (in arriving at the Poisson limit, a large number of independent events is considered with the probability of each event being small 
  and likelihood of simultaneous events negligible).
By evaluating the complement of the probability of zero events occurring, the cumulative probability 
  of freezing can be defined as:
\begin{equation}
  P(\textrm{one or more events in time t}) = 1 - P^{*}(k=0, t)
\end{equation}
or:
\begin{equation}
  \ln(1 - P) = -rt 
\end{equation}
In \citeauthor{Bigg_1953b}, the following factorization of the right-hand-side $-rt$ term was postulated,
  incorporating a particle-size-related variable labeled $X$ \cite<notation following>[see discussion of eq.~(1) therein]{Vali_2019} and 
  the ambient temperature-related variable $T$
  (cf. eq.~(2) in \citeauthor{Bigg_1953b} or eq.~(1) in \citeA{Carte_1959}):
\begin{equation}\label{eq:lnpm2=int}
  \ln\left(1\!-\!P(X,t)\right)= -X\!\!\underbrace{\int\limits_{0}^{t}\!\!J_\textnormal{\scriptsize X}\!\left(T(t')\right) dt'}_{ n_\textnormal{\scriptsize X}(T)}
\end{equation}
where $T(t)$ denotes the ambient temperature evolution, and $J_\textnormal{\scriptsize X}(T)$ is the nucleation rate expressed per size $X$ \cite<for generalization to
  both surface and volume dependent rate, see discussion of eqs.~1--3 in>{Leonard_and_Im_1999}.
Commenting on the assumption inherent in the Poisson counting process, a prerequisite for application of the model is that
  the macroscopic size $X$ is large with respect to the microscopic length scale on which freezing is initiated, and that
  the freezing events are uncorrelated in space or time, essentially instantaneous, hence non-overlapping.
Hereinafter, the model is applied to atmospheric aerosol particles populating macroscopic spatial domains
  and described collectively by a probability density defined over $X$.

In Bigg's 1953 works, $X=V$ was conceptualized as the drop volume and $J_\textnormal{\scriptsize X}$ as the homogeneous nucleation rate.
Starting with \citeA{Mossop_1955} and \citeA{Langham_and_Mason_1958}, an alternative interpretation was provided
  leading to the presently commonly used model where $X=S$ is the immersed insoluble surface area and $J_\textnormal{\scriptsize S}$ is the heterogeneous (immersion) freezing rate
  \cite<which provides a connection to the classical nucleation theory formalism, see>[and references therein]{Knopf_and_Alpert_2023}.
Nevertheless, Bigg's model proved seminal, and it allows to capture the link between time-dependent and
  singular views of the process through the relation embodied in eq.~\ref{eq:lnpm2=int}.
Bigg's analysis includes a derivation of a temperature-dependent probability-of-freezing 
  (\citeauthor<theoretical curves in Figs~2~\&~3 in>{Bigg_1953a})
  in an exponentially-linear form
\begin{equation}\label{eq:n_X}
  n_{\textrm{\tiny X}}(T) = c_0 + c_1 \exp(c_2 \cdot T + c_3),
\end{equation}
  where $c_k$ ($k\in0,1,2,3$) are coefficients constrained 
  by vanishing of the probability of freezing at $T=0^\circ$C\footnote{Bigg's formula given here as eq.~\ref{eq:n_X} is used widely in large-scale models for representation of droplet freezing embracing the $X=V$ interpretation, which implies that the freezing probability
  is modeled as dependent on volume of water rather than on the surface of immersed particles.
Examples include: \citeA[see discussion of eq.~(26) where parameter uncertainty is attributed with a 10K freezing temperature uncertainty]{Wisner_et_al_1972}, 
\citeA[eq.~A.22]{Reisner_et_al_1998},
\citeA[section 4.2.2]{Khain_et_al_2000}, \citeA[discussion of eq.~44]{Seifert_and_Beheng_2006}, \citeA[passage between (A2) and (A3)]{Sullivan_et_al_2018} and \citeA{Barrett_and_Hoose_2022}.
Noteworthily, in a number of instances, the formula is given with a typo \cite<e.g.,>{Seifert_and_Beheng_2006} with a ``-1'' term included under the 
  exponent instead of next to it to match Bigg's argument of vanishing probability at zero temperature, i.e.
       ``$\exp(\textrm{const}\cdot\Delta T)-1=0$''.   
  Using the GitHub code search, we have confirmed that the typo does not appear in a FORTRAN implementation of the Seifert \& Beheng scheme shared across COSMO, UCLA-LES and WRF code bases,
    however we also found a model which inherited the typo from the paper (authors contacted, bugfix released).
}.
Even though, eq.~(\ref{eq:n_X}) lacks explicit dependence on time, the $c_k$ coefficients do in fact
  depend on the cooling rate.
This thus, in principle, renders $n_\textnormal{\scriptsize X}$ applicable only for the cooling rate for which
  the coefficients $c_k$ are fitted (see section~\ref{sec:embed}).
An intuitive parameter interpretation links the median ($P=\frac{1}{2}$) freezing temperature with the 
  size parameter, which was embraced in the measurement data analysis in \citeauthor{Bigg_1953b}.  

Building on the heterogeneous freezing interpretation, the time-integrated (hence time-independent) density $n_{\textrm{\tiny S}}(T)$ 
  as a function of temperature
  mathematically corresponds to what is presently referred to as the active sites density, a term used as early as in
  \citeA[Sect.~4 and references to earlier works therein]{Fletcher_1969} and more recently popularized in the
  modeling community with the ``IASSD'' \cite<ice-active surface site density,>{Conolly_et_al_2009} or ``INAS'' \cite<ice nucleation active surface site, e.g.,>{Hoose_and_Moehler_2012} acronyms.
The INAS parameterization was used also in \citeA{Shima_et_al_2020} with the following form:
  \begin{equation}\label{eq:INAS}
    n_\textnormal{\scriptsize S}(T)=n_0 \cdot \exp(a_{_\textnormal{\scriptsize INAS}}\cdot(T-T_{0^\circ\!C})+b_{_\textnormal{\scriptsize INAS}})
  \end{equation}
  where $n_0=1 \textrm{m}^{-2}$,
  $a_{_\textnormal{\scriptsize INAS}}=-0.517 \textrm{K}^{-1}$ and $b_{_\textnormal{\scriptsize INAS}}=8.934$ \cite<measurement fit from>{Niemand_et_al_2012}.
As~can be seen from the above outline of the derivation, these coefficients depend on the cooling rate.
This implicit dependence on cooling rate will be further discussed in section~\ref{sec:embed} herein.

\subsection{Monte-Carlo simulations of immersion freezing}\label{sec:models}

Numerical simulations of the immersion freezing process can be realized using Monte-Carlo techniques which
  imply random sampling of the particle attribute spectrum, and consequently a probabilistic look
  on the process (with each model run yielding a unique realization).
In the next two subsections, we outline two previously developed aerosol-microphysics-coupled 
  Monte-Carlo frameworks for representing immersion freezing
  using time-dependent \cite{Alpert_and_Knopf_2016} and singular \cite{Shima_et_al_2020} parameterizations.
Noteworthily, a Monte-Carlo technique has recently been introduced in \citeA{Frostenberg_et_al_2022}
  for representing immersion freezing but without an explicit link to aerosol properties.

We describe both aerosol-coupled models using particle-based cloud microphysics nomenclature revolving around 
  super-particles and their attributes.
One of the key super-particle attributes is its multiplicity (also termed weighting factor) which
  describes the number of real particles represented by a given super-particle.
Whenever a freezing event occurs, all particles represented by a given super-particle
  crystallize, and regardless of the employed
  immersion freezing scheme the INP represented by the 
  newly frozen super-particle are removed from the INP reservoir (and reinstated upon melting).
There is no splitting of super-particles, and the number of computational particles is constant to ensure practical realizability
  of such simulations.

\begin{figure*}[t]
  \begin{center}
    \includegraphics[width=5in]{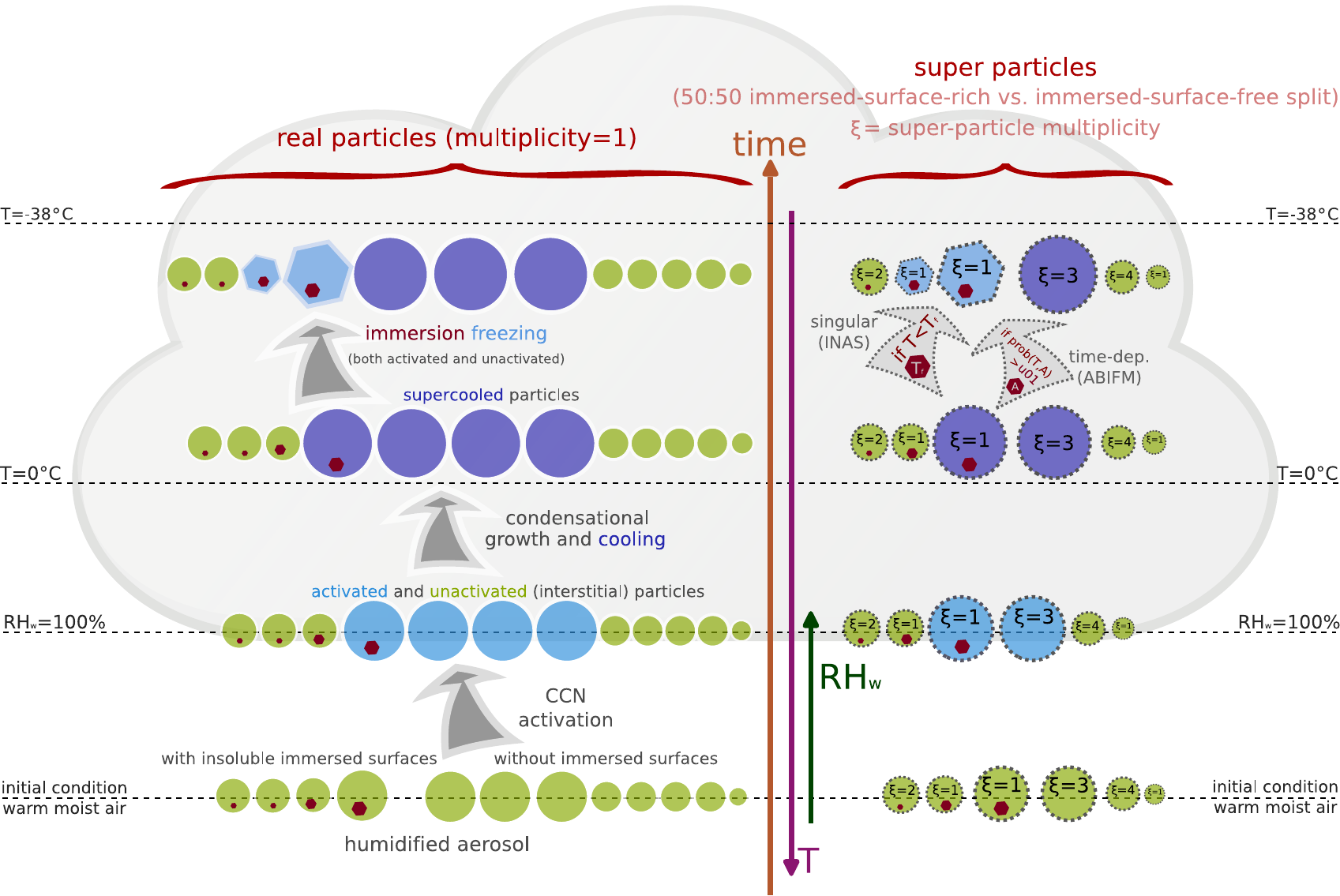}
  \end{center}
  \caption{\label{fig:schematic}
    Conceptual representation of the aerosol-cloud droplet activation and immersion freezing processes and their numerical implementation.
    Left: Depiction of the physical processes.
    Right: Numerical implementation using super-particles. 
    See section~\ref{sec:models} and \ref{sec:2D} for discussion.
  }
\end{figure*}

The concept of multiplicity is depicted in Fig.~\ref{fig:schematic}.
Without particle coalescence or fragmentation, the multiplicities for each super-particle are maintained throughout the simulation. 
A degree of freedom exists in the choice of initial multiplicities for super-particles.
Fig.~\ref{fig:schematic} depicts a case where only a subset of particles contains insoluble immersed surfaces, and
  the abundances of super-particles with and without insoluble immersed surfaces are equal.
In contrast, in the population of particles represented by super-particles, those without immersed surfaces are more abundant.
This is taken into account by assigning larger multiplicities $\xi$ to the more abundant super-particles.

\subsubsection{Time-dependent scheme using ABIFM parameterization}\label{sec:timedep}

The time-dependent Monte-Carlo model explored herein uses the stochastic water-activity based immersion freezing model \cite<ABIFM,>{Knopf_and_Alpert_2013}
  applied following \citeA{Alpert_and_Knopf_2016} where it was used in analysis of laboratory experiments.
Particle freezing is triggered by comparing a uniform random number in the interval (0,1) (shortened as u01 in the schematic) with the probability of 
  freezing evaluated at instantaneous ambient conditions in each timestep.
The probability is evaluated by assuming constant temperature within a timestep, thus
  eq.~(\ref{eq:lnpm2=int}) leads to (\citeauthor<cf.~eq~(1) in>{Bigg_1953b}):
\begin{equation}\label{eq:delta_t}
  P_i = 1 - \exp\left(-J_S(T) \cdot S_i \cdot \Delta t \right)\textrm{,}
\end{equation}
where $i$ denotes super-particle index, and $J_\textnormal{\scriptsize S}$ is expressed using the ABIFM parameterization \cite{Knopf_and_Alpert_2013}:
\begin{equation}\label{eq:ABIFM}
  \log_{10}\left(J_\textnormal{\scriptsize S}(T)\right) = c_{_\textnormal{\scriptsize ABIFM}} + m_{_\textnormal{\scriptsize ABIFM}} \cdot \Delta a_w \textrm{,}
\end{equation}
with $\Delta a_w$ denoting the difference between droplet's water activity and its value along the ice melting curve
here taken as (assuming pure water, ambient conditions at liquid saturation and a single equilibrium value of temperature for all phases):
\begin{equation}
  \left.\Delta a_w\right|_{RH\approx100\%} = 1 - \frac{p^\textnormal{\scriptsize ice}_s(T)}{p^\textnormal{\scriptsize liq}_s(T)}\textrm{,}
\end{equation}
where RH denotes the relative humidity and $p^\textnormal{\scriptsize ice}_s/p^\textnormal{\scriptsize liq}_s$ is the ratio of saturation vapor pressures over solid and liquid phases of water
  \cite{Koop_2002}.
The above assumption allows for comparison of ABIFM-based simulations with the INAS-based runs for the latter does not feature dependence on water activity.
The neglected effects of solutes on the water activity are readily representable using particle-based aerosol-cloud
  microphysics and using ABIFM; if represented they allow for simulation of immersion freezing of unactivated droplets (see Fig.~\ref{fig:schematic}).

Consequently, for a particle-based cloud microphysics model, each super-particle is required to include the immersed surface area $S_i$ of the modeled particle
  as an attribute (here, $S_i$ denotes the value for a single particle, not multiplied by the multiplicity).
Since, unlike in the case of coagulation, evaluation of the probability does not involve
  spatial concentration, its value can be applied for super-particles of any multiplicity
  \cite<see also the formulation of the probabilistic transport model in>{Curtis_et_al_2016}.
However, the longer the timestep and the larger the multiplicities (stemming from coarser size-spectral resolution), the larger will be
  the spread among different Monte-Carlo realizations of the process.

\subsubsection{Singular scheme using INAS parameterization}\label{sec:singular}

The singular formulation follows \citeA{Shima_et_al_2020}.
As in the case of the above time-dependent formulation, it also uses a probabilistic approach,
  however, here, it is limited to random attribute sampling at initialization,
  which reflects the lack of time dependence in $n_\textnormal{\scriptsize S}(T)$.
The randomly sampled attribute is the freezing temperature.
It is sampled from 
  a probability density function $p$ based on the cumulative $P(S,T)$ given in eq.~(\ref{eq:lnpm2=int})
  by employing a parameterized INAS density function (\ref{eq:INAS}).
This yields \cite<eq.~(1) in>{Shima_et_al_2020}:
\begin{equation}\label{eq:inas_pdf}
  \left.p(T)\right|_S = -S\frac{dn_\textnormal{\scriptsize S}(T)}{dT} \exp\left(-S \cdot n_\textnormal{\scriptsize S}(T)\right)\textrm{,}
\end{equation}
which can also be found in an approximated form in \citeauthor{Bigg_1953a} expressed using
  present notation through $p=\left.\partial_T P\right|_X\approx(1-P)\ln(1-P)$.
Effectively, such singular formulation may be regarded as Poissonian-in-space (where the nucleus surface $S$ stands
  for space, see eq.~\ref{eq:poissonian_in_space}) as opposed to the space-time Poissonian model (eq.~\ref{eq:lnpm2=int}) it is derived from.

In the singular formulation, the freezing of a super-particle is triggered by comparing
  the ambient temperature with the freezing temperature associated with a given
  particle at initialization.
Phase-change-triggering is thus deterministic.
Thus,~the immersed surface area $S$ does not need to be retained as a particle attribute
  and its values can be discarded after particle attribute initialization
  (or retained as an attribute which is not used by the singular freezing scheme).

In a way, this bears analogy to the simplified aerosol activation representation in particle-based
  microphysics models using the so-called Twomey CCN activation model \cite{Grabowski_et_al_2018}.
In such a model, the super-particles are assigned a critical supersaturation by sampling from
  a distribution at initialization, and the aerosol characteristics (such as mass or
  hygroscopicity) do not need to be retained as particle attributes.

The singular approach offers a significant performance gain, for there is no need to re-evaluate the probability of freezing or draw random numbers at every timestep.
This holds even if the immersed surface area is retained as particle attribute, whereas in so-called ``bin'' (non-super-particle) modeling approaches \cite<e.g.,>{Knopf_et_al_2023},
  resolving both the immersed surface area and the freezing-temeprature increases the dimensionality of the problem making the singular scheme significantly more
  expensive.

\subsection{INAS-embedded cooling-rate signature issue}\label{sec:embed}

\begin{figure}[t]
  \begin{center}
    \includegraphics[width=2.8in]{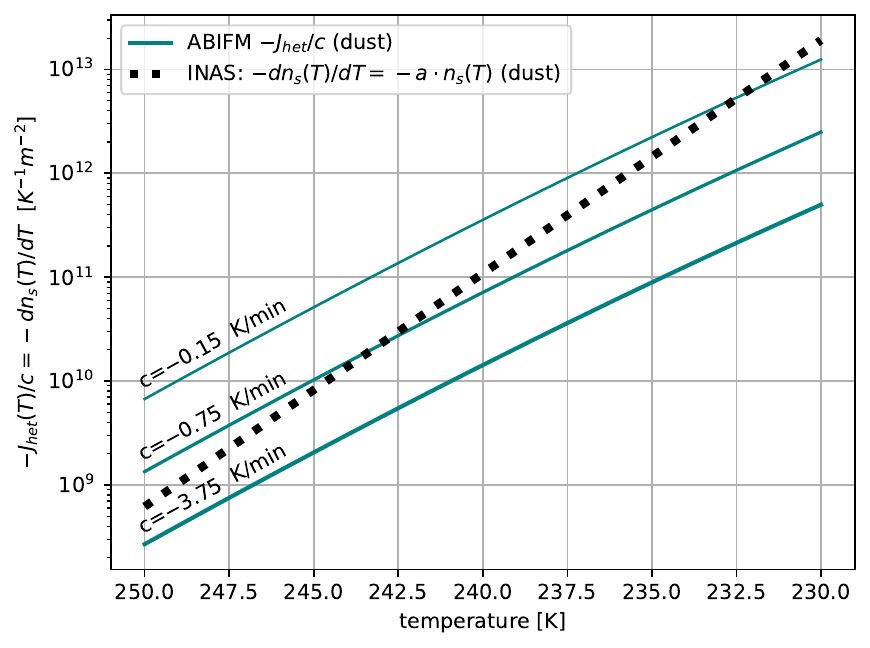}
  \end{center}
  \caption{\label{fig:theory}
    The INAS $n_\textnormal{\scriptsize s}(T)$ curve with parameters from \citeA{Niemand_et_al_2012} 
      (black filled squares) and
      a set of three ABIFM curves corresponding to three different cooling rates $c$ plotted (teal solid lines).
  }
\end{figure}

The performance gain and appealing simplicity of the singular scheme comes with numerous tradeoffs, though.
As noted hereinbefore discussing eq.~(\ref{eq:n_X}), and highlighted by several other authors
  \cite{Vali_1994,Murray_et_al_2011,Herbert_et_al_2014,Niedermeier_et_al_2015}, 
  the coefficients defining $n_\textnormal{\scriptsize S}(T)$ measurement fits embed a signature of the cooling
  rate characteristic of the laboratory freezing experiments the fits are based upon.
Noteworthily, coefficients derived for $J_\textnormal{\scriptsize het}$ and $n_\textnormal{\scriptsize S}$ parameterizations
  share several other sources of 
  uncertainties \cite<e.g.,>{Vali_and_Stansbury_1966,Murray_et_al_2011} stemming from: challenges in obtaining statistically significant
  sample of freezing events, characterization of ambient thermodynamic state, and characterization of the
  immersed surface area spectrum \cite{Alpert_and_Knopf_2016} -- all of which 
  apply to both methods of parameterising the INP activity.

Using the simple Poissonian model introduced above and 
  substituting a constant cooling/heating rate $c=dT/dt$ into eq.~(\ref{eq:lnpm2=int}) 
  yields (\citeauthor<cf. eq.~3 in>{Bigg_1953a}):
\begin{equation}\label{eq:poissonian_in_space}
  \ln(1-P(S,t))=-\frac{S}{c}\int\limits_{T_0}^{T_0+ct} \!\! J_\textnormal{\scriptsize het}(T') dT' = - S \cdot n_\textnormal{\scriptsize S}(T)\textrm{.}
\end{equation}
Taking temperature derivatives of both sides yields \cite<eqs~(1) and~(4) in>{Hoose_and_Moehler_2012}:
\begin{equation}\label{eq:Jhet_ns}
  -\frac{dn_\textnormal{\scriptsize S}(T)}{dT} = -a_{_\textnormal{\scriptsize INAS}} \cdot n_\textnormal{\scriptsize S}(T) = - \frac{1}{c} J_\textnormal{\scriptsize het}(T)\textrm{,}
\end{equation}
  which may be used to mathematically link the INAS and ABIFM parameterizations, namely the $n_\textnormal{\scriptsize S}(T)$ and the $J_\textnormal{\scriptsize het}(T)$ functions.
This is done without a link to the physics attributed to the central notions of both models (i.e., surface site density and nucleation rate)
  and implicitly assuming identical material type of particles in the modeled population.
We note also that this is only defined for a cooling process, whereas natural turbulent clouds subject parcels to chaotic cycles of cooling and warming.
Under such alternating cycles, this equation has limited practical modeling utility if the INP reservoir budget is to be closed \cite<cf.>{Knopf_et_al_2023},
  but it nonetheless serves as a useful benchmark.

For such a monotonic cooling process, Figure~\ref{fig:theory} presents the relevant mathematically-comparable quantities of ABIFM 
  and INAS immersion freezing parameterizations (for the same material type).
The terms $-dn_\textnormal{\scriptsize S}(T)/dT$ and $-J_\textnormal{\scriptsize het}(T)/c$ are plotted as ordinate in logarithmic scale
  with temperature on the abscissa.
Given the form of (\ref{eq:Jhet_ns}) in which the cooling rate is featured only 
  on one side of the equation, expectedly only one value of the cooling rate allows
  for a match between INAS and ABIFM.
An approximate match, within the 245--240 K temperature range, is depicted for $c=-0.75$~K/min.
The above analysis highlights that a discrepancy between singular
  and time-dependent simulations is expected.
It stems from the limited applicability of the INAS coefficients, which only apply to a single cooling rate
  matching the one characteristic of the lab experiment data to which the coefficients were fitted.
For this reason, INAS-type parameterization are considered applicable only for high cooling rate cases
  \cite<valid for updrafts of $\sim 1.5\!-\!3$~ms\textsuperscript{-1} as reported in>[p. 1.22]{Kanji_et_al_2017}.
Because mixing in Arctic stratiform clouds is typically driven by cloud-top cooling, downdrafts exceed updrafts that are typically weaker 
  than $1$~ms\textsuperscript{-1} as inferred from observation-constrained mean Doppler velocities in 
  \citeA{Avramov_et_al_2011}, \citeA[note opposite Doppler sign convention]{Fridlind_et_al_2012}
  and \citeA[Fig.~5, top panel in column 4]{Silber_2023}.

Early discussion of the relation of singular and time-dependent formulations can be found
  in \citeA{Fletcher_1958} and \citeA{Vali_and_Stansbury_1966}.
Recent works covering it include: \citeA[eq.~(6) therein]{Ervens_and_Feingold_2013}, \citeA[eq.~(25) therein]{Kubota_2019}
  and \citeA[eqs.~(1)-(2) in Supplement therein]{Cornwell_et_al_2021},
  albeit providing a less general relationship than here or in \citeA{Hoose_and_Moehler_2012}.
Depiction of the cooling-rate dependence of $n_s$ based on experimental data can be found in 
  \citeA[Fig. 5 \& 6]{Herbert_et_al_2014} and in \citeA[Fig.~3]{Niedermeier_et_al_2015}, 
  while the robustness of ABIFM $J_\textnormal{\scriptsize het}$ to the cooling rate is discussed in 
  \citeA{Knopf_and_Alpert_2013,Alpert_and_Knopf_2016}.

This line of research led to introduction of modified singular models
  that account for cooling rate dependence in INAS-like
  parameterizations \cite{Vali_1994, Murray_et_al_2011}.
While such parameterizations do address the issue highlighted in simulations presented
  herein, i.e., the experimental cooling rate ``signature`` in INAS fits,
  the lack of time-dependence in models employing an INAS fit renders any such 
  parameterization still less general compared with a nucleation-rate based approach.
From the herein presented particle-based perspective, this can be noted in the 
  assignment of $T_\textnormal{\scriptsize fz}$ to each super particle at initialization.
Such singular ansatz still entails prescription of a cooling rate, which limits
  application to flow regimes with a wide spectrum of cooling rates.
Moreover, any kind of singular approach will yield
  no nucleation in the limit of quiescent flow or within downdrafts where temperature likely increases along particle trajectory.
Another advantage of the time-dependent freezing event triggering is that the immersed surface area is not fixed for a particle
  and may be coupled with dynamically varying physical properties of the particles (e.g., coalescence-triggered increase in total immersed surface area or water-content dependence
  in case of computing the area covered by surfactants)
  as well as chemical characteristics of the solution (i.e., changes in the water activity and hence $\Delta a_w$ depending on the presence of solutes).

\subsection{Handling of freezing-related attributes upon particle collisions}

While in this study, particle collisions causing coalescence or fragmentation are not simulated, 
  the choice of super-particle attributes stemming from the formulation of the immersion freezing model
  has implications on the collisional growth/breakup handling.
In the case of above-presented singular probabilistic freezing model, the randomly sampled freezing temperatures are super-particle attributes.
In \citeA{Shima_et_al_2020}, it was proposed to select the higher of two freezing temperatures when assigning a freezing temperature for a freshly coalesced super-particle.
The time-dependent model inherently allows a more general treatment because the immersed surface is an extensive attribute and scales additively as any other extensive attribute upon coalescence.
In the case of collisions of particles with similarly sized immersions, likely having similar freezing temperatures in the singular scheme,
  the freshly coalesced particle has close-to-unchanged likelihood of freezing in the singular scheme,
  while in the time-dependent scheme the likelihood doubles by summing the surface areas of immersed material.
Analogous limitations apply to the Twomey CCN activation scheme \cite{Grabowski_et_al_2018} where activation supersaturation is used as an attribute
  leading to potentially unphysical attribute evolution upon coalescence or breakup.

It is an open question how to handle freezing-attribute value assignment upon super-particle breakup.
One of the challenges is linked with the insolubility of the immersed surfaces which conflicts with the equipartition assumption implied, e.g., in representation of hygroscopicity-related 
  aerosol attributes upon breakup.
Intuitively, only a subset of fragmented super-particles should retain an immersed surface, however representing it would require an additional super-particle to be introduced into the system
  which is essentially unfeasible in large-scale simulations for which fragmentation algorithms 
  are required to maintain constant super-particle count \cite{Lee_and_Matsoukas_2000,Kotalczyk_et_al_2017,deJong_et_al_2023_GMD}.

\subsection{Summary: probabilistic perspective on singular and time-dependent models}

From Sections \ref{sec:timedep} and \ref{sec:singular}, it follows that not only the time-dependent scheme 
  but also the singular scheme generate different realizations if the simulation is repeated. 
This is obvious for the time-dependent scheme because it simulates the time evolution with probabilistic sampling,
  but it may come as a surprise for the singular scheme since it is also referred to as ``deterministic'' in literature. 

To understand why the singular scheme also generates different realizations, we need to consider that the particle-based 
  modelling approach requires assignment of the initial particle attributes to a finite set of computational particles.
That is, in the case of the time-dependent scheme, we need to assign a surface area to each computational particle, 
  while in case of the singular scheme, we need to assign both a surface area and a freezing temperature. 
How can this be accomplished? 

This per-particle information could, in principle, come from single-particle measurements, but these do not exist in practice. 
We therefore adopt the standard approach of sampling to assign the initial particle attributes, which is explained in detail 
  in sections~\ref{sec:cool}, \ref{sec:2D} and \ref{sec:Lagrangian}.
This random initialization is the reason why the singular scheme also exhibits probabilistic features even though the freezing 
  process itself is entirely deterministic. 
We will provide more detail on the sampling process in Section 3. 
Figure~\ref{fig:sampling} summarizes the two approaches schematically.

The initial sampling of freezing temperatures for the singular scheme corresponds to assigning active sites to the computational particles.
Equation~(\ref{eq:inas_pdf}) tells us how to do this---the INAS density function $n_\textnormal{\scriptsize S}(T)$ is key here. 
An implication is that even if two particles have the exact same surface area, the number of active sites and their associated 
  freezing temperatures may be different, hence they will freeze at different temperatures. 

In summary, it is helpful to think of the initialization process and the time evolution process of the simulations as two separate issues. 
For the singular scheme, this results in a probabilistic assignment of freezing temperatures to each computational particle, 
  which in turn will result in a different outcome for repeated simulations even though the time evolution for this scheme is 
  strictly deterministic. 
For the time-dependent scheme, particles are initialized with randomly sampled surface areas but are otherwise identical, 
  but then evolve with a stochastic freezing process, necessarily resulting in different outcomes for each repeated simulation.

Ensemble means (and ensemble spread) from multiple realizations of the singular and the time-dependent models are expected to 
  match if: (i) the ambient cooling rate matches the conditions corresponding to the
  dataset from which the INAS coefficients were derived, (ii) the distribution of sampled immersed surface areas
  is consistent, (iii) the freezing temperatures are sampled from a distribution matching condition~(\ref{eq:Jhet_ns}) and
  (iv) auxiliary conditions for freezing triggering are consistent (RH$>$1 in this case). 
  
\section{Box-model simulations}\label{sec:0D}

\subsection{Simulation framework and initial condition attribute sampling}\label{sec:aero_params}

The box model simulations lack any spatial context, and can thus be referred to
  as zero-dimensional or pertaining to an unspecified volume of air.
The only ambient dynamics considered is an imposed temperature evolution.
  
The aerosol parameters are the median surface area \cite<here always set to the surface area of a sphere of diameter $0.74\,\upmu\textrm{m}$ corresponding
  to the large mode of the ``Desert Dust'' spectrum used in>{Knopf_et_al_2023} and the geometric standard deviation of a lognormal 
  spectrum of the immersed insoluble surface (unless indicated otherwise, set to an arbitrary value of $\sqrt[4]{e}$).
Two parameters defining the immersed material type are needed for both singular and time-dependent models, 
  and these are related with the coefficients in the INAS (eq.~\ref{eq:INAS}) and ABIFM (eq.~\ref{eq:ABIFM}) formulae, respectively. 
For INAS, we use the experimental fits from \citeA{Niemand_et_al_2012} which correspond to mineral dust (see discussion of eq.~\ref{eq:INAS} for values).
For ABIFM, we use coefficients for mineral dust given in \citeA[Table~2]{Alpert_and_Knopf_2016}:
  $m_{_\textnormal{\scriptsize ABIFM}}=22.62$ and $c_{_\textnormal{\scriptsize ABIFM}}=-1.35$ (note that $J_\textnormal{\scriptsize S}$ is expressed in cm\textsuperscript{-2} s\textsuperscript{-1} there).
Both fits were obtained using the same AIDA chamber experiment.
Note however, that procedures for deriving the coefficients for INAS and ABIFM fundamentally differ \cite{Rigg_et_al_2013}.
INAS coefficients are fitted to aggregated data in frozen fraction vs. temperature space.
ABIFM parameters are calculated from $J_\textnormal{\scriptsize het}$ vs. water activity space.
$J_\textnormal{\scriptsize het}$ is directly calculated from experimentally measured temperature and time of freezing, and immersed surface area.
Frozen fraction is thus not an input for fitting, but rather can be calculated from the derived $J_\textnormal{\scriptsize het}$
  values, and used for validation of the ABIFM parameters \cite<see discussion in>{Alpert_et_al_2011,Rigg_et_al_2013}.

\begin{figure}[H]
  \begin{center}
    \begin{adjustbox}{clip,trim=0cm 1.1cm 0cm 0cm}
      \includegraphics[width=2.255in]{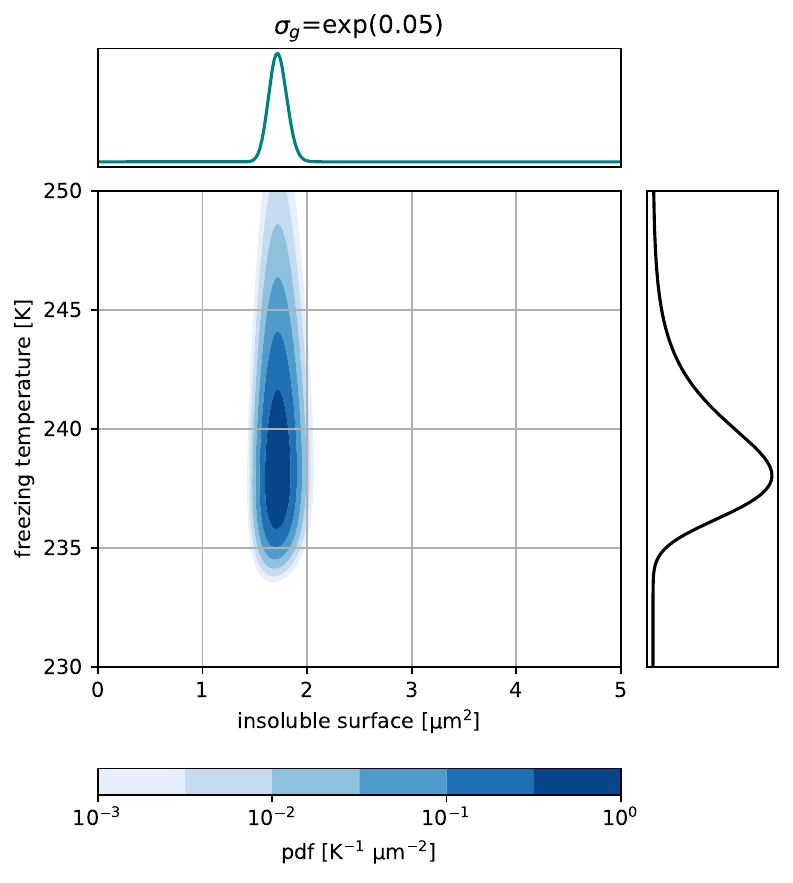}
    \end{adjustbox}\\
    \begin{adjustbox}{clip,trim=0cm 1.1cm 0cm 0cm}
      \includegraphics[width=2.255in]{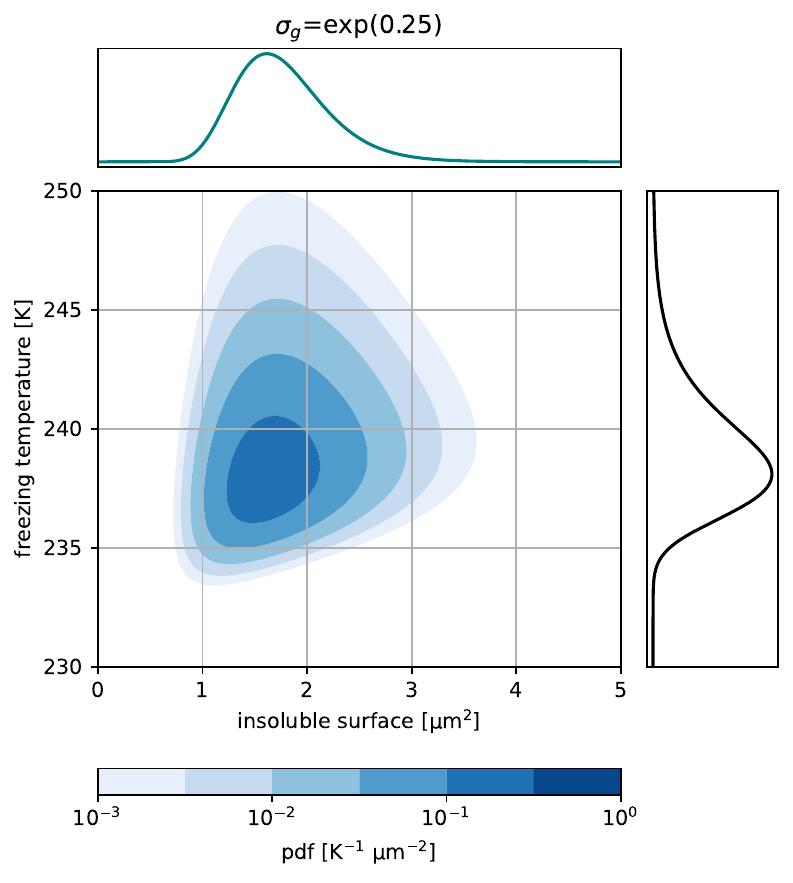}
    \end{adjustbox}\\
    \includegraphics[width=2.255in]{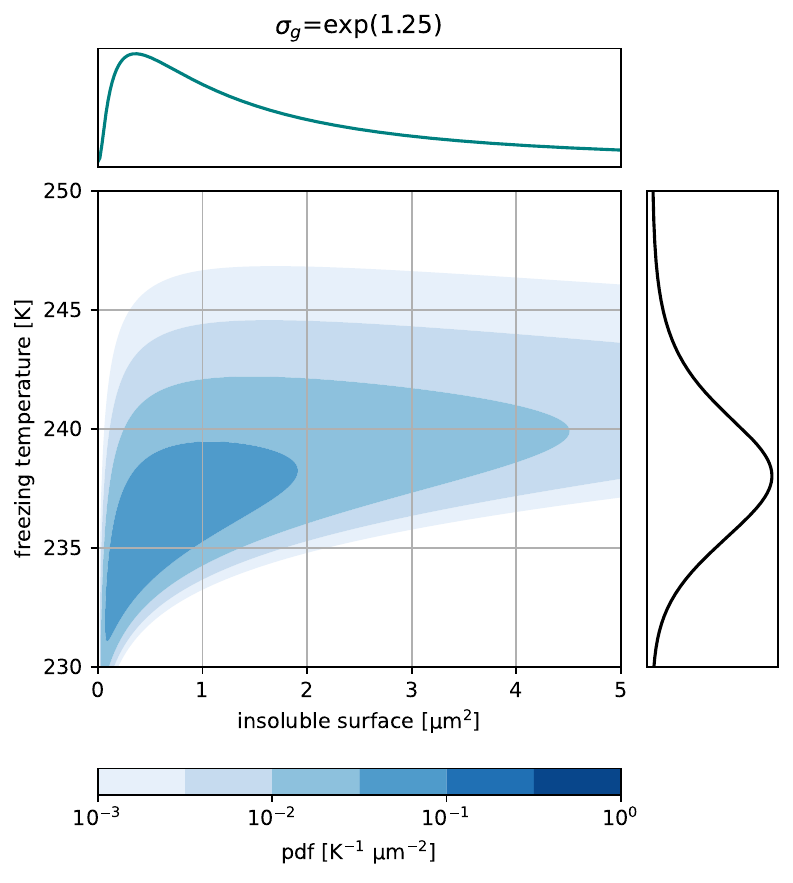}\\
  \end{center}
  \vspace{-1em}
  \caption{\label{fig:0d_pdf} 
    Two-dimensional probability densities as a function of immersed insoluble surface area $S$ and freezing temperature $T$, $p(S, T)$, 
      used for sampling the initial conditions for the box model simulations presented in Section~\ref{sec:width}, using geometric 
      standard deviation $\sigma_{\rm g} = \exp(0.05)$ (top), $\sigma_{\rm g} = \exp(0.25)$ (middle), and $\sigma_{\rm g} = \exp(1.25)$ (bottom). 
    The marginals with respect to $S$ and $T$ are shown above and on the right, respectively.
  }
  \vspace{-5em}
\end{figure}

For each temperature profile (discussed in sec.~\ref{sec:cool}) or each aerosol spectrum shape (sec.~\ref{sec:width}), a set of multiple simulations is performed---an ensemble of runs for the singular
  model and a second ensemble for the time-dependent model, with each member of the ensemble 
  using a different random number generator seed.
The sampling procedure for the initial conditions is as follows.
Every simulation uses super-particles of equal multiplicity each (for discussion of other approaches, see \ref{sec:Lagrangian}).
The value of multiplicity is determined by assigning each super-particle a consecutive quantile of the dry aerosol mass distribution.
To match the equal-multiplicity attribute-space sampling with freezing-related attributes, in the case of the time-dependent simulations, the values of the immersed surface attribute
  are initialized by inverting the cumulative lognormal distribution, and randomly sampling
  the resultant quantile function.
In the case of the singular simulations, the immersed surface areas are sampled in the same
  way, however, they are not used as particle attributes; instead the immersed surface area values
  are used to evaluate the cumulative freezing probability as a function of temperature,
  and this cumulative distribution is inverted into a quantile function $Q$ used for random sampling
  \cite<see discussion of eq.~1 in>{Shima_et_al_2020}:
\begin{equation}
  \left. Q_\textnormal{\scriptsize INAS}(p) \right|_S = 
      T_0 + \frac{1}{a_\textnormal{\scriptsize INAS}}\left[
        \ln\left(
            \frac{
                \ln(1 - p)
                + \exp(
                    -S n_0 \exp(-a_\textnormal{\scriptsize INAS} T_0 + b_\textnormal{\scriptsize INAS})
                )
            }{-S n_0}
        \right) - b_\textnormal{\scriptsize INAS}\right]
\end{equation}
  where $p \in [0,1]$.
Effectively, the above procedure corresponds to sampling a two-dimensional probability density function as in Figure~\ref{fig:0d_pdf} 
  (depicted for three different values of the geometric standard deviation) 
  and defined through the chain rule by:
\begin{equation}
    p_{(T,S)}(T, S) = p_{T|S}(T)\,\,p_S(S)\textrm{,}
\end{equation}  
  where $p_S(S)$ is a probability density function of the immersed insoluble surface (e.g., a lognormal distribution) and $p_{T|S}$ is the conditional probability 
  labeled in eq.~(\ref{eq:inas_pdf}) as $\left. p(T)\right|_S$.
Noteworthily, despite the singular scheme being associated with determinism \cite<e.g.>{Vali_et_al_2015}, 
  the hereby embraced particle-based formulation involves random sampling at initialization and thus leads to different
  realization in each simulation (see~Fig.~\ref{fig:sampling}). 

From a statistics perspective, Figure~\ref{fig:0d_pdf} presents dependency among two variables which can in turn 
  be described by two marginal distributions (plotted in Fig.~\ref{fig:0d_pdf} with teal and black curves) and a so-called {\em copula}, the latter containing solely
  the information on the dependency structure \cite<see, e.g.,>{Schoelzel_and_Friederichs_2008}.
From a cloud physics perspective, the figure depicts a two-dimensional probability density of which marginal
  distributions are commonly measurable quantities, namely: the size spectrum (per surface) and
  the freezing spectrum (per temperature).
In our example, both the INAS parameterization parameters as well as the parameter of 
  the lognormal size spectrum are based on laboratory measurements, while the dependency 
  structure is defined by the $n_\textnormal{\scriptsize X}$ function.

\subsection{Response of simulated frozen fraction to ambient cooling rates}\label{sec:cool}

\begin{figure*}[t]
  \begin{center}
    \includegraphics[width=5in]{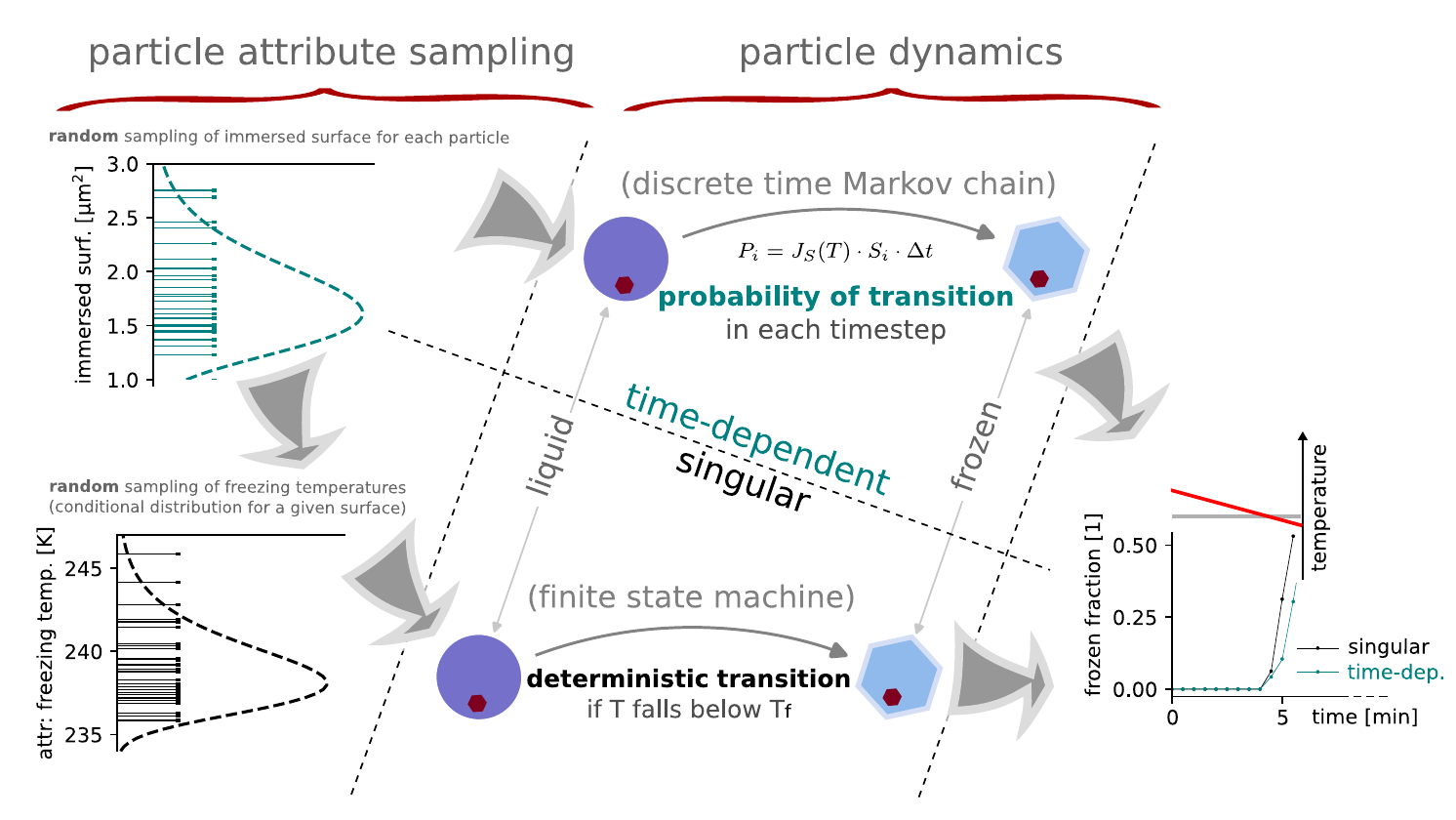}
  \end{center}
  \caption{\label{fig:sampling}
    Schematic of operation of time-dependent (top) and singular (bottom) immersion freezing models.
    Left: particle attribute sampling strategy at initialization.
    Middle: evolving particle dynamics of super-particles during the simulations.
    Right: aggregation of results as frozen fractions as a function of time.
    Gray line on the temperature plot indicates ice melting point.
  }
  \vspace{-3em}
\end{figure*}

In order to highlight the differences between the singular INAS-based and the time-dependent ABIFM-based schemes,
  here we present a set of simulations in which the two models are driven by a set of diverse
  idealized temperature profiles, with all other settings unchanged \cite<approach akin to the analysis depicted in Fig.~3 in>{Kaercher_and_Marcolli_2021}.
The particle attribute sampling, particle dynamics and analysis logic are schematically
  illustrated in Fig.~\ref{fig:sampling}. 
The leftmost section outlines sampling of immersed surface areas for the time-dependent scheme, and
  of the freezing temperatures for the singular scheme.
The middle section depicts how freezing is triggered: using a path-independent probability of transition
  evaluated in each timestep for the time-dependent scheme (as in a discrete-time Markov chain),
  or using a deterministic transition for the singular scheme (as in a finite state machine).
The right section illustrates that the same procedure is used for both schemes to 
  derive the fraction of frozen particles.

\begin{figure*}[t]
  \begin{center}
    \includegraphics[width=.98\textwidth]{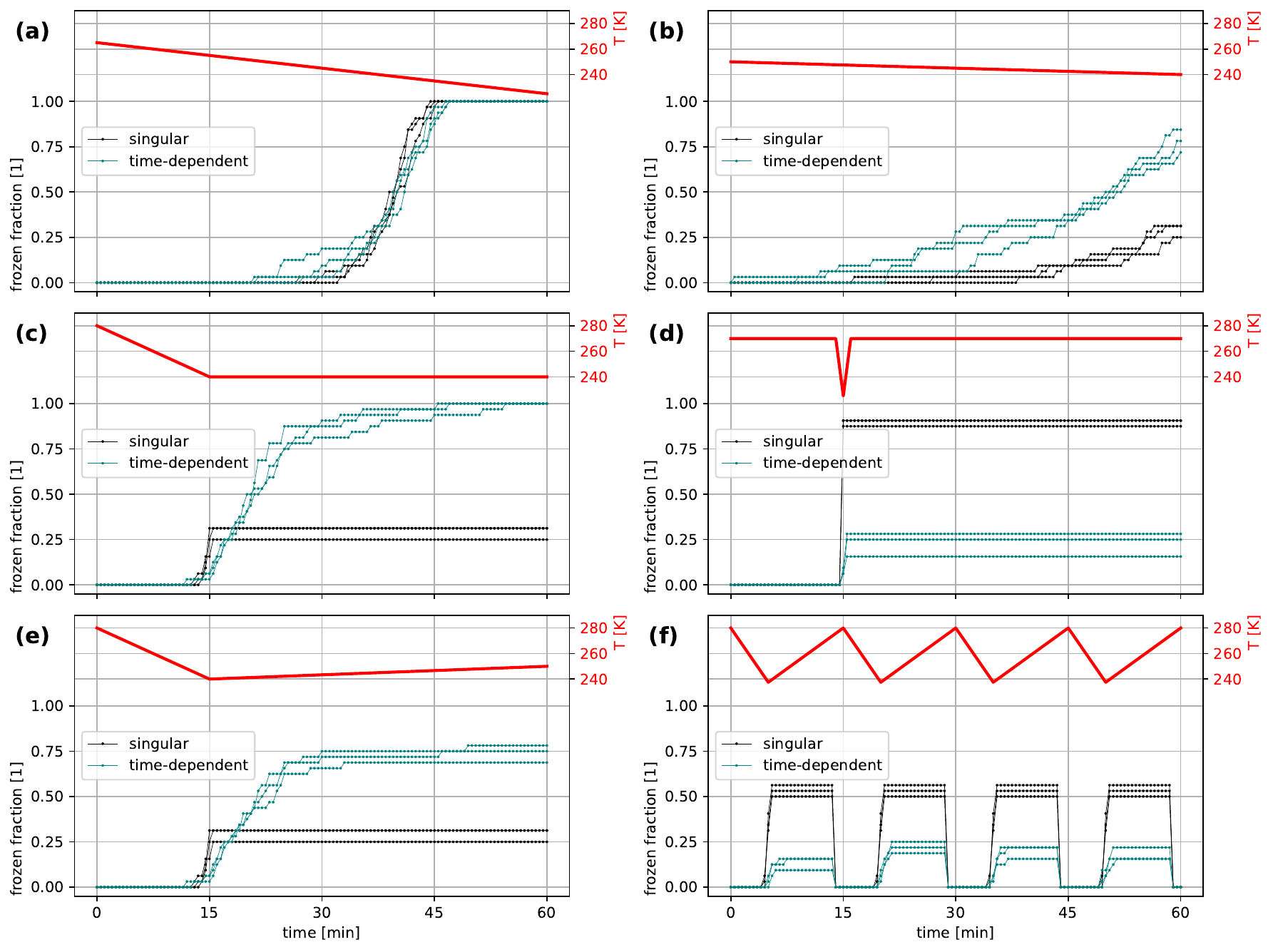}
  \end{center}
  \vspace{-2em}
  \caption{\label{fig:thought_experiments} 
    Temporal evolution of prescribed temperature and resulting frozen fractions calculated using the singular scheme 
      and the time-dependent scheme.
    Three realizations are shown for each case.
    (a) Linear temperature decrease, with temperature gradient comparable to the experimental conditions that were used to derive the parameterizations. 
    (b) Linear temperature decrease, temperature gradient lower than in case (a). 
    (c) Temperature drop followed by constant temperature. 
    (d) Impulse-like temperature drop. 
    (e) Linear temperature decrease followed by linear temperature increase. 
    (f) Repeated freezing cycles.
  }
\end{figure*}

The results are depicted in Figure~\ref{fig:thought_experiments} and
  are based on simulations performed with 32 super-particles, each simulation covering a time period
  of one hour divided into 120 timesteps. 
The immersed insoluble surface area is sampled from a lognormal spectrum (see Sect.~\ref{sec:aero_params} for parameters). 
The ensemble size is set to three realizations per model. 
To emphasize differences in the temporal evolution of the
  systems depending on the model choice, we present the simulated frozen fraction as a function of time
  rather than as a function of temperature.
Note that immersion freezing is the only process modeled here; while in a natural environment, the frozen 
  fraction evolution would be governed by an interplay of immersion freezing with other processes (particle sedimentation, diffusional growth).

Panel (a) in Fig.~\ref{fig:thought_experiments} features a linear temperature gradient of $-\frac{2}{3}\textrm{~K/min}$
  which is comparable to the cooling rate in the AIDA chamber experiment on which both
  INAS and ABIFM parameterization fits were based \cite[Fig.~2]{Niemand_et_al_2012}.
The temperature profile is given by the red line with axis markings on the right-hand side.
The teal and black line-connected points correspond to the time-dependent and singular simulations, respectively,
  and are plotted as frozen fraction vs. time.
The results from the two models match well, with the discrepancies being comparable in magnitude 
  to the differences between realizations for a particular scheme.

Panel (b) in Fig.~\ref{fig:thought_experiments} complements the analysis using a lower magnitude of cooling rate with $c=-\frac{1}{6}\textrm{~K/min}$.
Here, the discrepancy between singular and time-dependent simulation is pronounced and clearly larger than the
  inter-realization spread.
The singular scheme triggers fewer freezing events compared to the time-dependent one.
Compared to the results in panel~(a), as expected, the singular scheme reaches the same frozen fraction at
  $240\,\textrm{K}$ in both panels---regardless of the ambient cooling rate, while the time-dependent scheme is sensitive
  to the temperature time series.

Panel (c) in Fig.~\ref{fig:thought_experiments} depicts how the two schemes respond to a steep temperature 
  drop with a temperature gradient of $-\frac{8}{3}\textrm{~K/min}$ followed by a constant-temperature leg.
First, complementing the results plotted in panel~(b), during the initial steep temperature 
  drop, the singular scheme triggers more freezing than the time-dependent one.
Second, during the constant-temperature leg, the two models differ qualitatively as the singular scheme
  yields no freezing at all and the time-dependent scheme continues to trigger freezing up until
  the frozen fraction reaches unity.

The scenario depicted in panel~(d) in Fig.~\ref{fig:thought_experiments} is constructed with a twofold aim.
First, it confirms that during constant-temperature legs at the temperature of $270\textrm{~K}$ where
  freezing is unlikely, neither of the schemes triggers any increase in the frozen fraction.
Second, it can be observed that the impulse-like drop in temperature around $t=15\textrm{~min}$ 
  leads to almost four-fold higher frozen fraction in the singular simulation than in the time-dependent one.
The behavior depicted in the latter corresponds to a kinetic limitation, connected with the nucleation rate $J_\textnormal{\scriptsize het}$
  which limits the amount of ice produced, even though a low temperature is reached for a short instance (with the length of this instance
  multiplying $J_\textnormal{\scriptsize het}$, see eqs.~(\ref{eq:lnpm2=int}) and (\ref{eq:delta_t})).

\begin{figure*}[t]
  \begin{center}
    \includegraphics[width=5.5in]{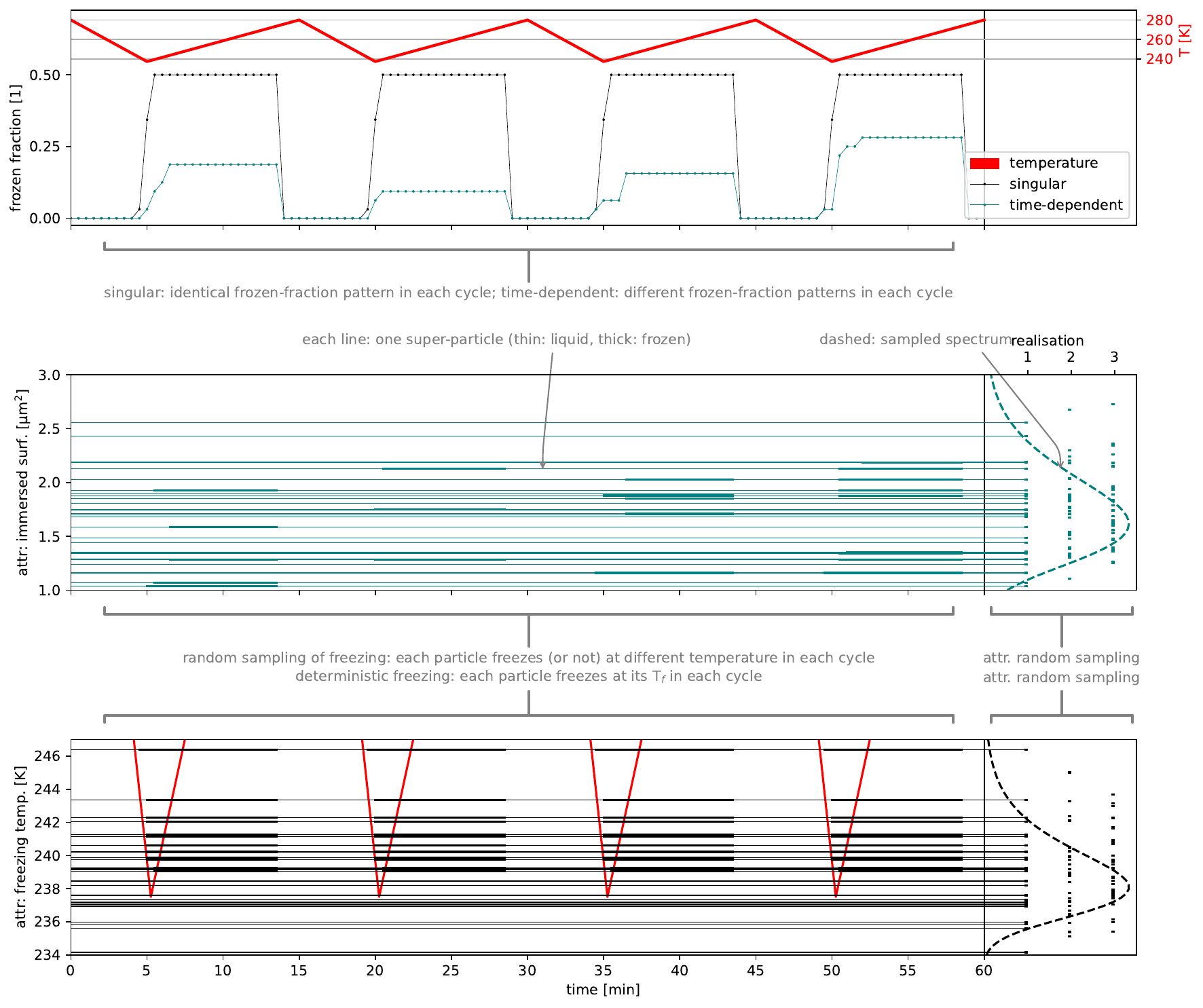}
  \end{center}
  \caption{\label{fig:realisations}
    Detailed view of one singular and one time-dependent realizations depicted in panel~(f) of Fig.~\ref{fig:thought_experiments}.
    Top: Prescribed temperature and calculated frozen fractions for repeated freezing cycles using the singular scheme and the time-dependent scheme,
      showing one realization for each case. 
    Middle: Evolution of the attribute immersed surface area for the time-dependent scheme.
    Bottom: Evolution of the attribute freezing temperature for the singular scheme (with temperature profile in red). 
    Each line represents one super particle with the thick line representing a frozen state and the thin line representing a liquid state.
  }
  \vspace{-3em}
\end{figure*}

Panel (e) in Fig.~\ref{fig:thought_experiments} presents results from a simulation in which a sharp initial
  temperature drop as in panel~(c) is followed by a steady slow increase in temperature.
For the INAS-based scheme, this scenario does not differ from the one in panel~(c) as freezing can only
  be triggered in singular schemes while the temperature gradient is negative.
For the ABIFM-based time-dependent scheme, there is no such limitation, and while the droplets remain supercooled and the temperature is
  low enough for the freezing probability to be non-negligible, a freezing-while-warming 
  behavior is observed (see also top panel in Fig.~\ref{fig:realisations}).
Note this is contrary to typical paradigms. 
For example, CCN activation in a downdraft is not expected to occur.
This reflects one of the fundamental differences which makes prediction of ice formation difficult.

Panel (f) in Fig.~\ref{fig:thought_experiments}, complements the analysis with a scenario composed of 
  repeated freezing cycles \cite<in a way in the spirit of refreezing studies of INP as in, e.g.,>{Fornea_et_al_2009,Wright_et_al_2013,Kaufmann_et_al_2017}.
As in panel~(d), each sharp drop in temperature leads the singular scheme to triggering much more
  freezing events than the time-dependent one.
As in panel~(e), freezing-while-warming can be observed only for the time-dependent scheme.
After the temperature reaches back to $0^\circ\textrm{C}$, instantaneous melting is
  happening (for simplicity, as in \citeA{Shima_et_al_2020}, the phase change back into 
  liquid state is represented without any inertia).
Subsequent freezing cycles follow exactly the same pattern for the singular scheme, while the time-dependent
  simulation features different realizations in each cycle. 
This depicts where the stochastic nature of immersion freezing proceeds: for the singular scheme, only
  the initial sampling of nucleus sizes is probabilistic, while the time-dependent scheme 
  additionally performs Monte-Carlo triggering in each time step.

\begin{figure}[H]
  \begin{center}
    {\bf (a)}$\!\!\!\!$
    \raisebox{-4.5cm}{
      \begin{adjustbox}{clip,trim=0cm 1.1cm 0cm 0cm}
        \includegraphics[width=2.42in]{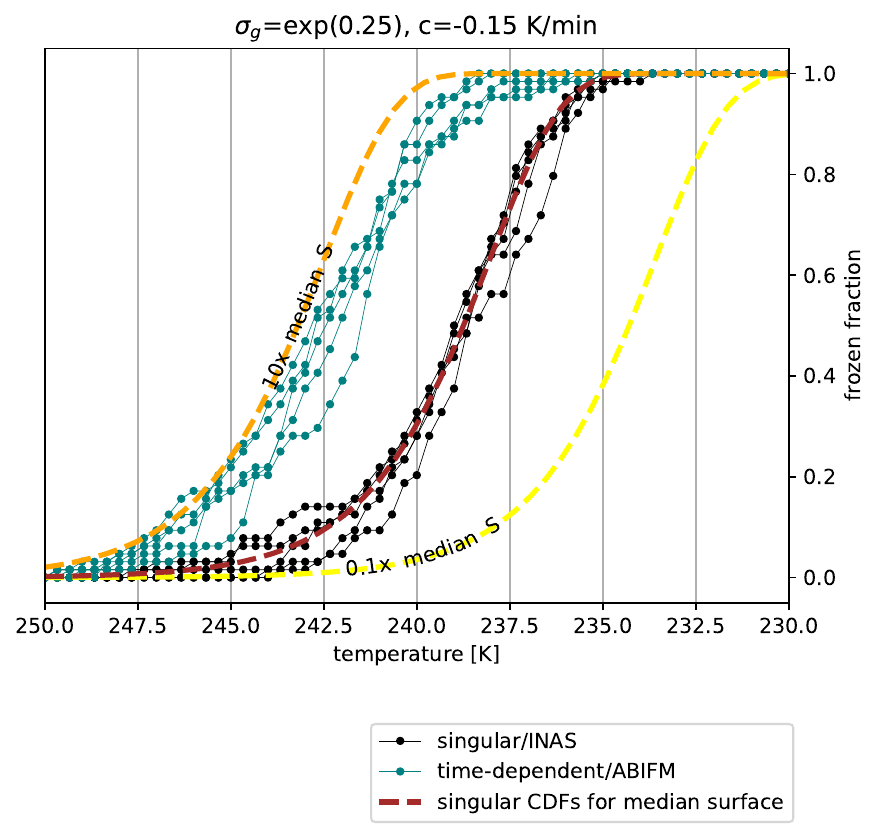}
      \end{adjustbox}
    }\\
    {\bf (b)}$\!\!\!\!$
    \raisebox{-4.5cm}{
      \begin{adjustbox}{clip,trim=0cm 1.1cm 0cm 0cm}
        \includegraphics[width=2.42in]{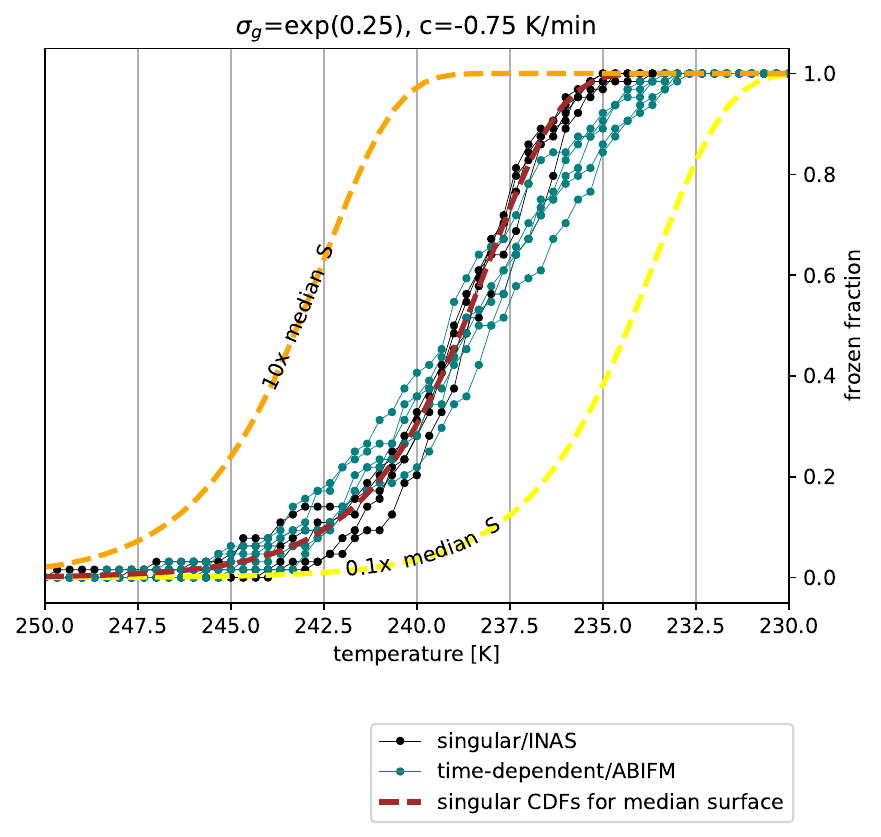}
      \end{adjustbox}
    }\\
    {\bf (c)}$\!\!\!\!$
    \raisebox{-5.7cm}{
      \includegraphics[width=2.42in]{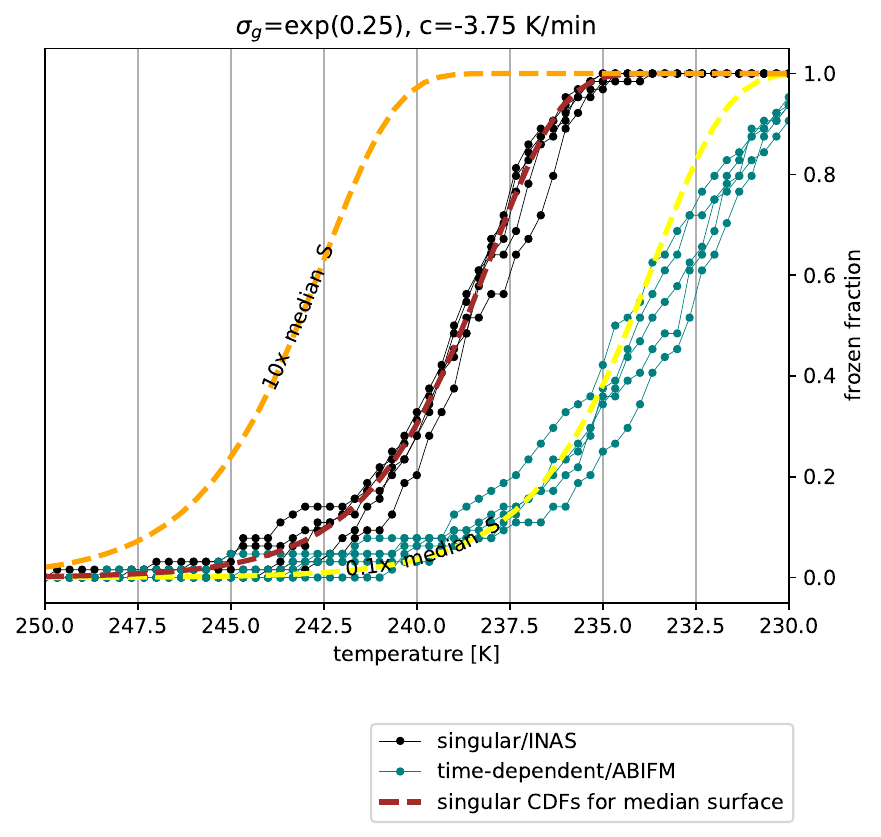}
    }
  \end{center}
  \caption{\label{fig:0d_box_cooling_rate}
      Frozen fraction as a function of
      temperature comparing simulations using singular (INAS, black
      markers) and time-dependent (ABIFM, teal symbols)
      parameterizations. For reference, the broken lines indicate the
      analytic cumulative distribution functions corresponding to
      monodisperse particle populations with surface areas equal to
      the median (red), 10 times the median (burgundy) and 1/10 of the
      median (yellow). 
      (a) cooling rate of -0.15 K/min, 
      (b) cooling rate of -0.75 K/min,
      (c) cooling rate of -3.75 K/min.
   }
\end{figure}

The simulation presented in panel~(f) of Fig.~\ref{fig:thought_experiments} is presented in more detail in 
  Figure~\ref{fig:realisations}.
The top graph is plotted in the same manner as panel (f), while the middle and bottom graphs depict
  the evolution of the model state vector (particle attributes).
All three plots share the abscissa scale.
The ordinate scales correspond to freezing attribute values: immersed surface for the time-dependent 
  scheme (middle graph), and the freezing temperature for the singular one (bottom graph).
Whenever a~line turns from thin to thick, freezing occurs; melting is indicated by 
  thinning of a given line.
Data plotted in teal correspond to the state vector of the time-dependent simulation for which different
  super-particles freeze in each cycle (due to the random sampling procedure).
In contrast, the state vector of the singular simulation depicted with black lines
  features a repeating pattern in which the same super-particles freeze in the same order 
  in each cycle.
The probability density functions used for attribute sampling are shown with dashed lines in the right sub-panels of 
  both the middle and bottom graphs, and correspond to the marginal distributions plotted in Fig.~\ref{fig:0d_pdf}.
While the state vector evolution is shown for single realization per model only, the initial sampling for subsequent 
  realizations is presented with unconnected dots.

\subsection{Response of simulated frozen fraction to immersed surface spectrum}\label{sec:width}

\begin{figure}[t]
  \begin{center}
    {\bf (a)}$\!\!\!\!$
    \raisebox{-4.5cm}{
      \begin{adjustbox}{clip,trim=0cm 1.1cm 0cm 0cm}
        \includegraphics[width=2.42in]{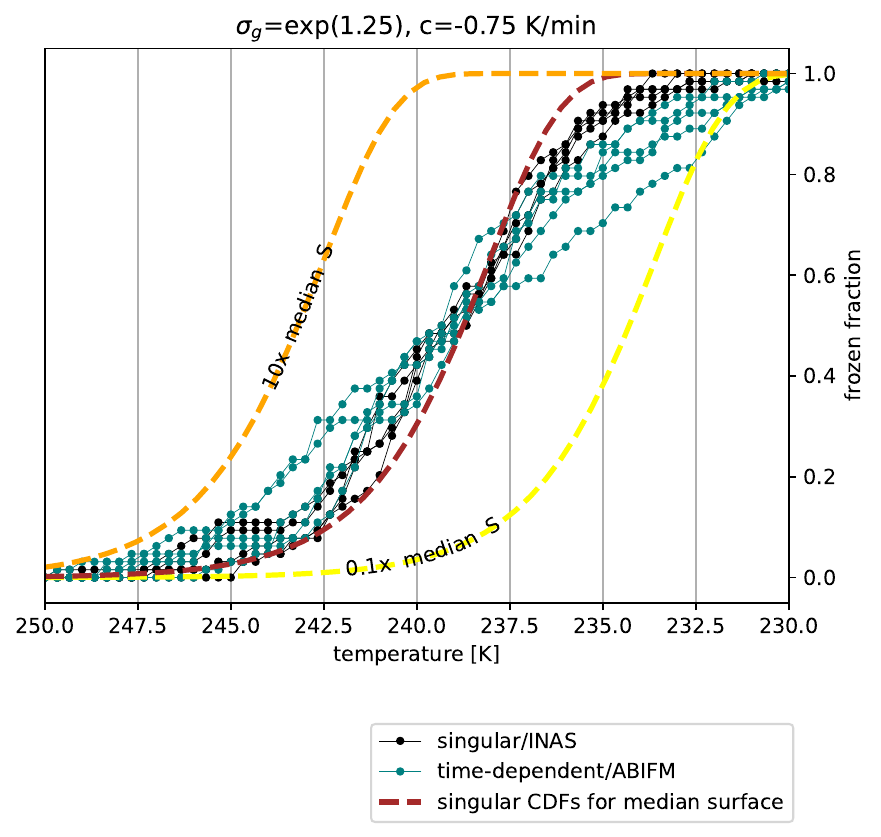}
      \end{adjustbox}
    }\\
    {\bf (b)}$\!\!\!\!$
    \raisebox{-5.7cm}{
      \includegraphics[width=2.42in]{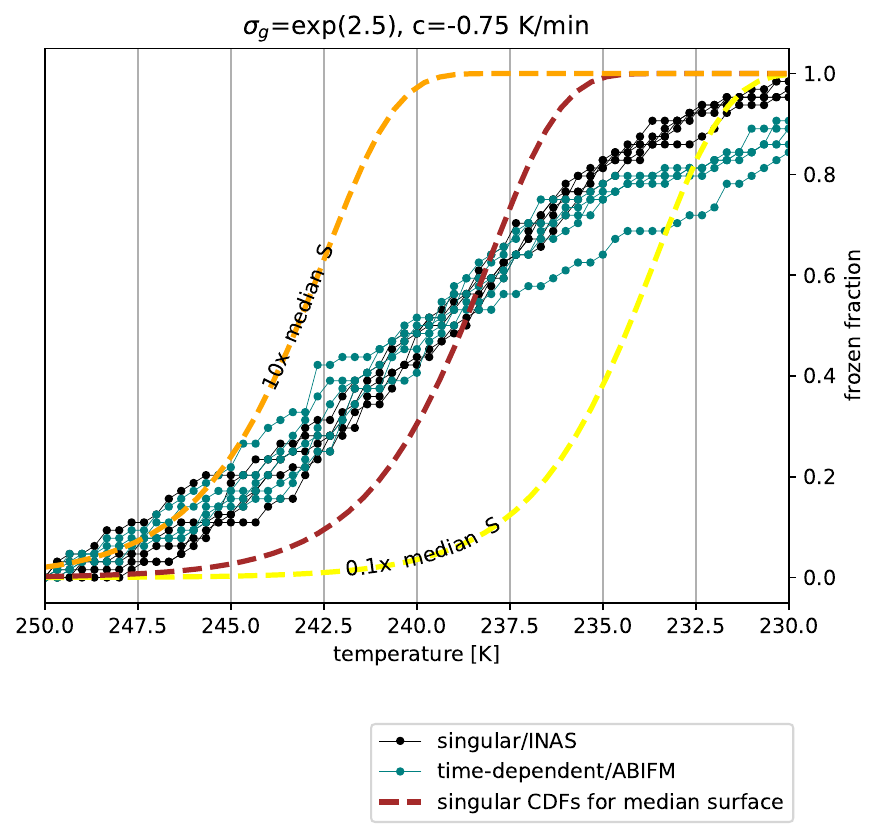}
    }
  \end{center}
  \caption{\label{fig:0d_box_sgeom} 
    Frozen fraction as a function of temperature comparing simulations using singular (INAS, black
      markers) and time-dependent (ABIFM, teal symbols) parameterizations.
    Diagrams constructed as in Fig.~\ref{fig:0d_box_cooling_rate}, see caption for details.    
    Panel (a): geometric standard deviation of the immersed nuclei spectrum $\sigma_{\rm g} = \exp(1.25)$.
    Panel (b): $\sigma_{\rm g} = \exp(2.5)$.
  }
\end{figure}

After exploring the impact of cooling rate on the frozen fraction evolution, we now extend the discussion to cover also 
  the impact of the surface area
  of the immersed particles (Fig.~\ref{fig:0d_box_cooling_rate}) and its spectrum width (Fig.~\ref{fig:0d_box_sgeom}).
In Figure~\ref{fig:0d_box_cooling_rate}, two sets of simulation are presented for 
  three different cooling rates: $c=-0.15$~K/min, $c=-0.75$~K/min and $c=-3.75$~K/min.
The lognormal immersed surface area distribution has a width defined by $\sigma_\textnormal{\scriptsize g}=\exp(1.5)$.
The simulations are carried out with 64 super-particles,
  five ensemble members are plotted.
The panels depict the fraction of frozen particles as a function of the temperature.
Monte-Carlo simulation results are plotted with connected dots: teal for the time-dependent
  simulations and black for simulations using the singular scheme.
The spread across different realizations originates from the probabilistic treatment.
In addition, analytic cumulative distribution functions corresponding to monodisperse
  particle populations 
  are plotted with dashed lines for three
  values of the median surface diameter: $0.074\,\upmu\textrm{m}^2$ (yellow, $0.1\times$ label), 
  $0.74\,\upmu\textrm{m}^2$ (brown) and $7.4\,\upmu\textrm{m}^2$ (orange, $10\times$ label).

When plotted as a function of temperature, the frozen-fraction
  profiles are insensitive to the cooling rate for the singular scheme
  (black connected dots) and match the INAS-derived cumulative
  probability distribution for the median surface (red broken line).
The time-dependent simulations (teal connected dots) match the theoretical INAS cumulative
  curve only for the case of $c=-0.75$~K/min (roughly corresponding to the AIDA chamber
  conditions used for the derivation of this parameterization).
This is in line with the remarks above
  pointing out the role of the cooling rate for the coefficients
  defining time-integrated quantities based on the Poissonian model of freezing events.
It is also in line with the behavior depicted in panels~(a)-(c) in Fig.~\ref{fig:thought_experiments} herein and
  in previous results \cite[Fig.~5 therein]{Alpert_and_Knopf_2016}.
The singular scheme yields lower ice concentrations than the time-dependent scheme
  for the slower cooling case, and higher ice concentrations for
  faster cooling. 
The behavior depicted in figure~\ref{fig:0d_box_cooling_rate} can be reconciled with the expected behavior of the
  two models in the limiting cases.
For an instantaneous temperature drop, the time-dependent scheme will yield negligible ice concentrations (infinitely lower ice concentrations than the singular scheme).
In the opposite limit, for a constant temperature in the supercooled regime though above the freezing temperatures of all droplets, the singular scheme would not
  yield any ice and hence would have infinitely lower concentrations than the 
  time-dependent simulation.
  
Comparing the results obtained for the three different cooling rates plotted in Figure~\ref{fig:0d_box_cooling_rate}, it is evident that for the presented
  case, a discrepancy associated with a five-fold increase/decrease in the cooling rate is roughly 
  comparable to that originating from a tenfold change in the median immersed surface area.
The 25-fold decrease in the magnitude of cooling rate corresponding to the change from $c=-3.75$~K/min (panel c) to $c=-0.15$~K/min (panel a) results in over 5K shift of the
  frozen fraction curve, i.e., over 2K shift for a tenfold decrease (\citeauthor<cf.>[sect. 2~(iii)]{Bigg_1953a}). 
  
Figure~\ref{fig:0d_box_sgeom} presents two sets of simulations (five singular and
  five time-dependent) for the following values of the immersed surface spectrum
  width expressed through the geometric standard deviation $\sigma_\textnormal{\scriptsize g}$:
  $\exp(1.25)$ and $\exp(2.5)$ (i.e., increasingly larger than in simulations summarized in Fig.~\ref{fig:0d_box_cooling_rate} where $\sigma_\textnormal{\scriptsize g}=\exp(0.25)$ was used). 
In all simulations, the cooling rate is set to $c=-0.75$~K/min, which is the value
  corresponding to the best fit across INAS and ABIFM parameterizations depicted in section~\ref{sec:embed}.
Hence, regardless of the geometric standard deviation value, the singular and the
  time-dependent simulations agree with each other. However, compared to
 the monodisperse analytic frozen-fraction profiles, increasing $\sigma_\textnormal{\scriptsize g}$ causes a flattening of the frozen fraction profiles.

The significance of the polydispersity of immersion freezing nuclei has been previously highlighted, e.g., in \citeA{Alpert_and_Knopf_2016} where
  analogous Monte-Carlo simulations were employed to quantify the bias in laboratory
  data analyses stemming from assuming monodispersity where actual spectra were
  polydisperse.
Noteworthily, the simulations presented in \citeA{Shima_et_al_2020} and \citeA{Abade_and_Albuquerque_2024}, using the singular particle-based scheme,
  were only performed with monodisperse spectra. 
Our analysis confirms that both the INAS- and ABIFM-based formulations explored herein are 
  capable of capturing the impacts of polydispersity of the immersed surface spectrum.

\section{2D prescribed-flow super-particle simulations}\label{sec:2D}

To extend the discussion beyond simple box-model considerations, we present an analysis of a set of idealized flow-coupled
  particle-based simulations.
The simulations are performed in two spatial dimensions (2D) and are driven by a prescribed flow field.
The goal is 
  to qualitatively explore the macroscopic impact of the choice of either the singular or the time-dependent immersion freezing model in a system characterized by a range of cooling rates;
  to depict challenges stemming from simultaneous simulation of aerosol-constrained CCN activation processes; and 
  to relate the preceding discussion with model resolution parameters characteristic of flow-coupled applications (such as LES).
We use a minimal framework consisting of 
  (i)~an Eulerian fluid-flow component solving for conservation of water vapor and heat in the domain and
  (ii)~a Lagrangian super-particle component solving for transport, diffusional growth/evaporation and immersion freezing of the particles.
Despite the prescribed-flow simplification, the two components are bidirectionally coupled:
  (i)~particle transport is driven by the fluid flow, diffusional growth, evaporation and immersion freezing is driven by ambient thermodynamic conditions;
  (ii)~vapor and heat budget solved by the Eulerian component feature sink/source terms representing uptake/release of heat and moisture by the particles.
Collisions, ice sedimentation and ice diffusional growth are not represented in the simulations. 
Unlike in preceding box-model simulations and unlike in analytically solvable setups, each super-particle is exposed to different cooling-rate history 
  which stems from random sampling of particle locations at initialisation (the flow is laminar, particle trajectories are not perturbed).

\begin{figure*}
  \includegraphics[width=\textwidth]{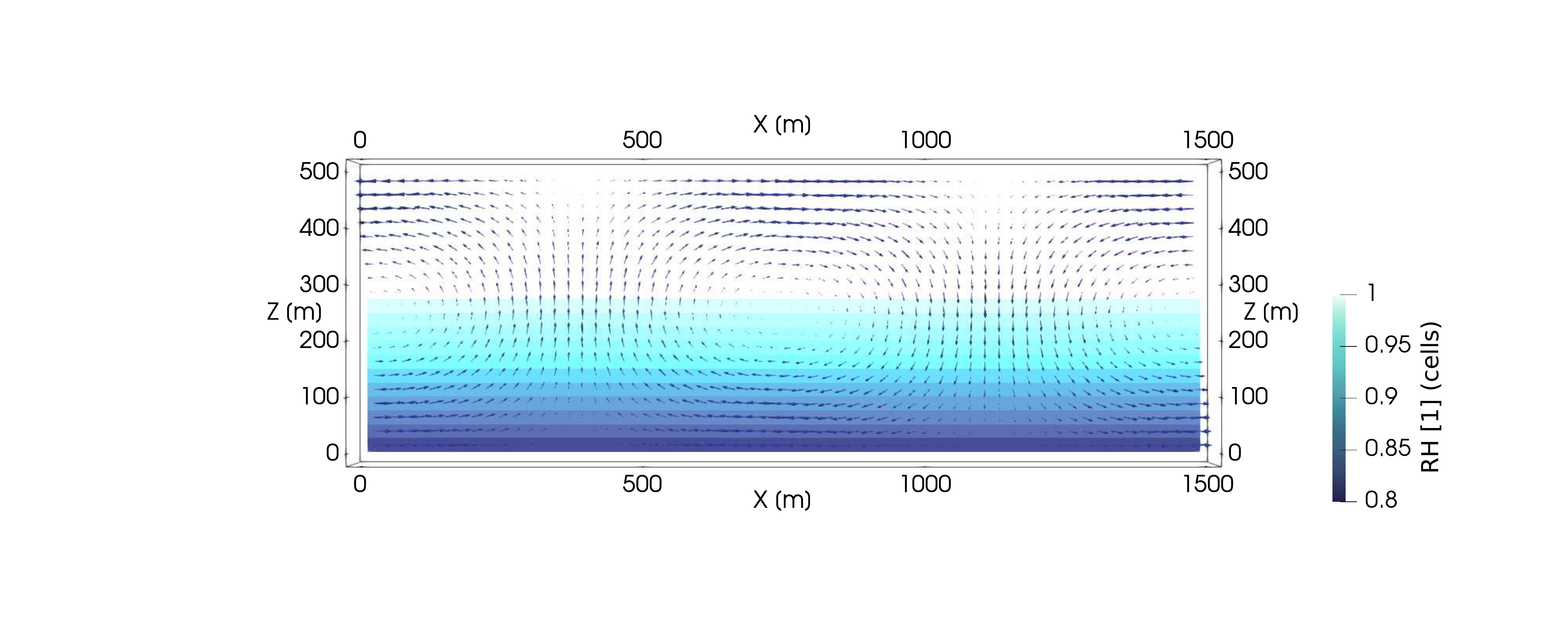}
  \vspace{-6em}
  \caption{\label{fig:arrows}
    Flow field pattern visualized with arrows depicting air velocity.
    Initial relative humidity field plotted with colored filled cells.
  }
  \vspace{-3em}
\end{figure*}

The concept of using such kinematic simulation framework for cloud microphysics schemes development
  can be traced back to the work of \citeA[section 3C]{Kessler_1969} and 
  was subsequently embraced in multiple studies \cite{Gedzelman_and_Arnold_1993,Gedzelman_and_Arnold_1994,Szumowski_et_al_1998,Grabowski_1998,Grabowski_1999,Morrison_and_Grabowski_2007,Slawinska_et_al_2009,Rasinski_et_al_2011,Szabo_2011,Lebo_and_Morrison_2013,Muhlbauer_et_al_2013,Sulia_et_al_2013,Arabas_et_al_2015,Yang_et_al_2015,Jaruga_and_Pawlowska_2018,Schmeller_and_Geresdi_2019}.
The setup employed here, which mimics a stratiform cloud deck and features
  periodic horizontal boundary condition and vanishing flow at vertical boundaries, was introduced in 
  \citeA{Morrison_and_Grabowski_2007} and later adopted for particle-based simulations
  in \citeA{Arabas_et_al_2015}.
Herein, we modify the setup parameters to roughly resemble thermodynamic and aerosol conditions
  of an Arctic mixed-phase cloud.
The dry-air potential temperature and water-vapor mixing ratio fields are initialized with 
  constant values throughout the domain: $\theta=(289-33.3)=255.7\textrm{ K}$ and $q_\textnormal{\scriptsize v}=(7.5-6.66)=0.84\textrm{ g/kg}$ 
  \cite<where the minuends are the values from the original setup of>[used for warm-rain simulations, and the subtrahends are arbitrarily 
  chosen for the relative humidity profile to roughly match]{Morrison_and_Grabowski_2007}.
The dry-air density profile is initialized by integrating the hydrostatic equilibrium
  equation for dry air.
These conditions result in water supersaturation in the upper part of the domain (demonstrated in Figure~\ref{fig:arrows})
  and supercooled conditions throughout the domain (hence melting never occurs in the simulation).
The setup parameters are further outlined below, for details on the simulation framework, see \ref{sec:Eulerian} and \ref{sec:Lagrangian}
  for Eulerian and Lagrangian components, respectively.

We note that in the case of dominant control of dynamics by the liquid-phase properties and radiative processes in Arctic stratiform clouds 
  \cite<e.g.,>{Silber_et_al_2020}, the main effect of ice is a relatively slow glaciation of coupled or decoupled layers as a whole via net sedimentation of     
  vapor-grown crystals \cite<e.g.,>{Morrison_et_al_2011,Fridlind_et_al_2012}. 
Thus, the approach introduced in \citeA{Morrison_and_Grabowski_2007} serves here to represent such typical quasi-steady states in Arctic     
  stratiform clouds with weak vertical motions $<$1~ms\textsuperscript{-1}.
Under such conditions, the strongest parcel-wise cooling rates will follow moist adiabats at cloud base.

The domain extents are set to $X=1500$~m and $Z=500$~m discretized on a 60$\times$20 grid ($\Delta x = \Delta z = 25\textrm{m}$).
The domain is populated with 32$\times$60$\times$20 super-particles with random spatial locations (i.e., 32 per grid cell on average,
  38400 of super-particles in total in the domain).

The aerosol composition, sizes and concentrations are specified as in the mixed-phase particle-based simulations reported in \citeA[sect.~6.1.3]{Shima_et_al_2020}.
The super-particles are split into two subpopulations: soluble particles with insoluble immersed surfaces ($1000$ per liter) and immersed-surface-free soluble particles 
  ($315$ per cubic centimeter).
Both subpopulations have the same soluble mass spectrum defined by lognormal dry-radius
  distribution with two modes having concentrations of $270\textrm{ cm}^{-3}$ and $45\textrm{ cm}^{-3}$,
  geometric mean radii of $0.03$~$\upmu$m and $0.14$~$\upmu$m and geometric standard deviations of $1.28$ and $1.75$, respectively.
The hygroscopicity parameter for the soluble masses is set to $\kappa=0.61$ corresponding to ammonium bisulfate \cite[Table~1]{Petters_and_Kreidenweis_2007}.
While the $1000$ per liter concentration of immersed-surface-rich particles matches the setup in \citeA{Shima_et_al_2020}, instead of a monodisperse population, 
  here a lognormal spectrum of immersed surfaces is used with the same geometric mean surface and geometric standard deviation as used for the box-model 
  simulations presented above \cite<i.e., corresponding to the surface of a sphere with diameter of $0.74$~$\upmu$m and with geometric standard deviation of $2.55$
  following the large mode of the "Desert Dust" spectrum used in>{Knopf_et_al_2023}.

All given concentrations are interpreted as corresponding to the standard atmosphere conditions at zero height (i.e., $T=15^\circ$C, $p=1013.25$~hPa and $\textrm{RH}=0$) labeled here as STP.
The actual volume concentrations have a vertical gradient due to the stratification of the 
  dry-air density.

The INAS fit parameters used to compute freezing temperatures for the singular scheme are taken
  from \citeA{Niemand_et_al_2012} and thus correspond to mineral dust.
Consistently, for the runs with the time-dependent scheme, the ABIFM parameters for mineral dust 
  are used (see sec.~\ref{sec:0D}).

In the present study, both subpopulations are set to have equal super-particle counts.
However, those subpopulations effectively represent contrasting concentrations of immersed-surface-rich 
  (out of which only a subset may take part in freezing) 
  and immersed-surface-free particles of $1$ and $315$ particles per cubic centimeter, respectively.
This results in multiplicities ranging from $1.2\times10^{6}$ to $1.7\times10^{10}$.
For comparison, in the Arctic-stratiform-cloud modeling study of \citeA{Fridlind_et_al_2012}, the concentration discrepancy between INPs and non-INP aerosol 
  was set at $1.7$ per liter (all taking part in freezing) vs. 352 per cubic centimeter.
The employed super-particle sampling strategy is thus  motivated by the challenge of resolving rare particles (and rare freezing events) with relatively low number of super-particles.
The concept is depicted in the schematic in Fig.~\ref{fig:schematic} and its implementation is outlined in \ref{sec:Lagrangian}.
Note that without employing the subpopulation split, taking an indicative LES grid cell volume of (100m)\textsuperscript{3},
  100 super-particles per grid cell, and employing common constant-multiplicity sampling of the an aerosol distribution
  with concentration of 1000~cm\textsuperscript{3} results in each super-particle having multiplicity of 10\textsuperscript{13}.
Such setup would result in most grid cells having no INP representation at all.

The simulations are performed with either singular or time-dependent representations
  of immersion freezing described in section~\ref{sec:models}.
For both schemes, freezing is set to be contingent on ambient vapor supersaturation,
  and applies to all particles, regardless of their wet size (i.e., both aerosol-sized
  and droplet-sized particles may freeze).
While the ABIFM scheme is capable of resolving the influence of dissolved soluble material
  on immersion freezing activity, and the particle-based representation
  of particle attributes can be set to track the relevant aerosol-composition and water-content
  parameters, for simplicity and for allowing comparison with INAS results, it is not taken into account.

With the main goal of the simulations being to explore and compare characteristics of the
  immersion freezing schemes, no other ice-phase processes are enabled.
Upon freezing, the super-particles start to act as tracers, vapor transfer to or from the
  ice surface is not represented, hence neither is the Wegener--Bergeron--Findeisen (WBF) process. 
Only condensation/evaporation-related latent heat exchange with the environment is simulated.

\begin{figure*}
  ~~~~~~~~~~~~~~~~~~~~~~~~~~~~~~~~~~~
  $t=600$~s 
  ~~~~~~~~~~~~~~~~~
  $t=1800$~s 
  ~~~~~~~~~~~~~
  $t=6000$~s \\
  {\bf (a)}$\!\!\!\!$
  \raisebox{-4.5cm}{
    \begin{adjustbox}{clip,trim=0cm .5cm 9.6cm 1.25cm}  % l b r t
      \includegraphics[width=6.1875in]{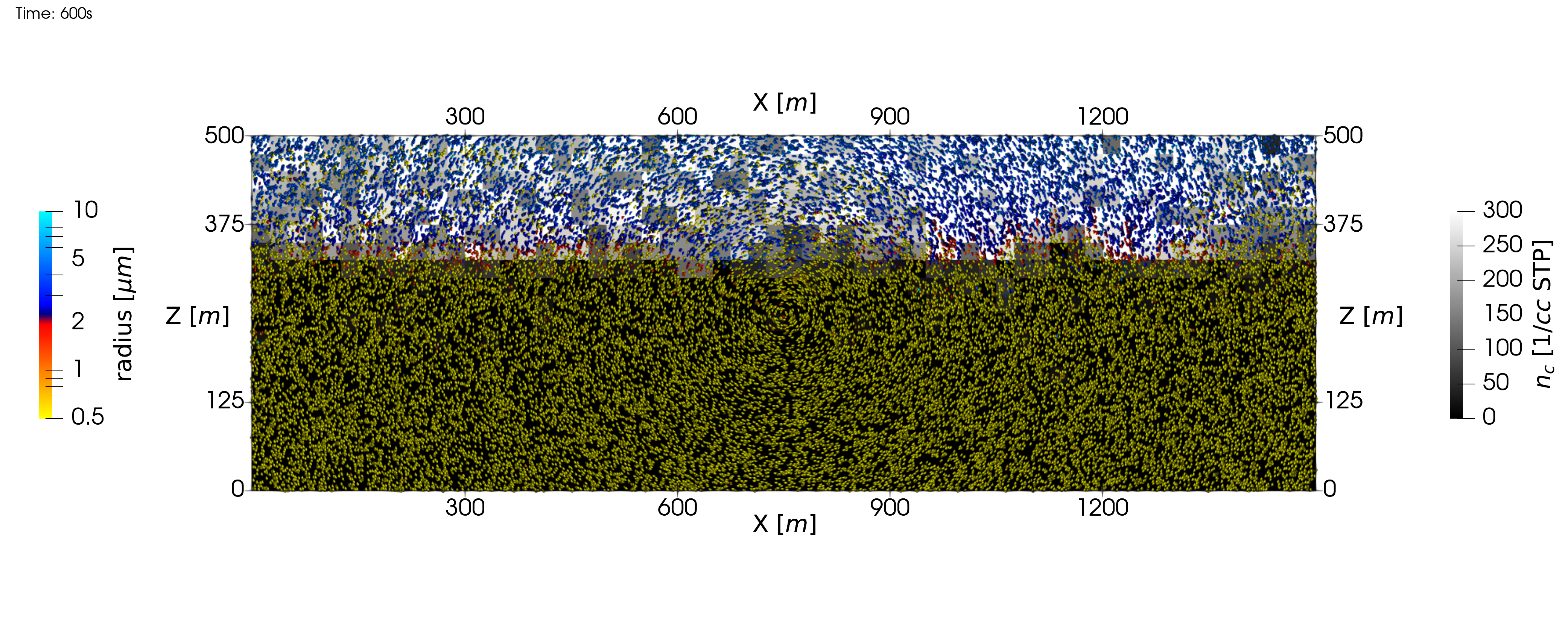}
    \end{adjustbox}
    \begin{adjustbox}{clip,trim=6.222cm .5cm 6cm 1.25cm}  % l b r t
      \includegraphics[width=6.1875in]{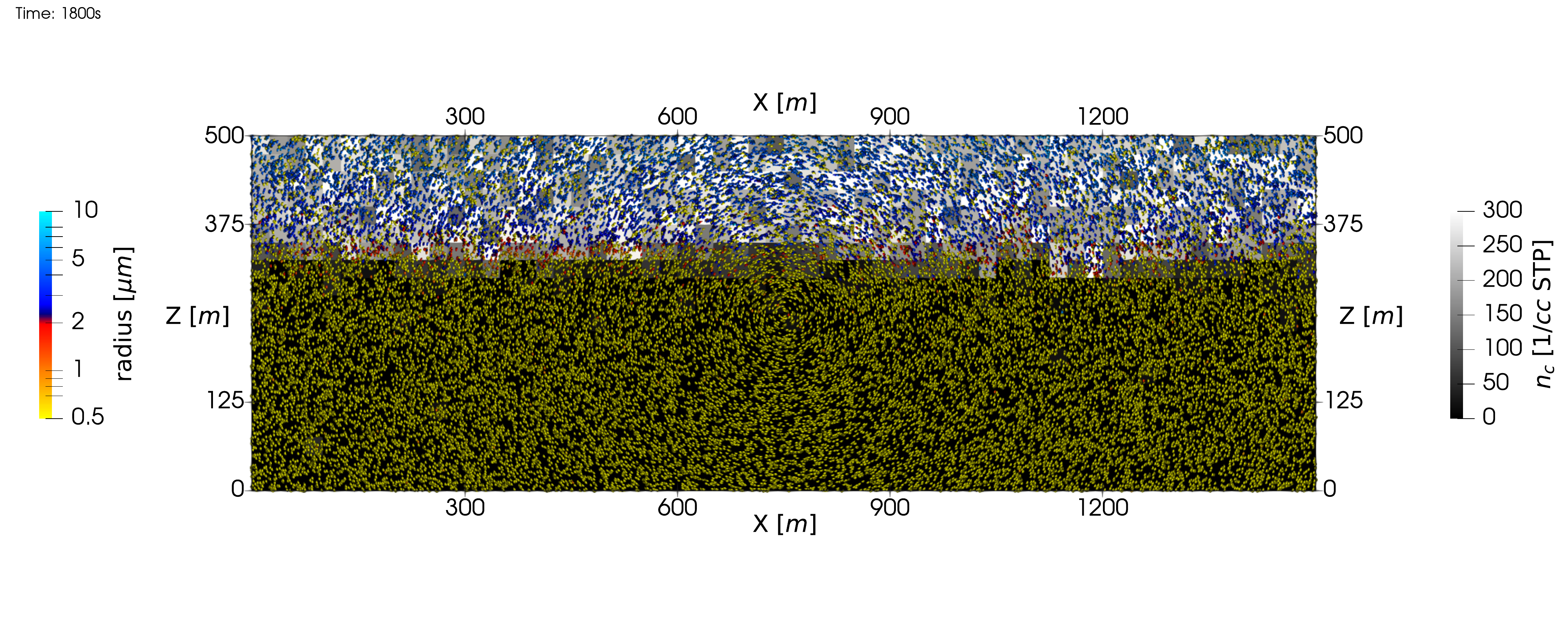}
    \end{adjustbox}
    \begin{adjustbox}{clip,trim=9.666cm .5cm 0cm 1.25cm}  % l b r t
      \includegraphics[width=6.1875in]{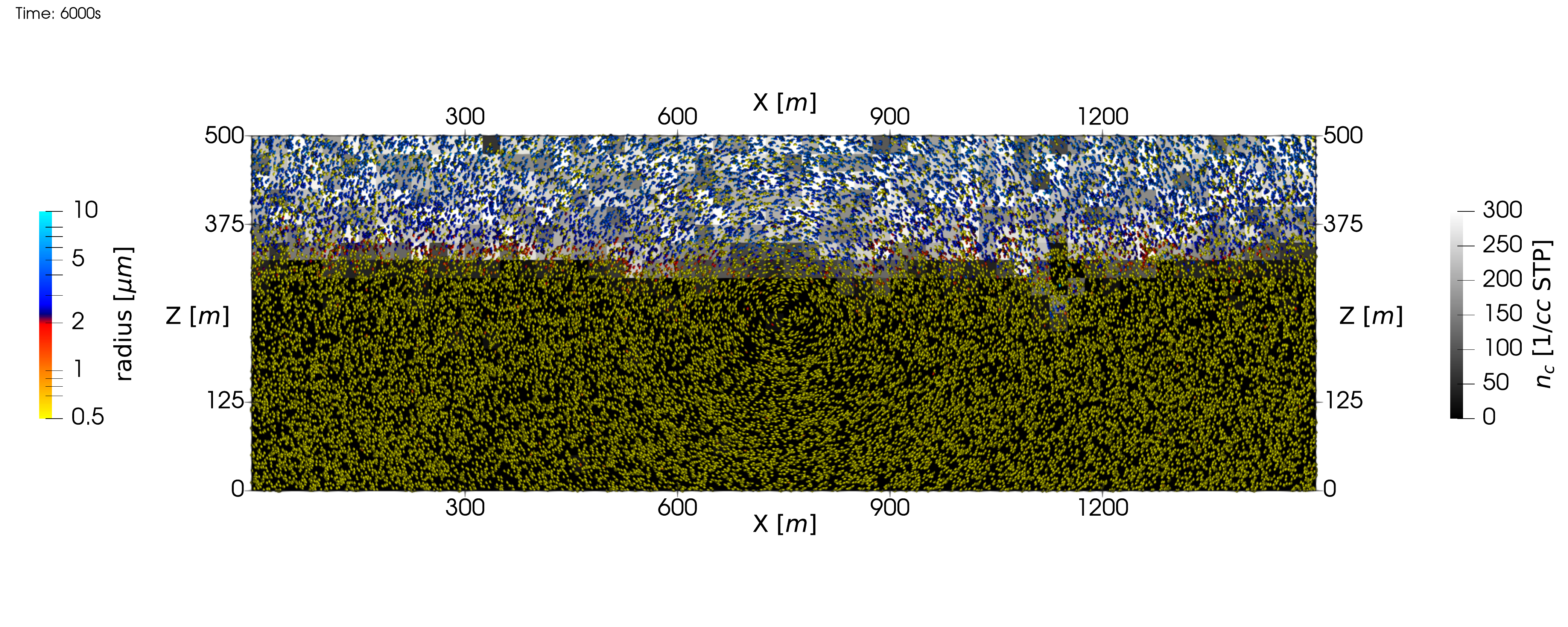}
    \end{adjustbox}
  }
  
  {\bf (b)}$\!\!\!\!$
  \raisebox{-4.5cm}{
    \begin{adjustbox}{clip,trim=0cm .5cm 9.6cm 1.25cm}  % l b r t
      \includegraphics[width=6.1875in]{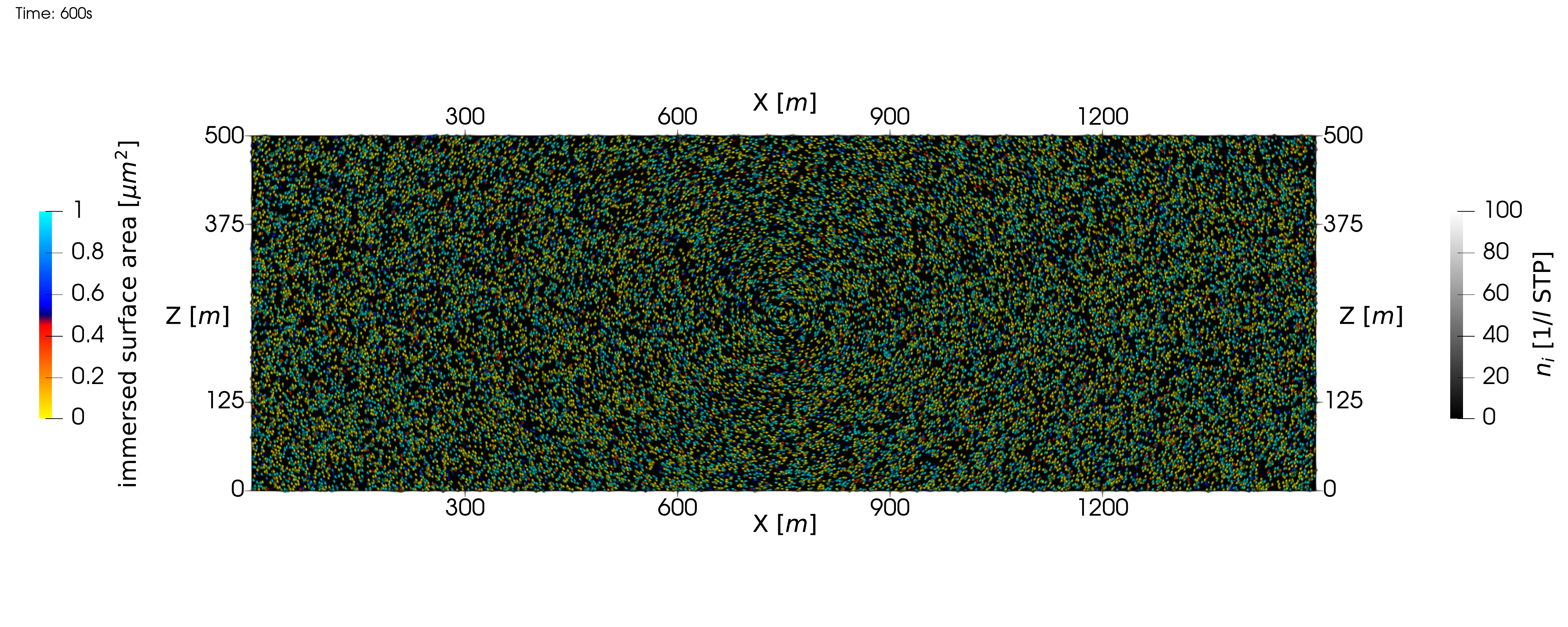}
    \end{adjustbox}
    \begin{adjustbox}{clip,trim=6.222cm .5cm 6cm 1.25cm}  % l b r t
      \includegraphics[width=6.1875in]{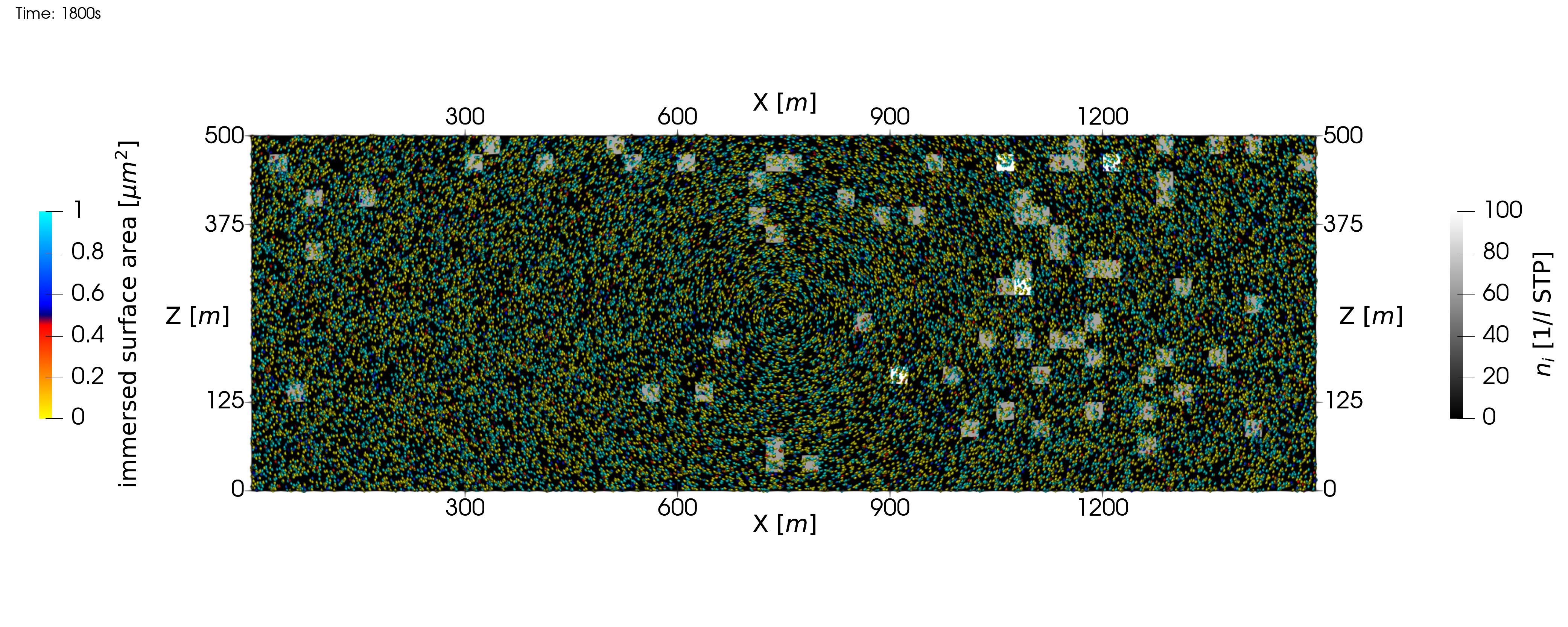}
    \end{adjustbox}
    \begin{adjustbox}{clip,trim=9.666cm .5cm 0cm 1.25cm}  % l b r t
      \includegraphics[width=6.1875in]{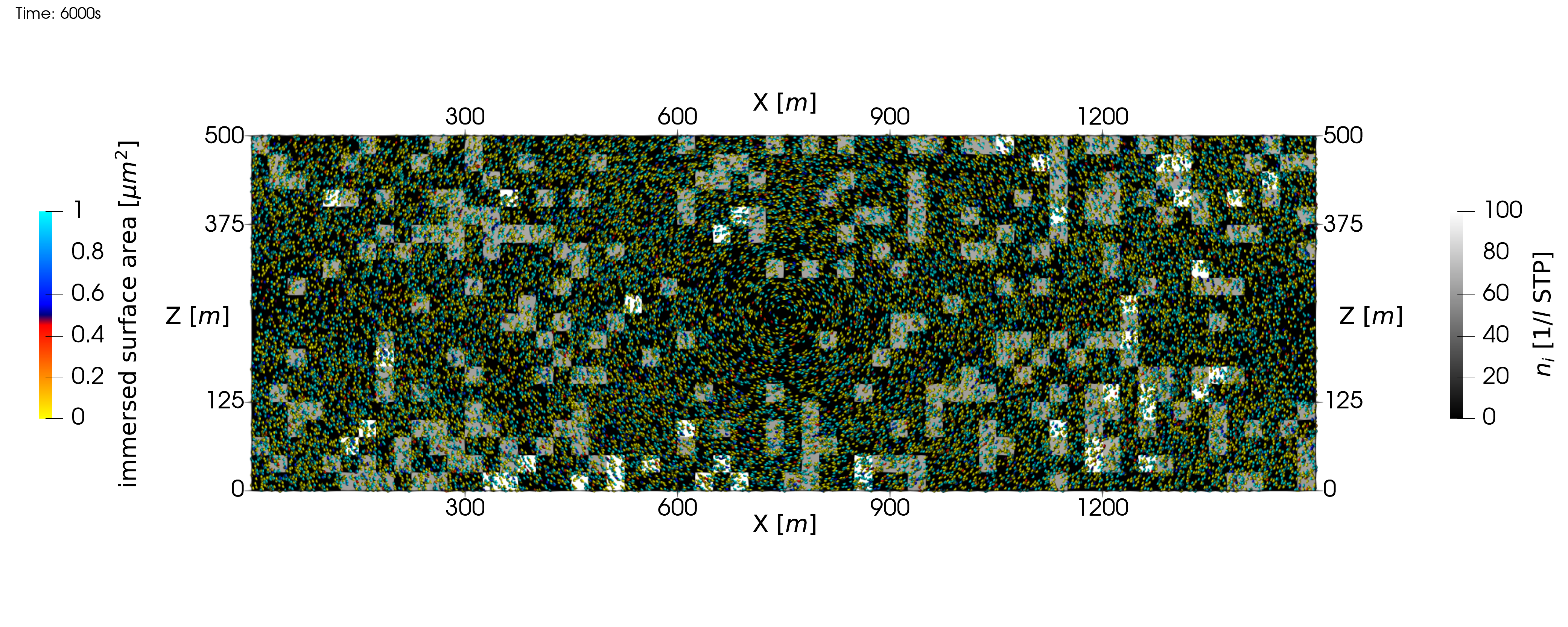}
    \end{adjustbox}
  }
  
  {\bf (c)}$\!\!\!\!$
  \raisebox{-4.5cm}{
    \begin{adjustbox}{clip,trim=0cm .5cm 9.6cm 1.25cm}  % l b r t
      \includegraphics[width=6.1875in]{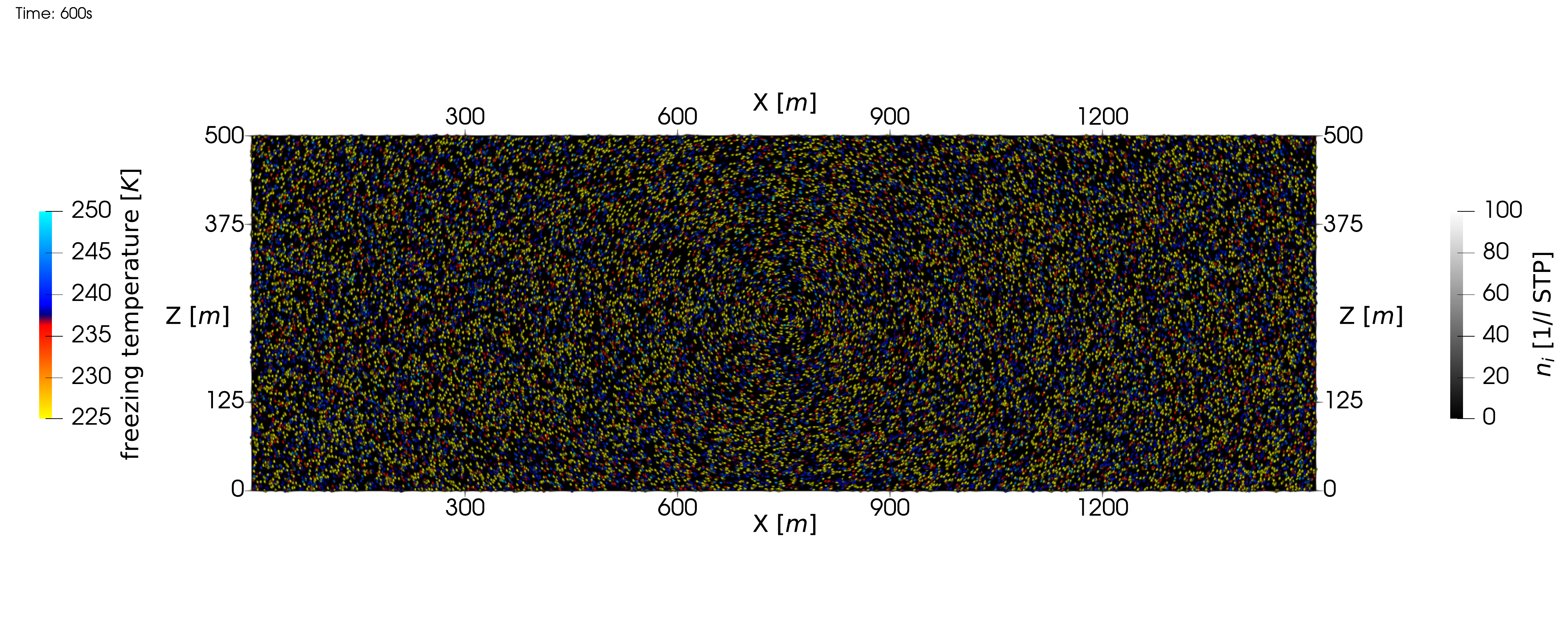}
    \end{adjustbox}
    \begin{adjustbox}{clip,trim=6.222cm .5cm 6cm 1.25cm}  % l b r t
      \includegraphics[width=6.1875in]{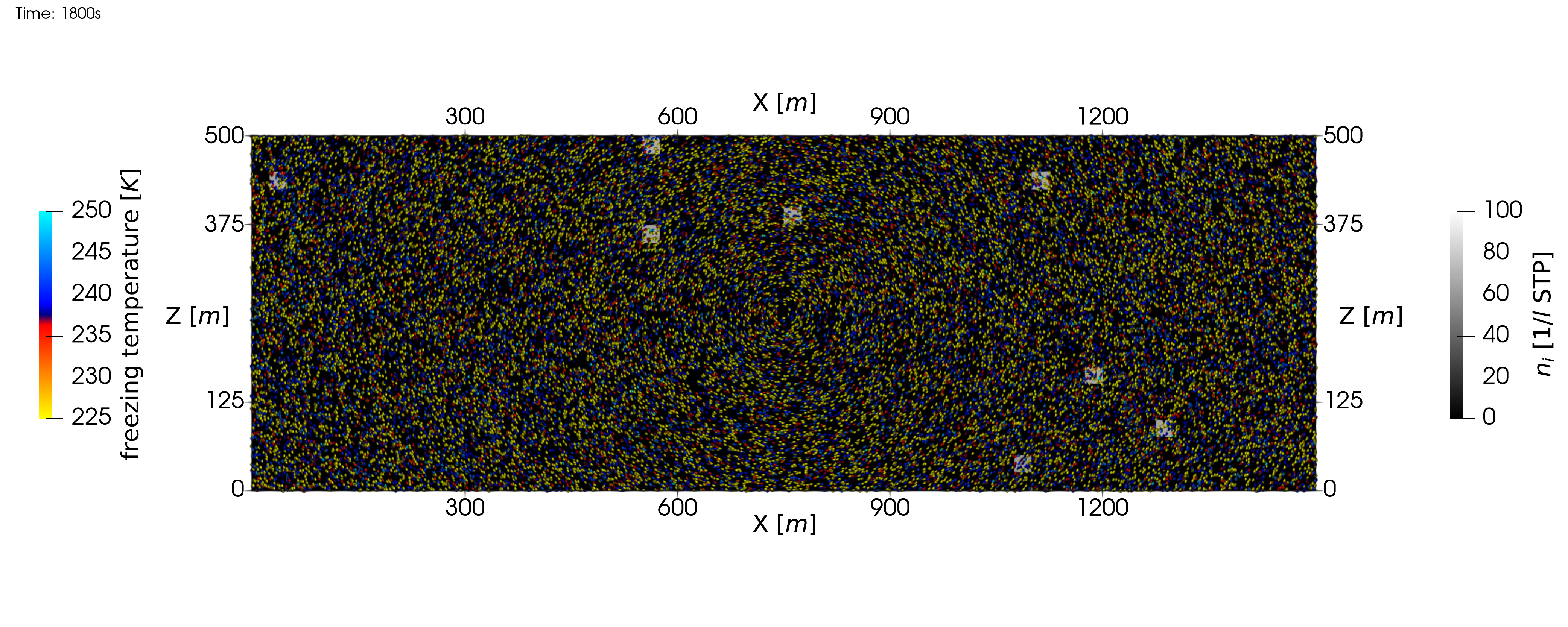}
    \end{adjustbox}
    \begin{adjustbox}{clip,trim=9.666cm .5cm 0cm 1.25cm}  % l b r t
      \includegraphics[width=6.1875in]{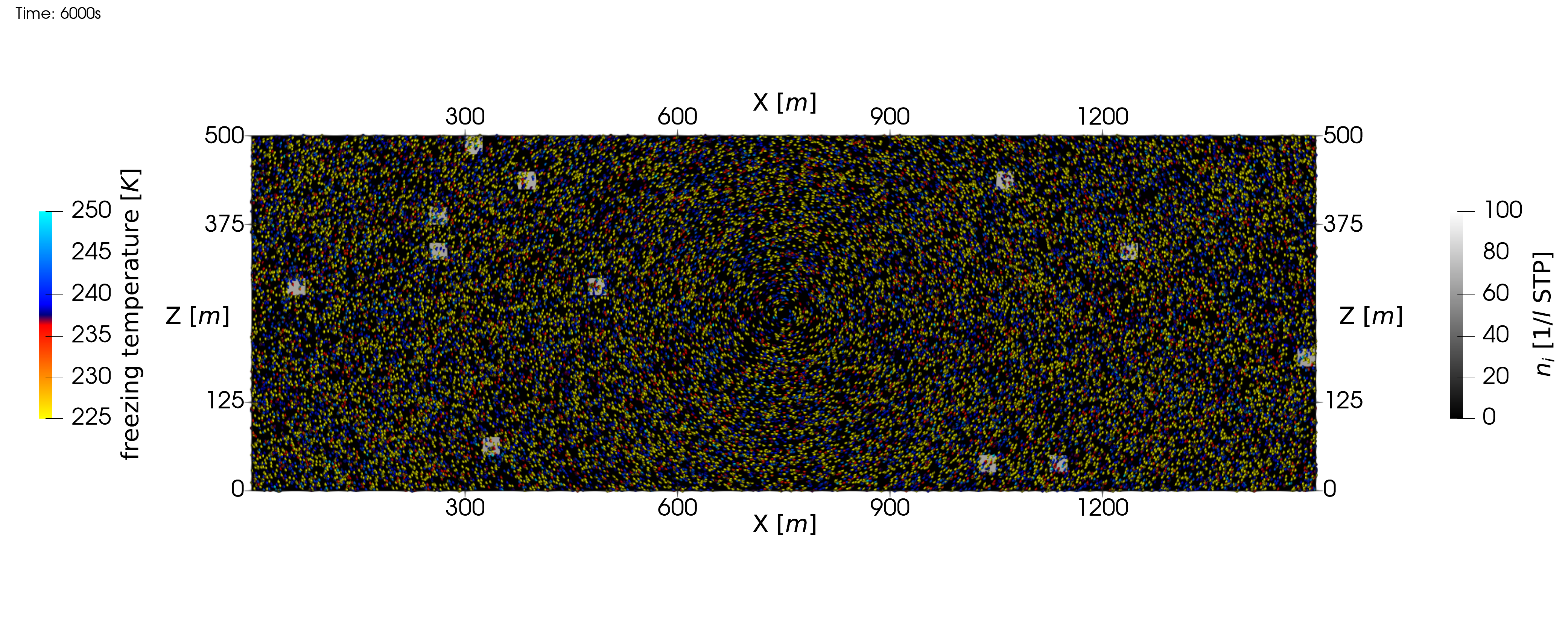}
    \end{adjustbox}
  }
  \caption{\label{fig:2d_scene}
Renderings of snapshots of the prescribed-flow simulations at $t=600$s, $t=1800$s and $t=6000$s, 
  with super-droplets depicted as solid spheres colored
  by: (a) (wet) radius (from yellow: aerosol to blue: droplets),
      (b) immersed surface area (for time-dependent scheme, values of zero correspond to INP-void super-particles) and
      (c) freezing temperature (for singular scheme, INP-void particles plotted with yellow colour).
In each row of plots, the domain is cut in three parts, and each part is shown for different time step.
Concentrations of droplets $n_\textnormal{\scriptsize c}$ and ice crystals $n_\textnormal{\scriptsize i}$ within grid cells are
  indicated with grayscale shading.
Animated renderings of the whole length of all discussed simulations are available as electronic supplement to the paper
  at \url{https://doi.org/10.5281/zenodo.13904749}.
  }
\end{figure*}

The wet radii of the particles are initialized by solving for equilibrium size at 
  ambient humidity or at RH=100\% -- whichever is smaller (i.e. solving for 
  wet radius matching zero growth rate). 
Due to the initial supersaturation in the domain, the simulations are carried out with an initial
  spinup period of 10 minutes during which freezing is disabled.
During the spinup period, the initially activated droplets formed under unrealistically high supersaturation
  are first deactivated in the downdraft region of the domain, and then reactivated in realistic 
  supersaturation within the updraft region.
As a result of the spinup, the cloud deck becomes abundant with interstitial aerosol particles while
  the size spectrum of the activated particles ceases to be related to the initial unphysically high
  supersaturation.
The spinup procedure is analogous to that described in \citeA[sect.~2.2]{Arabas_et_al_2015} 
  where sedimentation and coalescence were switched off during the spin-up. 

Figure~\ref{fig:2d_scene} depicts snapshots of the simulation state at 600~s, 1800~s (i.e. 1200~s after
  the spinup period) and 6000~s.
The figure is meant to depict the simulation resolution--both the grid of the Eulerian component and the 
  size of the population of super-particles of the Lagrangian component.
Each super-particle in the domain is rendered as a macroscopic sphere with its color corresponding
  to: (a) the value of the wet radius attribute of the super particle,
  (b) the immersed surface area (for the time-dependent scheme) and
  (c) the freezing temperature (for the singular scheme).
Concentration of liquid and ice particles is depicted with greyscale shading of grid cells.
In panel (a), the cloud base is visible at an altitude of around 350~m above which most of the super-particles
  have sizes indicating activation into cloud droplets; interstitial aerosol is visible in the cloud layer.
In panels (b) and (c), the yellow-colored particles correspond to INP-void super droplets; distribution of 
  ice concentration throughout the domain is due to lack of representation in the simulations of ice growth/shrinkage
  mechanisms; immersion freezing is the only ice process represented, there are supercooled conditions throughout the domain,
  hence, once frozen, particles remain as frozen tracers.

\begin{figure*}
  \begin{center}
    \includegraphics[width=2.5in]{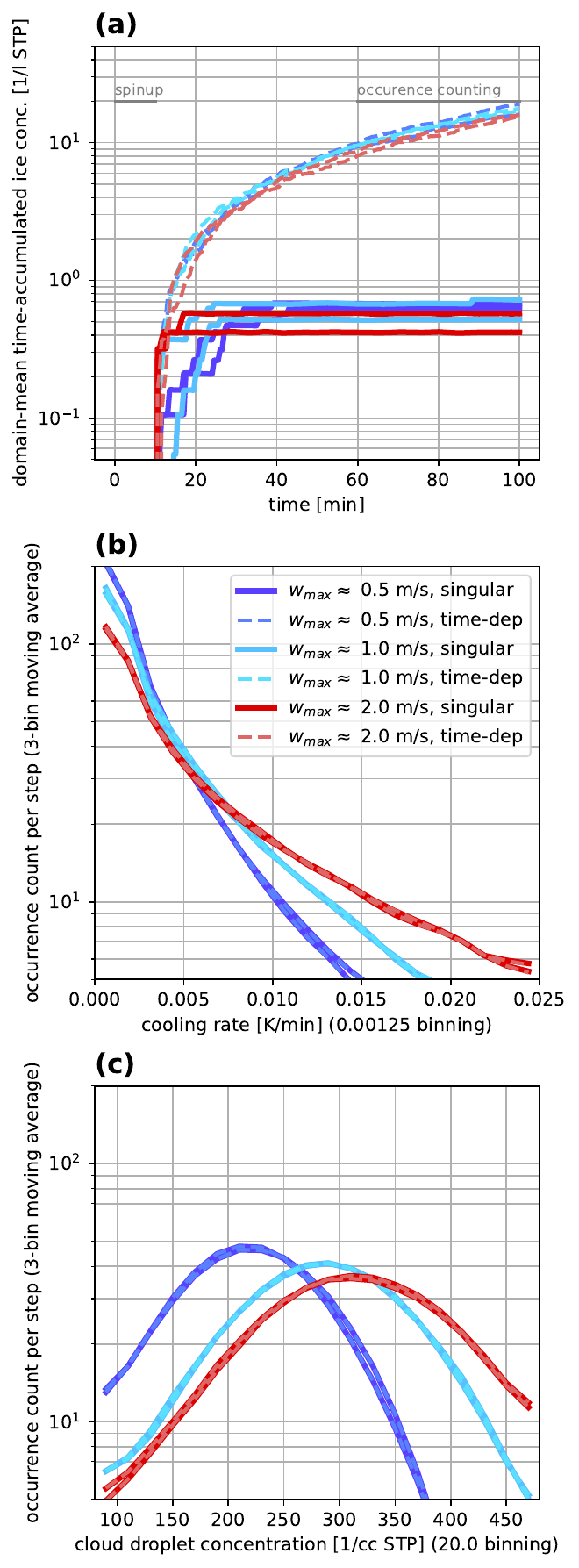}
  \end{center}
  \caption{\label{fig:2d_freqs}
    Aggregated results from twelve prescribed-flow simulations, performed for three different stream-function amplitudes $A$ and 
      for both singular (solid lines) and time-dependent (dashed lines) immersion freezing representations, for two different values
      of the random seed each.
    Panel (a) depicts the time evolution of ice concentration, with the initial spinup period (freezing disabled) indicated with
      gray line in the upper part of the plot.
    Panels (b) and (c) are based on histograms constructed from the simulation output over full domain in the final 40 minutes
      of the simulation (the period over which binning is performed is indicated in the upper part of panel (a)).
  }
\end{figure*}

In order to assess the impact of differing ambient cooling rates on the modeled efficiency
  of immersion freezing, a set of 2$\times$6 simulations is analyzed (six settings, each repeated twice for different random seed).
Figure~\ref{fig:2d_freqs} summarizes the simulations.
The simulations are run for three different settings of the eddy frequency parameter $A$ 
  (different line colors in Fig.~\ref{fig:2d_freqs}) 
  and with either singular (solid lines) or time-dependent (dashed lines) 
  representation of immersion freezing.
The different eddy frequencies correspond to maximal air velocities of ca.
  $2$, $1$ and $0.5$~m/s (see legend in panel b in Fig.~\ref{fig:2d_freqs}).
Each line in the plots depicts a single realization of the system evolution, there are 
  two realizations for each setting shown.
In principle, these are the ensemble-mean quantities that can be a basis for conclusions on the system behavior
  (and the inter-realization spread can be a basis for uncertainty estimation).

Panel (a) in Fig.~\ref{fig:2d_freqs} presents the time evolution of the domain-mean time-accumulated ice concentration.
The whole-domain averaging implies counting cells with no cloud or no ice (see gray shading in Fig.~\ref{fig:2d_scene})
  (local in-cloud droplet concentrations are depicted in panel~(c) discussed below).
The somewhat unphysical choice of domain-wide statistics does not affect the comparison between the singular and time-dependent schemes.
Given the simplistic treatment of ice particles as tracers after freezing (no ice growth or ice sedimentation physics is featured)
  any subsetting would anyhow not result in more realistic concentration values for this setup.

The 10-min period of spinup when freezing is disabled is indicated with a gray bar at the top of the plot.
After the spinup concludes, it is evident that singular
  simulations (solid lines) consistently predict an order of magnitude lower concentrations
  of ice than the time-dependent simulations.
Such differences are notable because precipitation and desiccation rates can be expected to scale with ice number concentration 
  \cite<e.g.,>{Fridlind_et_al_2012}.
Furthermore, the singular curves level off within 30 minutes after the end of spinup, while the time-dependent
  curves exhibit continuous upward trend.
This is in line with the model formulation as the singular scheme limits attainable ice concentrations
  to those corresponding to the number of particles with freezing temperatures above the minimal temperature in the
  domain, while in the time-dependent scheme there is no such limitation and the concentration of ice
  monotonically increases with time.
It is also evident, in agreement with the box-model results, that the different ambient cooling
  rates (see discussion of panel b below) stemming from different air velocities trigger differences in singular scheme behavior
  visible in the first minutes after spinup, while the time-dependent freezing rates are more robust to the flow regime.

Plots in panels~(b) and~(c) are constructed by analyzing occurrence counts of values of selected parameters in the domain.
To ensure that transient spin-up effects at the beginning of the simulation are excluded from the statistics
  (e.g., concentration of interstitial aerosol stabilises only after a full eddy revolution),
  counting is carried out over a period of 40 min from 60 till 100 min of the simulation as indicated
  with a gray bar at the top of panel (a).
The occurrence counts are recorded in bins and plotted using 3-bin moving average for plot clarity.

Panel (b) in Fig.~\ref{fig:2d_freqs} depicts frequency of occurrence of cooling rate values in the domain.
Since the super-particle model assumes uniform thermodynamic state within each grid cell \cite<see discussion in>[sect.~5.1.3]{Arabas_et_al_2015},
  the particles encounter changes in ambient temperature only when crossing grid cell boundaries;
  however, the derived per-grid-cell cooling rate reported in the plots is a multiplicity-weighted average of cooling rates for
  all super particles within each grid cell (while the cooling rates for each particle are computed taking into account ambient
  state properties of both the current and the previous cell for each particle).
This approach is consistent with the stochastic super-particle transport representation
  proposed in \citeA{Curtis_et_al_2016} in which particle positions within a cell are not part of the model state at all.
It is evident from the plotted data that even for a simple flow regime governed by the single-eddy stream
  function (eq.~\ref{eq:stream}), there is a range of cooling rates encountered in the domain.
The magnitudes of the cooling rates grow with the eddy spinning rate.
The simulated magnitudes of up to $0.025$ K/min
  are smaller (by more than an order of magnitude) than those for which agreement among ABIFM
  and INAS parameterizations is expected given the cooling rates observed in experiments.
Only in a few instances (note the logarithmic scale for occurrence counts),
  the cooling rate magnitudes exceed the $|c|>0.01$~K/min threshold above which the 
  validity of the singular approximation 
  was explored in \citeA[see the last row of panels in Figs.~1~\&~2 therein]{Wright_et_al_2013},
  and in general the values are well beyond the $1-2$ K/min INAS applicability range reported in \citeA{Kanji_et_al_2017}.

Figure \ref{fig:2d_freqs}c depicts analogous occurrence statistics for values of the concentration of particles
  larger than 1~$\upmu$m in diameter. 
Similarly as in the case of the cooling rates, the choice of immersion freezing parameterization
  has a negligible effect on the presented data, as expected.
The vast difference in ice concentrations (few per liter) and droplet concentrations (few hundreds per cubic centimeter)
  makes the different outcomes of freezing for singular and time-dependent schemes not discernible
  from the data plotted in panel~(c) (note that no ice growth mechanisms are represented here, in particular there is no
  representation of the WBF process).
The gradual shift towards larger droplet concentrations with increasing air velocities is consistent
  with the expectation of CCN activation in more vigorous updrafts resulting in higher concentration
  of activated droplets.
The spread in concentration values is consistent with the range of vertical velocities encountered 
  at cloud base (analogous to the range of cooling rates depicted in panel~b).
Quantitatively, in the case of the slow-spinning eddy (maximal velocity of ca. $0.5$~m/s), the
  occurrence counts peak roughly at 200-250 per cubic cm (with domain-mean aerosol concentration of $316$ per cubic cm),
  while in the case of the fast-spinning eddy, almost all particles activate into cloud droplets.
This confirms that the setup captures the influence of the flow regime on aerosol
  activation efficiency.

\section{Summary, limitations and future directions}\label{sec:summary}

Here, we contrasted two probabilistic representation of immersion freezing for aerosol-cloud interaction models: the singular scheme 
  embracing the INAS concept proposed in \citeA{Shima_et_al_2020} 
  and the alternative time-dependent scheme based on the ABIFM parameterization detailed in \citeA{Knopf_and_Alpert_2013}.
We depicted that both schemes can be used in particle-based constant-state-vector-size microphysics frameworks.
Both approaches involve Monte-Carlo-type random sampling of the attribute space and both admit arbitrary initial attribute distributions (including polydisperse immersed surface spectra).
The singular scheme employs deterministic triggering of freezing (finite state machine), while the time-dependent scheme employs Monte-Carlo triggering (discrete time Markov chain).
Both schemes keep track of the INP reservoir budget.
We explored these two contrasting approaches, using both box-model and flow-coupled simulations, aiming
  at portraying the trade-offs inherent to the singular
  description in particle-based cloud microphysics modeling.

The key takeaways from the herein discussion, confirming and extending findings from earlier works, are:

%  \item{
1. Both the singular and time-dependent schemes can be viewed as constituting numerical representations of the same Poissonian model of the heterogeneous nucleation.
    However, integrating out the time for a given temperature profile needed to formulate the singular parameterizations embeds the cooling rate characteristics
      within the parameterization coefficients; as a result, the INAS-based and ABIFM-based simulations match only for one specific cooling rate which 
      is characteristic for the laboratory measurements employed for fitting the INAS coefficients.
    Only parameterizations featuring
      time dependence can be robust to ambient flow regimes which might substantially differ from laboratory settings or 
      feature flow patterns precluding nucleation with the singular scheme (downdrafts or quiescent flow).
    Whereas time-dependent and singular parameterizations derived from the same laboratory data provide a match to that data in a laboratory scenario, 
      they can result in order-of-magnitude and greater differences in ice formation on timescales that are more typical of natural mixed-phase clouds \cite<see also>{Knopf_et_al_2023}.
%  }
%  \item{

2. Despite the time-dependent approach being computationally costlier than the singular approach when cast in the particle-based framework, it provides the added value
    of robustness to flow regimes; moreover since the time-dependent approach features immersed insoluble surface area as the
    freezing-relevant super-particle attribute (as opposed to freezing temperature for the singular approach), it opens up
    possibilities of online coupling of the immersion freezing scheme with aerosol physio-chemical dynamics that lead to
    evolution of the immersed surface. Being an extensive attribute, the immersed surface dynamics are also more readily
    representable upon particle collisions.
%  }
%  \item{

3. For both singular and time-dependent schemes, one of the key challenges from the viewpoint of particle-based
      mixed-phase cloud microphysics modeling, is the sampling of attribute space in which rarely occurring INPs (outnumbered by CCN)
      need to be consciously represented by admitting much lower multiplicities for immersed-surface-rich super-particles.
    Furthermore, neglecting polydispersity of the immersed surface area spectrum was shown to have an effect on the frozen fraction predictions
      comparable to multi-fold changes in cooling rate or median surface values.
%  }

\appendix

\section{Convergence properties of the ABIFM/Monte-Carlo scheme}\label{sec:convergence}

\begin{figure*}[t]
  \begin{center}
    \includegraphics[width=5in]{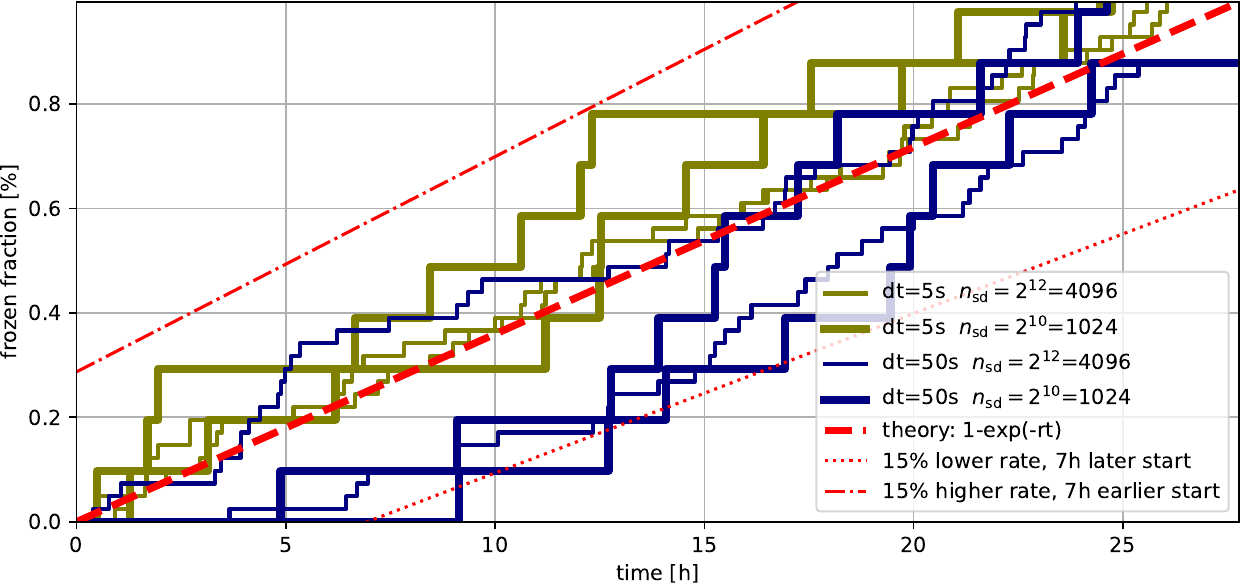}
  \end{center}
  \vspace{-2em}
  \caption{\label{fig:appendix_x=time}
    Eight runs of the test case used in the convergence analysis with two different timestep lengths $dt$,
      two different super-particle counts $n_{sd}$ and two different random number generator seeds (two realizations plotted for each combination of $dt$ and $n_{sd}$).
    Red dashed lines represent theoretical solutions: thick dashed line corresponds to the rate used in numerical integration, 
      thin lines depict how the solution changes with 15\% higher or 15\% lower rate, and with an earlier or later start.
  }
  \vspace{-1em}
\end{figure*}

\begin{figure*}[t]
  \begin{center}
  \includegraphics[width=2.5in]{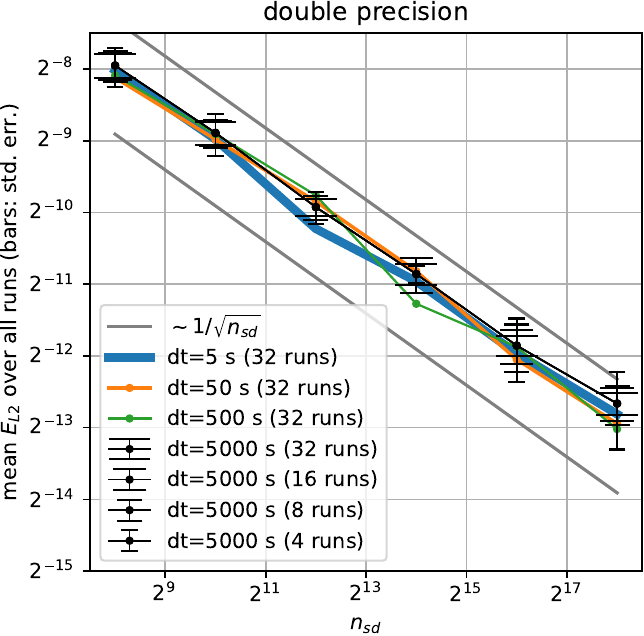}
  ~~~
  \includegraphics[width=2.5in]{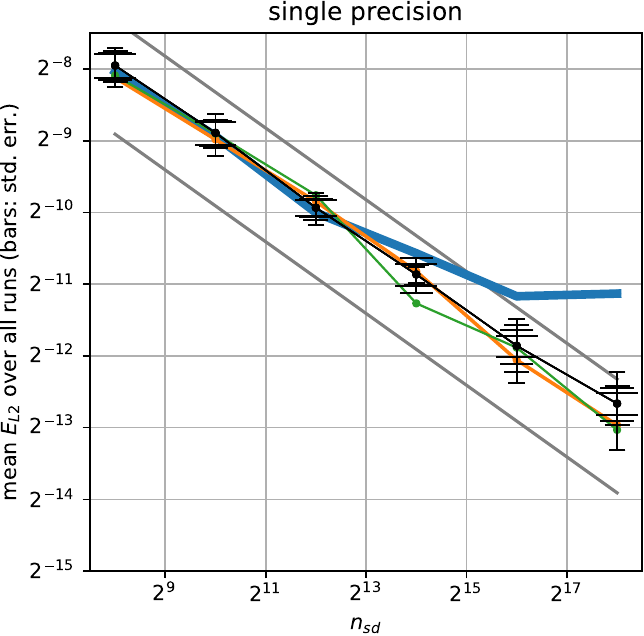}
  \end{center}
  \caption{\label{fig:appendix_slopes}
    Mean values of the L2 error measure plotted as a function of the number of super droplets $n_\textnormal{\scriptsize sd}$, for double- (left panel) and single-precision (right panel).
    Different line widths and colors correspond to timestep length settings.
    Error bars plotted for the longest timestep only depict standard error values computed for 4, 8, 16 and 32 realizations (different random number seeds).
    Gray guidelines indicate theoretical inverse square-root convergence.
  }
\end{figure*}

The fidelity of the time-dependent Monte-Carlo scheme depends on three resolution-related
  aspects, namely: 
  (i) temporal resolution (i.e., the timestep length),
  (ii)~size-spectral resolution (i.e., the number of super-particles) and 
  (iii) ensemble size (i.e., the number of realizations simulated).
To depict the above and quantify the convergence, we are analyzing the departures from
  expected analytic solution of exponential decay for a simple case of constant 
  event rate $r=1\times 10^{-7}$s\textsuperscript{-1} (i.e., corresponding to constant $T$ and constant 
  $X$ in eq.~\ref{eq:lnpm2=int}).
In such case, since evaluation of the probability integral in eq.~\ref{eq:lnpm2=int} is exact regardless of the timestep,
  the temporal resolution is not expected to play a role (while,
  in a realistic simulation, the rate changes in time, hence the smaller the timestep
  the more accurate is the representation).
For the particle population sampling, from the law of large numbers, an inverse square root
  scaling of the error is expected with the number of independent samples (super-particles).
Analogous scaling is expected with the size of the simulation realizations ensemble
  \cite<see, e.g. a recent discussion in the context of particle-based modeling in>[sect.~2.6 and 3.2.3 therein]{Nordam_et_al_2023}.

To quantify the departure from the expected analytical solution, we use a simple L2 root-mean-square norm defined as $E_{L2}=\sqrt{\overline{\left(A[i] - E[i]\right)^2}}$
where $E[i]$ are the ``expected'' values (analytic solution: $1-\exp(-rt)$), $A[i]$ are the ``actual'' numerical solution results, and the mean is computed over all timesteps indexed by $i$.

Figure~\ref{fig:appendix_x=time} depicts the dataset using a small subset of the simulations the analysis is based on: 
  eight sample simulations carried out with two different
  random number generator seeds, two different timesteps and two different numbers of
  super-particles.
Time is given on the abscissa and the ordinate represent the frozen fraction.
The thick-dashed-red line in the plot depicts the analytic solution (see caption for details).
Each freezing event results in the entirety of particles represented by a single super-droplet to become frozen, 
  hence the step-wise evolution of the frozen fraction. 
The more super droplets (thin lines), the smaller is their multiplicity, and the smaller is the magnitude of the steps.
Since constant-multiplicity sampling is employed, each step is of the same "height".

The whole dataset consists of $32 \times 6 \times 4 \times 2 = 1536$ simulations performed for: 32 different seeds, 6 different
  numbers of super-particles (powers of two from $2^{8}$ through $2^{18}$, 4 different timestep
  settings, and single or double floating point precision. 
Figure~\ref{fig:appendix_slopes} depicts the computed error measures plotted in base-2 logarithmic scale as a function of base-2 logarithm of the number of super-particles.
In addition, for the simulations with longest timesteps (5000 seconds), overlaid error bars depict values of the standard error computed from 4, 8, 16 and 32 realizations (different random seeds).
The plot depicting computations in double precision shows that, as expected, the errors are essentially insensitive to the timestep length (due to assumed constant decay rate). The slope of the dependence of the error on super-particle count matches the theoretical inverse square root lines. The realization spread expectedly diminishes with the number of runs considered.

The single-precision calculations in general match, with the intriguing exception of results for shortest timesteps and largest numbers of super particles.
This is attributed to the accumulation of arithmetic errors in computation of the frozen fraction (the more super particles, the more numbers being summed up; the smaller the multiplicities and shorter the timesteps, the more contrasting the individual vs. total numbers are).
The captured effect of error accumulation is reported here as a cautionary remark in the context of ongoing discussion of introduction of reduced-precision (i.e., below single precision) arithmetics in atmospheric modeling \cite<e.g.,>{Banderler_et_al_2023}, including in super-particle simulations \cite{Matsushima_et_al_2023}.

\section{Kinematic framework: Eulerian component}\label{sec:Eulerian}

The conservation of moisture and heat solved by the Eulerian component of the simulation
  is expressed through two heterogeneous advection equations in a stratified incompressible flow 
  (of a compressible fluid---dry air):
  \begin{eqnarray}
    \label{eq:cont_th}
    \partial_t \theta_\textnormal{\scriptsize d} + \frac{1}{\rho_\textnormal{\scriptsize d}} \nabla \cdot (\vec{u} \rho_\textnormal{\scriptsize d} \theta_\textnormal{\scriptsize d}) &\!\!\!=\!\!\!& -\frac{\theta_\textnormal{\scriptsize d} l_\textnormal{\scriptsize v}(T)}{c_\textnormal{\scriptsize pd} T}\dot{q}_v\\
    \label{eq:cont_qv}
    \partial_t q_\textnormal{\scriptsize v} + \frac{1}{\rho_\textnormal{\scriptsize d}} \nabla \cdot (\vec{u} \rho_\textnormal{\scriptsize d} q_\textnormal{\scriptsize v}) &\!\!\!=\!\!\!& \dot{q}_v \\
    {\color{white}\frac{1}{1}}\nabla \cdot (\vec{u}\rho_\textnormal{\scriptsize d}) &\!\!\!=\!\!\!& 0\textrm{,}
  \end{eqnarray}
  where $c_\textnormal{\scriptsize pd}$ is the dry-air specific heat at constant pressure, $\theta_d=T(1000 \textrm{ hPa} / p_\textnormal{\scriptsize d})^{R_\textnormal{\scriptsize d}/c_\textnormal{\scriptsize pd}}$ is the dry-air potential temperature 
  and $q_v=\rho_\textnormal{\scriptsize v}/\rho_\textnormal{\scriptsize d}$ is the water vapor mixing ratio (with $p_\textnormal{\scriptsize d}$, 
  $\rho_\textnormal{\scriptsize v}$ and $R_\textnormal{\scriptsize d}$ denoting partial pressure of dry air, water vapor density
  and gas constant for dry air, respectively).
The density profile with altitude $\rho_d(z)$ is assumed constant in time and constant in the horizontal dimension.
The velocity field $\vec{u}$ is also constant in time and is defined by:
\begin{eqnarray}
  u_\textnormal{\scriptsize x} &=& -\frac{1}{\rho_\textnormal{\scriptsize d}} \partial_z \psi \\
  u_\textnormal{\scriptsize z} &=& \frac{1}{\rho_\textnormal{\scriptsize d}} \partial_x \psi
  \textrm{,}
\end{eqnarray}
where $\psi$ is a stream function constructed to mimic a single eddy spanning the periodic domain (see example in Fig.~\ref{fig:arrows}) and specified in the following form
  \cite<for discussion, see also>[eq.~2.3]{Maxey_and_Corrsin_1986}:
\begin{equation}\label{eq:stream}
  \psi(x,z) = -A \frac{X}{\pi} \sin\left(\pi\frac{z}{Z}\right) \cos\left(2\pi\frac{x}{X}\right)\textrm{,}
\end{equation}
  where $A$ is a constant, with the dimension of density times velocity, controlling the eddy frequency
  (and hence air velocities); $X$ and $Z$ are domain extents in horizontal and vertical directions, respectively.
The choice of $\theta_\textnormal{\scriptsize d}$, $q_\textnormal{\scriptsize v}$, $\rho_\textnormal{\scriptsize d}$ state variable triplet to describe moist air
  is discussed in \citeA[Appendix~A therein]{Arabas_et_al_2015}.
The advective terms in the transport equations \ref{eq:cont_th} and \ref{eq:cont_qv} are solved numerically using the MPDATA solver 
  \cite<see,~e.g.,>[for an overview]{Smolarkiewicz_2006} using the PyMPDATA 
  package \cite{Bartman_et_al_2022_PyMPDATA}.
The timestep for the Eulerian component (MPDATA) is set to $2.5$~s.

\section{Kinematic framework: Lagrangian component and attribute sampling}\label{sec:Lagrangian}

The source terms in eqs.~\ref{eq:cont_th} and \ref{eq:cont_qv} are evaluated by summing over
  the changes in super-particle wet volumes within a given timestep as in eqs. (25) and (26) in \citeA{Arabas_et_al_2015}.
The numerics of particle attribute dynamics are solved as follows:
  particle displacement is integrated with an implicit-in-space
  scheme \cite<see sect.~5.1.2 in>{Arabas_et_al_2015}; condensational growth/evaporation (including CCN activation and deactivation) 
  is integrated with an explicit-in-time/implicit-in-supersaturation scheme 
  \cite<see sect.~5.1.3 in>{Arabas_et_al_2015}.
The Lagrangian component uses adaptive substepping for condensational growth/evaporation and for particle transport; freezing is calculated using $2.5$~s timestep and double-precision arithmetics (see \ref{sec:convergence} for discussion).

There are multiple possible sampling strategies for aerosol attributes to be applied in context
  of particle-based cloud microphysics modeling.
First, the multiplicity can either vary across the super-particle population or can be set 
  constant (or constant across super-particle subpopulations).
Second, there is degree of freedom in the specification of the multidimensional probability 
  density across particle attribute space.

In \citeA{Unterstrasser_et_al_2017}, an analysis was presented highlighting how the choice of
  constant-multiplicity versus variable-multiplicity sampling has a profound effect on the 
  performance of coagulation algorithms.
A sampling strategy particularly relevant to the present work,
  where concentration of particles that drive the population evolution is small 
  relative to the background density, was presented in \citeA{DeVille_et_al_2019}.
In \citeA{Dziekan_and_Pawlowska_2017} and \citeA{Li_et_al_2022}, the different multiplicity 
  choices were analyzed in the particular context of sampling for the so-called ``lucky droplet''
  system with contrasting sizes of particles.
In \citeA[]{Shima_et_al_2020}, a uniform sampling per subpopulation
  (immersed-surface-rich vs. immersed-surface-free particles) was used, while in Section~5.3 therein alternative strategies
  are discussed.
\citeA{Shima_et_al_2020} pointed out that the sampling strategy analyses of \citeA{Unterstrasser_et_al_2017} and
  \citeA{Dziekan_and_Pawlowska_2017} were limited to zero-dimensional coagulation-only setups,
  and generally it is expected 
  that in higher-dimensional attribute space a uniform sampling
  strategy potentially reinforced with a quasi-random numbers is optimal.
Noteworthily, simulations involving processes such as condensation and freezing do increase
  the dimensionality of the attribute space by incorporating such particle properties
  as hygroscopicity, soluble mass and insoluble surface, and particle habit information.
Clearly the challenge in representation of INP is very relevant as these particles are often 
  rare in terms of concentration compared to other aerosol, while the fidelity
  of the representation of their collisions also influences the potential for representing
  the contact-freezing mechanism.

Here, the super-droplets are initialized in pairs, with both super-particles within a pair sharing 
  their location in space and their soluble-substance dry radius.
Locations in space are shuffled from a uniform distribution separately in horizontal and vertical
  dimensions.
Soluble-substance dry radii are sampled from the lognormal distribution 
  by inverting its cumulative distribution function (to compute quantiles)
  and assigning each super-particle pair with an equal multiplicity.
Within each super-particle pair, one super-droplet belongs to the immersed-surface-rich subpopulation, while
  the other to the immersed-surface-free subpopulation.
The multiplicities are split among super-particle within each pair according to the 
  immersed-surface-free to immersed-surface-rich concentration ratio.
Super-particles belonging to the immersed-surface-rich subpopulation are initialized by sampling from the
  lognormal distribution of immersed surface areas by inverting its cumulative distribution
  (consistently with constant multiplicity sampling of the soluble mass distribution).
In the case of simulations with the time-dependent scheme, the insoluble immersed surface is used as the 
  particle attribute.
In the case of singular simulations, the insoluble immersed surface is used to evaluate the freezing 
  temperature which is used as the particle attribute (i.e., insoluble immersed surface area is not retained as super-particle attribute).
Super-particles belonging to the immersed-surface-free subpopulation have their freezing attribute (either
  freezing temperature or immersed insoluble surface) set to zero (Kelvins or meters squared) precluding triggering of freezing.

\section*{Code availability}

All simulations performed for the study and discussed in the text were carried out using free and open-source 
  Python packages PySDM \cite{Bartman_et_al_2022_PySDM,deJong_et_al_2023_JOSS} and PyMPDATA
  \cite{Bartman_et_al_2022_PyMPDATA}.
All presented figures can be reproduced using Jupyter notebooks shipped in the PySDM-examples package
  available on PyPI.org (\url{https://pypi.org/p/PySDM-examples}) 
  as well as persistently archived at Zenodo (DOI:10.5281/zenodo.6321262).
The notebooks can be executed in the cloud using platforms such as Google Colab, ARM JupyterHub or anonymously through mybinder.org (links provided within each notebook).
In addition, PySDM-examples includes notebooks reproducing using PySDM several simulation plots from 
  \citeA{Alpert_and_Knopf_2016}.

\section*{Author contributions}

NR and DAK conceptualized and supervised the project.
SA carried out software development,
  performed the simulations and analyses.
Simulation setups, analysis workflows, and figures were conceptualized by SA, MW and NR.
JHC, IS, AMF and DAK provided feedback on methodology and analyses results throughout all
  stages of the project.
SA wrote the first draft of the manuscript.
All authors provided further feedback and contributed to the present version.

\section*{Acknowledgments}

This study was supported by the Atmospheric System Research Program sponsored by the US Department of Energy (DOE), Office of Science, Office of Biological and Environmental Research (OBER), Climate and Environmental Sciences Division (CESD), US DOE grants no. DE-SC0021034 (SA, IS, DAK \& NR) and no. DE-SC0022130 (JHC \& MW).
AMF was supported by the NASA Modeling, Analysis, and Prediction Program
SA acknowledges support from the Polish National Science Centre (grant no. 2020/39/D/ST10/01220).
SA is grateful to Shin-ichiro Shima and Thomas Nagler for helpful comments and exchanges on the project.
We thank three anonymous reviewers for their feedback on the initial version of the manuscript.

\bibliography{\jobname}

\begin{thebibliography}{}

\bibitem [\protect \citeauthoryear {%
Abade%
\ \BBA {} Albuquerque%
}{%
Abade%
\ \BBA {} Albuquerque%
}{%
{\protect \APACyear {2024}}%
}]{%
Abade_and_Albuquerque_2024}
\APACinsertmetastar {%
Abade_and_Albuquerque_2024}%
\begin{APACrefauthors}%
Abade, G.%
\BCBT {}\ \BBA {} Albuquerque, D.%
\end{APACrefauthors}%
\unskip\
\newblock
\APACrefYearMonthDay{2024}{}{}.
\newblock
{\BBOQ}\APACrefatitle {Persistent mixed-phase states in adiabatic cloud parcels
  under idealised conditions} {Persistent mixed-phase states in adiabatic cloud
  parcels under idealised conditions}.{\BBCQ}
\newblock
\APACjournalVolNumPages{Q. J. Royal Meteorol. Soc.}{}{}{}.
\newblock
\begin{APACrefDOI} \doi{10.1002/qj.4775} \end{APACrefDOI}
\PrintBackRefs{\CurrentBib}

\bibitem [\protect \citeauthoryear {%
Abade%
, Grabowski%
\BCBL {}\ \BBA {} Pawlowska%
}{%
Abade%
\ \protect \BOthers {.}}{%
{\protect \APACyear {2018}}%
}]{%
Abade_et_al_2018}
\APACinsertmetastar {%
Abade_et_al_2018}%
\begin{APACrefauthors}%
Abade, G.%
, Grabowski, W.%
\BCBL {}\ \BBA {} Pawlowska, H.%
\end{APACrefauthors}%
\unskip\
\newblock
\APACrefYearMonthDay{2018}{}{}.
\newblock
{\BBOQ}\APACrefatitle {Broadening of Cloud Droplet Spectra through Eddy
  Hopping: Turbulent Entraining Parcel Simulations} {Broadening of cloud
  droplet spectra through eddy hopping: Turbulent entraining parcel
  simulations}.{\BBCQ}
\newblock
\APACjournalVolNumPages{J. Atmos. Sci.}{75}{}{}.
\newblock
\begin{APACrefDOI} \doi{10.1175/JAS-D-18-0078.1} \end{APACrefDOI}
\PrintBackRefs{\CurrentBib}

\bibitem [\protect \citeauthoryear {%
Alpert%
, Aller%
\BCBL {}\ \BBA {} Knopf%
}{%
Alpert%
\ \protect \BOthers {.}}{%
{\protect \APACyear {2011}}%
}]{%
Alpert_et_al_2011}
\APACinsertmetastar {%
Alpert_et_al_2011}%
\begin{APACrefauthors}%
Alpert, P.%
, Aller, J.%
\BCBL {}\ \BBA {} Knopf, D.%
\end{APACrefauthors}%
\unskip\
\newblock
\APACrefYearMonthDay{2011}{}{}.
\newblock
{\BBOQ}\APACrefatitle {Ice nucleation from aqueous NaCl droplets with and
  without marine diatoms} {Ice nucleation from aqueous nacl droplets with and
  without marine diatoms}.{\BBCQ}
\newblock
\APACjournalVolNumPages{Atmos. Chem. Phys.}{11}{}{}.
\newblock
\begin{APACrefDOI} \doi{10.5194/acp-11-5539-2011} \end{APACrefDOI}
\PrintBackRefs{\CurrentBib}

\bibitem [\protect \citeauthoryear {%
Alpert%
\ \BBA {} Knopf%
}{%
Alpert%
\ \BBA {} Knopf%
}{%
{\protect \APACyear {2016}}%
}]{%
Alpert_and_Knopf_2016}
\APACinsertmetastar {%
Alpert_and_Knopf_2016}%
\begin{APACrefauthors}%
Alpert, P.%
\BCBT {}\ \BBA {} Knopf, D.%
\end{APACrefauthors}%
\unskip\
\newblock
\APACrefYearMonthDay{2016}{}{}.
\newblock
{\BBOQ}\APACrefatitle {Analysis of isothermal and cooling-rate-dependent
  immersion freezing by a unifying stochastic ice nucleation model} {Analysis
  of isothermal and cooling-rate-dependent immersion freezing by a unifying
  stochastic ice nucleation model}.{\BBCQ}
\newblock
\APACjournalVolNumPages{Atmos. Chem. Phys.}{16}{}{}.
\newblock
\begin{APACrefDOI} \doi{10.5194/acp-16-2083-2016} \end{APACrefDOI}
\PrintBackRefs{\CurrentBib}

\bibitem [\protect \citeauthoryear {%
Andreae%
\ \BBA {} Rosenfeld%
}{%
Andreae%
\ \BBA {} Rosenfeld%
}{%
{\protect \APACyear {2008}}%
}]{%
Andreae_and_Rosenfeld_2008}
\APACinsertmetastar {%
Andreae_and_Rosenfeld_2008}%
\begin{APACrefauthors}%
Andreae, M.%
\BCBT {}\ \BBA {} Rosenfeld, D.%
\end{APACrefauthors}%
\unskip\
\newblock
\APACrefYearMonthDay{2008}{}{}.
\newblock
{\BBOQ}\APACrefatitle {Aerosol–cloud–precipitation interactions. Part 1.
  The nature and sources of cloud-active aerosols}
  {Aerosol–cloud–precipitation interactions. part 1. the nature and sources
  of cloud-active aerosols}.{\BBCQ}
\newblock
\APACjournalVolNumPages{Earth-Sci. Rev.}{89}{}{13--41}.
\newblock
\begin{APACrefDOI} \doi{10.1016/j.earscirev.2008.03.001} \end{APACrefDOI}
\PrintBackRefs{\CurrentBib}

\bibitem [\protect \citeauthoryear {%
Arabas%
, Jaruga%
, Pawlowska%
\BCBL {}\ \BBA {} Grabowski%
}{%
Arabas%
\ \protect \BOthers {.}}{%
{\protect \APACyear {2015}}%
}]{%
Arabas_et_al_2015}
\APACinsertmetastar {%
Arabas_et_al_2015}%
\begin{APACrefauthors}%
Arabas, S.%
, Jaruga, A.%
, Pawlowska, H.%
\BCBL {}\ \BBA {} Grabowski, W.%
\end{APACrefauthors}%
\unskip\
\newblock
\APACrefYearMonthDay{2015}{}{}.
\newblock
{\BBOQ}\APACrefatitle {libcloudph++ 1.0: a single-moment bulk, double-moment
  bulk, and particle-based warm-rain microphysics library in {C++}}
  {libcloudph++ 1.0: a single-moment bulk, double-moment bulk, and
  particle-based warm-rain microphysics library in {C++}}.{\BBCQ}
\newblock
\APACjournalVolNumPages{Geosci. Model Dev.}{8}{}{}.
\newblock
\begin{APACrefDOI} \doi{10.5194/gmd-8-1677-2015} \end{APACrefDOI}
\PrintBackRefs{\CurrentBib}

\bibitem [\protect \citeauthoryear {%
Avramov%
\ \protect \BOthers {.}}{%
Avramov%
\ \protect \BOthers {.}}{%
{\protect \APACyear {2011}}%
}]{%
Avramov_et_al_2011}
\APACinsertmetastar {%
Avramov_et_al_2011}%
\begin{APACrefauthors}%
Avramov, A.%
, Ackerman, A.%
, Fridlind, A.%
, van Diedenhoven, B.%
, Botta, G.%
, Aydin, K.%
\BDBL {}Wolde, M.%
\end{APACrefauthors}%
\unskip\
\newblock
\APACrefYearMonthDay{2011}{}{}.
\newblock
{\BBOQ}\APACrefatitle {Towards ice formation closure in Arctic mixed-phase
  boundary layer clouds during {ISDAC}} {Towards ice formation closure in
  arctic mixed-phase boundary layer clouds during {ISDAC}}.{\BBCQ}
\newblock
\APACjournalVolNumPages{J. Geophys. Res.}{116}{}{D00T08}.
\newblock
\begin{APACrefDOI} \doi{10.1029/2011JD015910} \end{APACrefDOI}
\PrintBackRefs{\CurrentBib}

\bibitem [\protect \citeauthoryear {%
Banderier%
, Zeman%
, Leutwyler%
, Rüdisühli%
\BCBL {}\ \BBA {} Schär%
}{%
Banderier%
\ \protect \BOthers {.}}{%
{\protect \APACyear {2023}}%
}]{%
Banderler_et_al_2023}
\APACinsertmetastar {%
Banderler_et_al_2023}%
\begin{APACrefauthors}%
Banderier, H.%
, Zeman, C.%
, Leutwyler, D.%
, Rüdisühli, S.%
\BCBL {}\ \BBA {} Schär, C.%
\end{APACrefauthors}%
\unskip\
\newblock
\APACrefYearMonthDay{2023}{}{}.
\newblock
{\BBOQ}\APACrefatitle {Reduced floating-point precision in regional climate
  simulations: An ensemble-based statistical verification} {Reduced
  floating-point precision in regional climate simulations: An ensemble-based
  statistical verification}.{\BBCQ}
\newblock
\APACjournalVolNumPages{EGUsphere [preprint]}{}{}{}.
\newblock
\begin{APACrefDOI} \doi{10.5194/egusphere-2023-2263} \end{APACrefDOI}
\PrintBackRefs{\CurrentBib}

\bibitem [\protect \citeauthoryear {%
Barrett%
\ \BBA {} Hoose%
}{%
Barrett%
\ \BBA {} Hoose%
}{%
{\protect \APACyear {2023}}%
}]{%
Barrett_and_Hoose_2022}
\APACinsertmetastar {%
Barrett_and_Hoose_2022}%
\begin{APACrefauthors}%
Barrett, A.%
\BCBT {}\ \BBA {} Hoose, C.%
\end{APACrefauthors}%
\unskip\
\newblock
\APACrefYearMonthDay{2023}{}{}.
\newblock
{\BBOQ}\APACrefatitle {Microphysical Pathways Active within Thunderstorms and
  Their Sensitivity to {CCN} Concentration and Wind Shear} {Microphysical
  pathways active within thunderstorms and their sensitivity to {CCN}
  concentration and wind shear}.{\BBCQ}
\newblock
\APACjournalVolNumPages{J. Geophys. Res. Atmos.}{128}{}{}.
\newblock
\begin{APACrefDOI} \doi{10.1029/2022JD036965} \end{APACrefDOI}
\PrintBackRefs{\CurrentBib}

\bibitem [\protect \citeauthoryear {%
Bartman%
, Banaśkiewicz%
\BCBL {}\ \protect \BOthers {.}}{%
Bartman%
, Banaśkiewicz%
\BCBL {}\ \protect \BOthers {.}}{%
{\protect \APACyear {2022}}%
}]{%
Bartman_et_al_2022_PyMPDATA}
\APACinsertmetastar {%
Bartman_et_al_2022_PyMPDATA}%
\begin{APACrefauthors}%
Bartman, P.%
, Banaśkiewicz, J.%
, Drenda, S.%
, Manna, M.%
, Olesik, M.%
, Rozwoda, P.%
\BDBL {}Arabas, S.%
\end{APACrefauthors}%
\unskip\
\newblock
\APACrefYearMonthDay{2022}{}{}.
\newblock
{\BBOQ}\APACrefatitle {{PyMPDATA} v1: {Numba}-accelerated implementation of
  {MPDATA} with examples in {Python}, {Julia} and {Matlab}} {{PyMPDATA} v1:
  {Numba}-accelerated implementation of {MPDATA} with examples in {Python},
  {Julia} and {Matlab}}.{\BBCQ}
\newblock
\APACjournalVolNumPages{J. Open Source Soft.}{}{}{}.
\newblock
\begin{APACrefDOI} \doi{10.21105/joss.03896} \end{APACrefDOI}
\PrintBackRefs{\CurrentBib}

\bibitem [\protect \citeauthoryear {%
Bartman%
, Bulenok%
\BCBL {}\ \protect \BOthers {.}}{%
Bartman%
, Bulenok%
\BCBL {}\ \protect \BOthers {.}}{%
{\protect \APACyear {2022}}%
}]{%
Bartman_et_al_2022_PySDM}
\APACinsertmetastar {%
Bartman_et_al_2022_PySDM}%
\begin{APACrefauthors}%
Bartman, P.%
, Bulenok, O.%
, Górski, K.%
, Jaruga, A.%
, Łazarski, G.%
, Olesik, M.%
\BDBL {}Arabas, S.%
\end{APACrefauthors}%
\unskip\
\newblock
\APACrefYearMonthDay{2022}{}{}.
\newblock
{\BBOQ}\APACrefatitle {{PySDM} v1: particle-based cloud modelling package for
  warm-rain microphysics and aqueous chemistry} {{PySDM} v1: particle-based
  cloud modelling package for warm-rain microphysics and aqueous
  chemistry}.{\BBCQ}
\newblock
\APACjournalVolNumPages{J. Open Source Soft.}{}{}{}.
\newblock
\begin{APACrefDOI} \doi{10.21105/joss.03219} \end{APACrefDOI}
\PrintBackRefs{\CurrentBib}

\bibitem [\protect \citeauthoryear {%
Bauer%
, Ault%
\BCBL {}\ \BBA {} Prather%
}{%
Bauer%
\ \protect \BOthers {.}}{%
{\protect \APACyear {2013}}%
}]{%
Bauer_et_al_2013}
\APACinsertmetastar {%
Bauer_et_al_2013}%
\begin{APACrefauthors}%
Bauer, S.%
, Ault, A.%
\BCBL {}\ \BBA {} Prather, K.%
\end{APACrefauthors}%
\unskip\
\newblock
\APACrefYearMonthDay{2013}{}{}.
\newblock
{\BBOQ}\APACrefatitle {Evaluation of aerosol mixing state classes in the
  {GISS}-model{E}-{MATRIX} climate model using single particle mass
  spectrometry measurements} {Evaluation of aerosol mixing state classes in the
  {GISS}-model{E}-{MATRIX} climate model using single particle mass
  spectrometry measurements}.{\BBCQ}
\newblock
\APACjournalVolNumPages{J. Geophys. Res.}{118}{}{9834--9844}.
\newblock
\begin{APACrefDOI} \doi{10.1002/jgrd.50700} \end{APACrefDOI}
\PrintBackRefs{\CurrentBib}

\bibitem [\protect \citeauthoryear {%
Bauer%
\ \protect \BOthers {.}}{%
Bauer%
\ \protect \BOthers {.}}{%
{\protect \APACyear {2008}}%
}]{%
Bauer_et_al_2008}
\APACinsertmetastar {%
Bauer_et_al_2008}%
\begin{APACrefauthors}%
Bauer, S.%
, Wright, D.%
, Koch, D.%
, Lewis, E.%
, McGraw, R.%
, Chang, L\BHBI S.%
\BDBL {}Ruedy, R.%
\end{APACrefauthors}%
\unskip\
\newblock
\APACrefYearMonthDay{2008}{}{}.
\newblock
{\BBOQ}\APACrefatitle {{MATRIX} ({M}ulticonfiguration {A}erosol {TR}acker of
  m{IX}ing state): {A}n aerosol microphysical module for global atmospheric
  models} {{MATRIX} ({M}ulticonfiguration {A}erosol {TR}acker of m{IX}ing
  state): {A}n aerosol microphysical module for global atmospheric
  models}.{\BBCQ}
\newblock
\APACjournalVolNumPages{Atmos. Chem. Phys.}{8}{}{6603--6035}.
\newblock
\begin{APACrefDOI} \doi{10.5194/acp-8-6003-2008} \end{APACrefDOI}
\PrintBackRefs{\CurrentBib}

\bibitem [\protect \citeauthoryear {%
Bellouin%
\ \protect \BOthers {.}}{%
Bellouin%
\ \protect \BOthers {.}}{%
{\protect \APACyear {2020}}%
}]{%
Bellouin_et_al_2020}
\APACinsertmetastar {%
Bellouin_et_al_2020}%
\begin{APACrefauthors}%
Bellouin, N.%
, Quaas, J.%
, Gryspeerdt, E.%
, Kinne, S.%
, Stier, P.%
, Watson-Parris, D.%
\BDBL {}Stevens, B.%
\end{APACrefauthors}%
\unskip\
\newblock
\APACrefYearMonthDay{2020}{}{}.
\newblock
{\BBOQ}\APACrefatitle {Bounding Global Aerosol Radiative Forcing of Climate
  Change} {Bounding global aerosol radiative forcing of climate change}.{\BBCQ}
\newblock
\APACjournalVolNumPages{Rev. Geophys.}{58}{}{}.
\newblock
\begin{APACrefDOI} \doi{10.1029/2019RG000660} \end{APACrefDOI}
\PrintBackRefs{\CurrentBib}

\bibitem [\protect \citeauthoryear {%
Bigg 1953b}{%
Bigg 1953b}{%
{\protect \APACyear {1953}}%
}]{%
Bigg_1953b}
\APACinsertmetastar {%
Bigg_1953b}%
\begin{APACrefauthors}%
Bigg, E.%
\end{APACrefauthors}%
\unskip\
\newblock
\APACrefYearMonthDay{1953}{}{}.
\newblock
{\BBOQ}\APACrefatitle {The formation of atmospheric ice crystals by the
  freezing of droplets} {The formation of atmospheric ice crystals by the
  freezing of droplets}.{\BBCQ}
\newblock
\APACjournalVolNumPages{Q. J. Royal Meteorol. Soc.}{}{}{}.
\newblock
\begin{APACrefDOI} \doi{10.1002/qj.49707934207} \end{APACrefDOI}
\PrintBackRefs{\CurrentBib}

\bibitem [\protect \citeauthoryear {%
Bigg 1953a}{%
Bigg 1953a}{%
{\protect \APACyear {1953}}%
}]{%
Bigg_1953a}
\APACinsertmetastar {%
Bigg_1953a}%
\begin{APACrefauthors}%
Bigg, E.%
\end{APACrefauthors}%
\unskip\
\newblock
\APACrefYearMonthDay{1953}{}{}.
\newblock
{\BBOQ}\APACrefatitle {The Supercooling of Water} {The supercooling of
  water}.{\BBCQ}
\newblock
\APACjournalVolNumPages{Proc. Phys. Soc. B}{66}{}{}.
\newblock
\begin{APACrefDOI} \doi{10.1088/0370-1301/66/8/309} \end{APACrefDOI}
\PrintBackRefs{\CurrentBib}

\bibitem [\protect \citeauthoryear {%
Brdar%
\ \BBA {} Seifert%
}{%
Brdar%
\ \BBA {} Seifert%
}{%
{\protect \APACyear {2018}}%
}]{%
Brdar_and_Seifert_2018}
\APACinsertmetastar {%
Brdar_and_Seifert_2018}%
\begin{APACrefauthors}%
Brdar, S.%
\BCBT {}\ \BBA {} Seifert, A.%
\end{APACrefauthors}%
\unskip\
\newblock
\APACrefYearMonthDay{2018}{}{}.
\newblock
{\BBOQ}\APACrefatitle {{McSnow}: A {Monte-Carlo} Particle Model for Riming and
  Aggregation of Ice Particles in a Multidimensional Microphysical Phase Space}
  {{McSnow}: A {Monte-Carlo} particle model for riming and aggregation of ice
  particles in a multidimensional microphysical phase space}.{\BBCQ}
\newblock
\APACjournalVolNumPages{J. Adv. Model. Earth Syst.}{10}{}{}.
\newblock
\begin{APACrefDOI} \doi{10.1002/2017MS001167} \end{APACrefDOI}
\PrintBackRefs{\CurrentBib}

\bibitem [\protect \citeauthoryear {%
Burrows%
\ \protect \BOthers {.}}{%
Burrows%
\ \protect \BOthers {.}}{%
{\protect \APACyear {2022}}%
}]{%
Burrows_et_al_2022}
\APACinsertmetastar {%
Burrows_et_al_2022}%
\begin{APACrefauthors}%
Burrows, S.%
, McCluskey, C.%
, Cornwell, G.%
, Steinke, I.%
, Zhang, K.%
, Zhao, B.%
\BDBL {}DeMott, P.%
\end{APACrefauthors}%
\unskip\
\newblock
\APACrefYearMonthDay{2022}{}{}.
\newblock
{\BBOQ}\APACrefatitle {Ice-Nucleating Particles That Impact Clouds and Climate:
  Observational and Modeling Research Needs} {Ice-nucleating particles that
  impact clouds and climate: Observational and modeling research needs}.{\BBCQ}
\newblock
\APACjournalVolNumPages{Rev. Geophys.}{60}{}{}.
\newblock
\begin{APACrefDOI} \doi{10.1029/2021RG000745} \end{APACrefDOI}
\PrintBackRefs{\CurrentBib}

\bibitem [\protect \citeauthoryear {%
Carte%
}{%
Carte%
}{%
{\protect \APACyear {1959}}%
}]{%
Carte_1959}
\APACinsertmetastar {%
Carte_1959}%
\begin{APACrefauthors}%
Carte, A.%
\end{APACrefauthors}%
\unskip\
\newblock
\APACrefYearMonthDay{1959}{}{}.
\newblock
{\BBOQ}\APACrefatitle {Probability of Freezing} {Probability of
  freezing}.{\BBCQ}
\newblock
\APACjournalVolNumPages{Proc. Phys. Soc.}{73}{}{}.
\newblock
\begin{APACrefDOI} \doi{10.1088/0370-1328/73/2/126} \end{APACrefDOI}
\PrintBackRefs{\CurrentBib}

\bibitem [\protect \citeauthoryear {%
Ceppi%
, Brient%
, Zelinka%
\BCBL {}\ \BBA {} Hartmann%
}{%
Ceppi%
\ \protect \BOthers {.}}{%
{\protect \APACyear {2017}}%
}]{%
Ceppi_et_al_2017}
\APACinsertmetastar {%
Ceppi_et_al_2017}%
\begin{APACrefauthors}%
Ceppi, P.%
, Brient, F.%
, Zelinka, M.%
\BCBL {}\ \BBA {} Hartmann, D.%
\end{APACrefauthors}%
\unskip\
\newblock
\APACrefYearMonthDay{2017}{}{}.
\newblock
{\BBOQ}\APACrefatitle {Cloud feedback mechanisms and their representation in
  global climate models} {Cloud feedback mechanisms and their representation in
  global climate models}.{\BBCQ}
\newblock
\APACjournalVolNumPages{WIREs Clim. Change.}{8}{}{}.
\newblock
\begin{APACrefDOI} \doi{10.1002/wcc.465} \end{APACrefDOI}
\PrintBackRefs{\CurrentBib}

\bibitem [\protect \citeauthoryear {%
Connolly%
\ \protect \BOthers {.}}{%
Connolly%
\ \protect \BOthers {.}}{%
{\protect \APACyear {2009}}%
}]{%
Conolly_et_al_2009}
\APACinsertmetastar {%
Conolly_et_al_2009}%
\begin{APACrefauthors}%
Connolly, P.%
, Möhler, O.%
, Field, P.%
, Saathoff, H.%
, Burgess, R.%
, Choularton, T.%
\BCBL {}\ \BBA {} Gallagher, M.%
\end{APACrefauthors}%
\unskip\
\newblock
\APACrefYearMonthDay{2009}{}{}.
\newblock
{\BBOQ}\APACrefatitle {Studies of heterogeneous freezing by three different
  desert dust samples} {Studies of heterogeneous freezing by three different
  desert dust samples}.{\BBCQ}
\newblock
\APACjournalVolNumPages{Atmos. Chem. Phys.}{9}{}{}.
\newblock
\begin{APACrefDOI} \doi{10.5194/acp-9-2805-2009} \end{APACrefDOI}
\PrintBackRefs{\CurrentBib}

\bibitem [\protect \citeauthoryear {%
Cornwell%
, McCluskey%
, DeMott%
, Prather%
\BCBL {}\ \BBA {} Burrows%
}{%
Cornwell%
\ \protect \BOthers {.}}{%
{\protect \APACyear {2021}}%
}]{%
Cornwell_et_al_2021}
\APACinsertmetastar {%
Cornwell_et_al_2021}%
\begin{APACrefauthors}%
Cornwell, G.%
, McCluskey, C.%
, DeMott, P.%
, Prather, K.%
\BCBL {}\ \BBA {} Burrows, S.%
\end{APACrefauthors}%
\unskip\
\newblock
\APACrefYearMonthDay{2021}{}{}.
\newblock
{\BBOQ}\APACrefatitle {Development of Heterogeneous Ice Nucleation Rate
  Coefficient Parameterizations From Ambient Measurements} {Development of
  heterogeneous ice nucleation rate coefficient parameterizations from ambient
  measurements}.{\BBCQ}
\newblock
\APACjournalVolNumPages{Geophys. Res. Lett.}{48}{}{}.
\newblock
\begin{APACrefDOI} \doi{10.1029/2021GL095359} \end{APACrefDOI}
\PrintBackRefs{\CurrentBib}

\bibitem [\protect \citeauthoryear {%
Curtis%
, Michelotti%
, Riemer%
, Heath%
\BCBL {}\ \BBA {} West%
}{%
Curtis%
\ \protect \BOthers {.}}{%
{\protect \APACyear {2016}}%
}]{%
Curtis_et_al_2016}
\APACinsertmetastar {%
Curtis_et_al_2016}%
\begin{APACrefauthors}%
Curtis, J.%
, Michelotti, M.%
, Riemer, N.%
, Heath, M.%
\BCBL {}\ \BBA {} West, M.%
\end{APACrefauthors}%
\unskip\
\newblock
\APACrefYearMonthDay{2016}{}{}.
\newblock
{\BBOQ}\APACrefatitle {Accelerated simulation of stochastic particle removal
  processes in particle-resolved aerosol models} {Accelerated simulation of
  stochastic particle removal processes in particle-resolved aerosol
  models}.{\BBCQ}
\newblock
\APACjournalVolNumPages{J. Comp. Phys.}{322}{}{}.
\newblock
\begin{APACrefDOI} \doi{10.1016/j.jcp.2016.06.029} \end{APACrefDOI}
\PrintBackRefs{\CurrentBib}

\bibitem [\protect \citeauthoryear {%
Curtis%
, Riemer%
\BCBL {}\ \BBA {} West%
}{%
Curtis%
\ \protect \BOthers {.}}{%
{\protect \APACyear {2024}}%
}]{%
Curtis_et_al_2024}
\APACinsertmetastar {%
Curtis_et_al_2024}%
\begin{APACrefauthors}%
Curtis, J.%
, Riemer, N.%
\BCBL {}\ \BBA {} West, M.%
\end{APACrefauthors}%
\unskip\
\newblock
\APACrefYearMonthDay{2024}{}{}.
\newblock
{\BBOQ}\APACrefatitle {Explicit stochastic advection algorithms for the
  regional scale particle-resolved atmospheric aerosol model {WRF-PartMC}
  (v1.0)} {Explicit stochastic advection algorithms for the regional scale
  particle-resolved atmospheric aerosol model {WRF-PartMC} (v1.0)}.{\BBCQ}
\newblock
\APACjournalVolNumPages{EGUsphere}{}{}{}.
\newblock
\begin{APACrefDOI} \doi{10.5194/egusphere-2024-825} \end{APACrefDOI}
\PrintBackRefs{\CurrentBib}

\bibitem [\protect \citeauthoryear {%
de Jong%
, Mackay%
, Bulenok%
, Jaruga%
\BCBL {}\ \BBA {} Arabas%
}{%
de Jong%
, Mackay%
\BCBL {}\ \protect \BOthers {.}}{%
{\protect \APACyear {2023}}%
}]{%
deJong_et_al_2023_GMD}
\APACinsertmetastar {%
deJong_et_al_2023_GMD}%
\begin{APACrefauthors}%
de Jong, E.%
, Mackay, J.%
, Bulenok, O.%
, Jaruga, A.%
\BCBL {}\ \BBA {} Arabas, S.%
\end{APACrefauthors}%
\unskip\
\newblock
\APACrefYearMonthDay{2023}{}{}.
\newblock
{\BBOQ}\APACrefatitle {Breakups are Complicated: An Efficient Representation of
  Collisional Breakup in the Superdroplet Method} {Breakups are complicated: An
  efficient representation of collisional breakup in the superdroplet
  method}.{\BBCQ}
\newblock
\APACjournalVolNumPages{Geosci. Model Dev.}{16}{}{}.
\newblock
\begin{APACrefDOI} \doi{10.5194/gmd-16-4193-2023} \end{APACrefDOI}
\PrintBackRefs{\CurrentBib}

\bibitem [\protect \citeauthoryear {%
de Jong%
, Singer%
\BCBL {}\ \protect \BOthers {.}}{%
de Jong%
, Singer%
\BCBL {}\ \protect \BOthers {.}}{%
{\protect \APACyear {2023}}%
}]{%
deJong_et_al_2023_JOSS}
\APACinsertmetastar {%
deJong_et_al_2023_JOSS}%
\begin{APACrefauthors}%
de Jong, E.%
, Singer, C.%
, Azimi, S.%
, Bartman, P.%
, Bulenok, O.%
, Derlatka, K.%
\BDBL {}Arabas, S.%
\end{APACrefauthors}%
\unskip\
\newblock
\APACrefYearMonthDay{2023}{}{}.
\newblock
{\BBOQ}\APACrefatitle {New developments in {PySDM} and {PySDM-examples} v2:
  collisional breakup, immersion freezing, dry aerosol initialization, and
  adaptive time-stepping} {New developments in {PySDM} and {PySDM-examples} v2:
  collisional breakup, immersion freezing, dry aerosol initialization, and
  adaptive time-stepping}.{\BBCQ}
\newblock
\APACjournalVolNumPages{J. Open Source Soft.}{8}{}{}.
\newblock
\begin{APACrefDOI} \doi{10.21105/joss.04968} \end{APACrefDOI}
\PrintBackRefs{\CurrentBib}

\bibitem [\protect \citeauthoryear {%
Després%
\ \protect \BOthers {.}}{%
Després%
\ \protect \BOthers {.}}{%
{\protect \APACyear {2012}}%
}]{%
Despres_et_al_2012}
\APACinsertmetastar {%
Despres_et_al_2012}%
\begin{APACrefauthors}%
Després, V.%
, Huffman, J.%
, Burrows, S.%
, Hoose, C.%
, Safatov, A.%
, Buryak, G.%
\BDBL {}Jaenicke, R.%
\end{APACrefauthors}%
\unskip\
\newblock
\APACrefYearMonthDay{2012}{}{}.
\newblock
{\BBOQ}\APACrefatitle {Primary biological aerosol particles in the atmosphere:
  a review} {Primary biological aerosol particles in the atmosphere: a
  review}.{\BBCQ}
\newblock
\APACjournalVolNumPages{Tellus B}{64}{}{}.
\newblock
\begin{APACrefDOI} \doi{10.3402/tellusb.v64i0.15598} \end{APACrefDOI}
\PrintBackRefs{\CurrentBib}

\bibitem [\protect \citeauthoryear {%
DeVille%
, Riemer%
\BCBL {}\ \BBA {} West%
}{%
DeVille%
\ \protect \BOthers {.}}{%
{\protect \APACyear {2019}}%
}]{%
DeVille_et_al_2019}
\APACinsertmetastar {%
DeVille_et_al_2019}%
\begin{APACrefauthors}%
DeVille, L.%
, Riemer, N.%
\BCBL {}\ \BBA {} West, M.%
\end{APACrefauthors}%
\unskip\
\newblock
\APACrefYearMonthDay{2019}{}{}.
\newblock
{\BBOQ}\APACrefatitle {Convergence of a Generalized Weighted Flow Algorithm for
  Stochastic Particle Coagulation} {Convergence of a generalized weighted flow
  algorithm for stochastic particle coagulation}.{\BBCQ}
\newblock
\APACjournalVolNumPages{J. Comp. Dyn.}{2019}{}{}.
\newblock
\begin{APACrefDOI} \doi{10.3934/jcd.2019003} \end{APACrefDOI}
\PrintBackRefs{\CurrentBib}

\bibitem [\protect \citeauthoryear {%
Dziekan%
\ \BBA {} Pawlowska%
}{%
Dziekan%
\ \BBA {} Pawlowska%
}{%
{\protect \APACyear {2017}}%
}]{%
Dziekan_and_Pawlowska_2017}
\APACinsertmetastar {%
Dziekan_and_Pawlowska_2017}%
\begin{APACrefauthors}%
Dziekan, P.%
\BCBT {}\ \BBA {} Pawlowska, H.%
\end{APACrefauthors}%
\unskip\
\newblock
\APACrefYearMonthDay{2017}{}{}.
\newblock
{\BBOQ}\APACrefatitle {Stochastic coalescence in {L}agrangian cloud
  microphysics} {Stochastic coalescence in {L}agrangian cloud
  microphysics}.{\BBCQ}
\newblock
\APACjournalVolNumPages{Atmos. Chem. Phys.}{17}{}{}.
\newblock
\begin{APACrefDOI} \doi{10.5194/acp-17-13509-2017} \end{APACrefDOI}
\PrintBackRefs{\CurrentBib}

\bibitem [\protect \citeauthoryear {%
Ervens%
\ \BBA {} Feingold%
}{%
Ervens%
\ \BBA {} Feingold%
}{%
{\protect \APACyear {2013}}%
}]{%
Ervens_and_Feingold_2013}
\APACinsertmetastar {%
Ervens_and_Feingold_2013}%
\begin{APACrefauthors}%
Ervens, B.%
\BCBT {}\ \BBA {} Feingold, G.%
\end{APACrefauthors}%
\unskip\
\newblock
\APACrefYearMonthDay{2013}{}{}.
\newblock
{\BBOQ}\APACrefatitle {Sensitivities of immersion freezing: Reconciling
  classical nucleation theory and deterministic expressions} {Sensitivities of
  immersion freezing: Reconciling classical nucleation theory and deterministic
  expressions}.{\BBCQ}
\newblock
\APACjournalVolNumPages{Geophys. Res. Lett.}{40}{}{}.
\newblock
\begin{APACrefDOI} \doi{10.1002/grl.50580} \end{APACrefDOI}
\PrintBackRefs{\CurrentBib}

\bibitem [\protect \citeauthoryear {%
Fletcher%
}{%
Fletcher%
}{%
{\protect \APACyear {1958}}%
}]{%
Fletcher_1958}
\APACinsertmetastar {%
Fletcher_1958}%
\begin{APACrefauthors}%
Fletcher, N.%
\end{APACrefauthors}%
\unskip\
\newblock
\APACrefYearMonthDay{1958}{}{}.
\newblock
{\BBOQ}\APACrefatitle {Time lag in ice crystal nucleation in the atmosphere.
  Part II theoretical} {Time lag in ice crystal nucleation in the atmosphere.
  part ii theoretical}.{\BBCQ}
\newblock
\APACjournalVolNumPages{Bull. Obs. Puy de Dôme}{1}{}{}.
\newblock
\begin{APACrefURL}
  \url{https://web.archive.org/web/*/https://www.phys.unsw.edu.au/music/people/publications/Fletcher1958b.pdf}
  \end{APACrefURL}
\PrintBackRefs{\CurrentBib}

\bibitem [\protect \citeauthoryear {%
Fletcher%
}{%
Fletcher%
}{%
{\protect \APACyear {1969}}%
}]{%
Fletcher_1969}
\APACinsertmetastar {%
Fletcher_1969}%
\begin{APACrefauthors}%
Fletcher, N.%
\end{APACrefauthors}%
\unskip\
\newblock
\APACrefYearMonthDay{1969}{}{}.
\newblock
{\BBOQ}\APACrefatitle {Active Sites and Ice Crystal Nucleation} {Active sites
  and ice crystal nucleation}.{\BBCQ}
\newblock
\APACjournalVolNumPages{J. Atmos. Sci.}{26}{}{}.
\newblock
\begin{APACrefDOI} \doi{10.1175/1520-0469(1969)026<1266:ASAICN>2.0.CO;2}
  \end{APACrefDOI}
\PrintBackRefs{\CurrentBib}

\bibitem [\protect \citeauthoryear {%
Fornea%
, Brooks%
, Dooley%
\BCBL {}\ \BBA {} Saha%
}{%
Fornea%
\ \protect \BOthers {.}}{%
{\protect \APACyear {2009}}%
}]{%
Fornea_et_al_2009}
\APACinsertmetastar {%
Fornea_et_al_2009}%
\begin{APACrefauthors}%
Fornea, A.%
, Brooks, S.%
, Dooley, J.%
\BCBL {}\ \BBA {} Saha, A.%
\end{APACrefauthors}%
\unskip\
\newblock
\APACrefYearMonthDay{2009}{}{}.
\newblock
{\BBOQ}\APACrefatitle {Heterogeneous freezing of ice on atmospheric aerosols
  containing ash, soot, and soil} {Heterogeneous freezing of ice on atmospheric
  aerosols containing ash, soot, and soil}.{\BBCQ}
\newblock
\APACjournalVolNumPages{J. Geophys. Res. Atmos.}{114}{D13}{}.
\newblock
\begin{APACrefDOI} \doi{10.1029/2009JD011958} \end{APACrefDOI}
\PrintBackRefs{\CurrentBib}

\bibitem [\protect \citeauthoryear {%
Fridlind%
\ \BBA {} Ackerman%
}{%
Fridlind%
\ \BBA {} Ackerman%
}{%
{\protect \APACyear {2018}}%
}]{%
Fridlind_and_Ackerman_2018}
\APACinsertmetastar {%
Fridlind_and_Ackerman_2018}%
\begin{APACrefauthors}%
Fridlind, A.%
\BCBT {}\ \BBA {} Ackerman, A.%
\end{APACrefauthors}%
\unskip\
\newblock
\APACrefYearMonthDay{2018}{}{}.
\newblock
{\BBOQ}\APACrefatitle {Simulations of Arctic Mixed-Phase Boundary Layer Clouds:
  Advances in Understanding and Outstanding Questions} {Simulations of arctic
  mixed-phase boundary layer clouds: Advances in understanding and outstanding
  questions}.{\BBCQ}
\newblock
\BIn{} C.~Andronache\ (\BED), \APACrefbtitle {Mixed-Phase Clouds: Observations
  and Modeling} {Mixed-phase clouds: Observations and modeling}\ (\BPGS\
  153--183).
\newblock
\APACaddressPublisher{}{Elsevier}.
\newblock
\begin{APACrefDOI} \doi{10.1016/B978-0-12-810549-8.00007-6} \end{APACrefDOI}
\PrintBackRefs{\CurrentBib}

\bibitem [\protect \citeauthoryear {%
Fridlind%
\ \protect \BOthers {.}}{%
Fridlind%
\ \protect \BOthers {.}}{%
{\protect \APACyear {2012}}%
}]{%
Fridlind_et_al_2012}
\APACinsertmetastar {%
Fridlind_et_al_2012}%
\begin{APACrefauthors}%
Fridlind, A.%
, Van~Diedenhoven, B.%
, Ackerman, A.%
, Avramov, A.%
, Mrowiec, A.%
, Morrison, H.%
\BDBL {}Shupe, M.%
\end{APACrefauthors}%
\unskip\
\newblock
\APACrefYearMonthDay{2012}{}{}.
\newblock
{\BBOQ}\APACrefatitle {A {FIRE-ACE/SHEBA} Case Study of Mixed-Phase Arctic
  Boundary Layer Clouds: Entrainment Rate Limitations on Rapid Primary Ice
  Nucleation Processes} {A {FIRE-ACE/SHEBA} case study of mixed-phase arctic
  boundary layer clouds: Entrainment rate limitations on rapid primary ice
  nucleation processes}.{\BBCQ}
\newblock
\APACjournalVolNumPages{J. Atmos. Sci.}{69}{}{}.
\newblock
\begin{APACrefDOI} \doi{10.1175/JAS-D-11-052.1} \end{APACrefDOI}
\PrintBackRefs{\CurrentBib}

\bibitem [\protect \citeauthoryear {%
Frostenberg%
\ \protect \BOthers {.}}{%
Frostenberg%
\ \protect \BOthers {.}}{%
{\protect \APACyear {2023}}%
}]{%
Frostenberg_et_al_2022}
\APACinsertmetastar {%
Frostenberg_et_al_2022}%
\begin{APACrefauthors}%
Frostenberg, H.%
, Welti, A.%
, Luhr, M.%
, Savre, J.%
, Thomson, E.%
\BCBL {}\ \BBA {} Ickes, L.%
\end{APACrefauthors}%
\unskip\
\newblock
\APACrefYearMonthDay{2023}{}{}.
\newblock
{\BBOQ}\APACrefatitle {The chance of freezing – a conceptional study to
  parameterize temperature-dependent freezing by including randomness of
  ice-nucleating particle concentrations} {The chance of freezing – a
  conceptional study to parameterize temperature-dependent freezing by
  including randomness of ice-nucleating particle concentrations}.{\BBCQ}
\newblock
\APACjournalVolNumPages{Atmos. Chem. Phys.}{23}{}{}.
\newblock
\begin{APACrefDOI} \doi{10.5194/acp-23-10883-2023} \end{APACrefDOI}
\PrintBackRefs{\CurrentBib}

\bibitem [\protect \citeauthoryear {%
Fröhlich-Nowoisky%
\ \protect \BOthers {.}}{%
Fröhlich-Nowoisky%
\ \protect \BOthers {.}}{%
{\protect \APACyear {2016}}%
}]{%
Froechlich_Nowoisky_et_al_2016}
\APACinsertmetastar {%
Froechlich_Nowoisky_et_al_2016}%
\begin{APACrefauthors}%
Fröhlich-Nowoisky, J.%
, Kampf, C.%
, Weber, B.%
, Huffman, J.%
, Pöhlker, C.%
, Andreae, M.%
\BDBL {}Pöschl, U.%
\end{APACrefauthors}%
\unskip\
\newblock
\APACrefYearMonthDay{2016}{}{}.
\newblock
{\BBOQ}\APACrefatitle {Bioaerosols in the Earth system: Climate, health, and
  ecosystem interactions} {Bioaerosols in the earth system: Climate, health,
  and ecosystem interactions}.{\BBCQ}
\newblock
\APACjournalVolNumPages{Atmos. Res.}{182}{}{346--376}.
\newblock
\begin{APACrefDOI} \doi{10.1016/j.atmosres.2016.07.018} \end{APACrefDOI}
\PrintBackRefs{\CurrentBib}

\bibitem [\protect \citeauthoryear {%
Gedzelman%
\ \BBA {} Arnold%
}{%
Gedzelman%
\ \BBA {} Arnold%
}{%
{\protect \APACyear {1993}}%
}]{%
Gedzelman_and_Arnold_1993}
\APACinsertmetastar {%
Gedzelman_and_Arnold_1993}%
\begin{APACrefauthors}%
Gedzelman, S.%
\BCBT {}\ \BBA {} Arnold, R.%
\end{APACrefauthors}%
\unskip\
\newblock
\APACrefYearMonthDay{1993}{}{}.
\newblock
{\BBOQ}\APACrefatitle {The Form of Cyclonic Precipitation and Its Thermal
  Impact} {The form of cyclonic precipitation and its thermal impact}.{\BBCQ}
\newblock
\APACjournalVolNumPages{Mon. Wea. Rev.}{121}{}{}.
\newblock
\begin{APACrefDOI} \doi{10.1175/1520-0493(1993)121<1957:TFOCPA>2.0.CO;2}
  \end{APACrefDOI}
\PrintBackRefs{\CurrentBib}

\bibitem [\protect \citeauthoryear {%
Gedzelman%
\ \BBA {} Arnold%
}{%
Gedzelman%
\ \BBA {} Arnold%
}{%
{\protect \APACyear {1994}}%
}]{%
Gedzelman_and_Arnold_1994}
\APACinsertmetastar {%
Gedzelman_and_Arnold_1994}%
\begin{APACrefauthors}%
Gedzelman, S.%
\BCBT {}\ \BBA {} Arnold, R.%
\end{APACrefauthors}%
\unskip\
\newblock
\APACrefYearMonthDay{1994}{}{}.
\newblock
{\BBOQ}\APACrefatitle {Modeling the isotopic composition of precipitation}
  {Modeling the isotopic composition of precipitation}.{\BBCQ}
\newblock
\APACjournalVolNumPages{J. Geophys. Res. Atmos.}{99}{D5}{}.
\PrintBackRefs{\CurrentBib}

\bibitem [\protect \citeauthoryear {%
Grabowski%
}{%
Grabowski%
}{%
{\protect \APACyear {1998}}%
}]{%
Grabowski_1998}
\APACinsertmetastar {%
Grabowski_1998}%
\begin{APACrefauthors}%
Grabowski, W.%
\end{APACrefauthors}%
\unskip\
\newblock
\APACrefYearMonthDay{1998}{}{}.
\newblock
{\BBOQ}\APACrefatitle {Toward Cloud Resolving Modeling of Large-Scale Tropical
  Circulations: A Simple Cloud Microphysics Parameterization} {Toward cloud
  resolving modeling of large-scale tropical circulations: A simple cloud
  microphysics parameterization}.{\BBCQ}
\newblock
\APACjournalVolNumPages{J. Atmos. Sci.}{55}{}{}.
\newblock
\begin{APACrefDOI} \doi{10.1175/1520-0469(1998)055<3283:TCRMOL>2.0.CO;2}
  \end{APACrefDOI}
\PrintBackRefs{\CurrentBib}

\bibitem [\protect \citeauthoryear {%
Grabowski%
}{%
Grabowski%
}{%
{\protect \APACyear {1999}}%
}]{%
Grabowski_1999}
\APACinsertmetastar {%
Grabowski_1999}%
\begin{APACrefauthors}%
Grabowski, W.%
\end{APACrefauthors}%
\unskip\
\newblock
\APACrefYearMonthDay{1999}{}{}.
\newblock
{\BBOQ}\APACrefatitle {A parameterization of cloud microphysics for long-term
  cloud-resolving modeling of tropical convection} {A parameterization of cloud
  microphysics for long-term cloud-resolving modeling of tropical
  convection}.{\BBCQ}
\newblock
\APACjournalVolNumPages{Atmos. Res.}{52}{}{}.
\newblock
\begin{APACrefDOI} \doi{10.1016/S0169-8095(99)00029-0} \end{APACrefDOI}
\PrintBackRefs{\CurrentBib}

\bibitem [\protect \citeauthoryear {%
Grabowski%
, Dziekan%
\BCBL {}\ \BBA {} Pawlowska%
}{%
Grabowski%
\ \protect \BOthers {.}}{%
{\protect \APACyear {2018}}%
}]{%
Grabowski_et_al_2018}
\APACinsertmetastar {%
Grabowski_et_al_2018}%
\begin{APACrefauthors}%
Grabowski, W.%
, Dziekan, P.%
\BCBL {}\ \BBA {} Pawlowska, H.%
\end{APACrefauthors}%
\unskip\
\newblock
\APACrefYearMonthDay{2018}{}{}.
\newblock
{\BBOQ}\APACrefatitle {Lagrangian condensation microphysics with {T}womey {CCN}
  activation} {Lagrangian condensation microphysics with {T}womey {CCN}
  activation}.{\BBCQ}
\newblock
\APACjournalVolNumPages{Geosci. Model Dev.}{}{}{}.
\newblock
\begin{APACrefDOI} \doi{10.5194/gmd-11-103-2018} \end{APACrefDOI}
\PrintBackRefs{\CurrentBib}

\bibitem [\protect \citeauthoryear {%
Grabowski%
\ \protect \BOthers {.}}{%
Grabowski%
\ \protect \BOthers {.}}{%
{\protect \APACyear {2019}}%
}]{%
Grabowski_et_al_2019}
\APACinsertmetastar {%
Grabowski_et_al_2019}%
\begin{APACrefauthors}%
Grabowski, W.%
, Morrison, H.%
, Shima, S.%
, Abade, G.%
, Dziekan, P.%
\BCBL {}\ \BBA {} Pawlowska, H.%
\end{APACrefauthors}%
\unskip\
\newblock
\APACrefYearMonthDay{2019}{}{}.
\newblock
{\BBOQ}\APACrefatitle {Modeling of Cloud Microphysics: Can We Do Better?}
  {Modeling of cloud microphysics: Can we do better?}{\BBCQ}
\newblock
\APACjournalVolNumPages{Bull. Am. Meteorol. Soc.}{100}{}{}.
\newblock
\begin{APACrefDOI} \doi{10.1175/BAMS-D-18-0005.1} \end{APACrefDOI}
\PrintBackRefs{\CurrentBib}

\bibitem [\protect \citeauthoryear {%
Herbert%
, Murray%
, Whale%
, Dobbie%
\BCBL {}\ \BBA {} Atkinson%
}{%
Herbert%
\ \protect \BOthers {.}}{%
{\protect \APACyear {2014}}%
}]{%
Herbert_et_al_2014}
\APACinsertmetastar {%
Herbert_et_al_2014}%
\begin{APACrefauthors}%
Herbert, R.%
, Murray, B.%
, Whale, T.%
, Dobbie, S.%
\BCBL {}\ \BBA {} Atkinson, J.%
\end{APACrefauthors}%
\unskip\
\newblock
\APACrefYearMonthDay{2014}{}{}.
\newblock
{\BBOQ}\APACrefatitle {Representing time-dependent freezing behaviour in
  immersion mode ice nucleation} {Representing time-dependent freezing
  behaviour in immersion mode ice nucleation}.{\BBCQ}
\newblock
\APACjournalVolNumPages{Atmos. Chem. Phys.}{14}{}{}.
\newblock
\begin{APACrefDOI} \doi{10.5194/acp-14-8501-2014} \end{APACrefDOI}
\PrintBackRefs{\CurrentBib}

\bibitem [\protect \citeauthoryear {%
Hoffmann%
\ \BBA {} Feingold%
}{%
Hoffmann%
\ \BBA {} Feingold%
}{%
{\protect \APACyear {2019}}%
}]{%
Hoffmann_et_al_2019}
\APACinsertmetastar {%
Hoffmann_et_al_2019}%
\begin{APACrefauthors}%
Hoffmann, F.%
\BCBT {}\ \BBA {} Feingold, G.%
\end{APACrefauthors}%
\unskip\
\newblock
\APACrefYearMonthDay{2019}{}{}.
\newblock
{\BBOQ}\APACrefatitle {Entrainment and Mixing in Stratocumulus: Effects of a
  New Explicit Subgrid-Scale Scheme for Large-Eddy Simulations with
  Particle-Based Microphysics} {Entrainment and mixing in stratocumulus:
  Effects of a new explicit subgrid-scale scheme for large-eddy simulations
  with particle-based microphysics}.{\BBCQ}
\newblock
\APACjournalVolNumPages{J. Atmos. Sci.}{76}{}{}.
\newblock
\begin{APACrefDOI} \doi{10.1175/JAS-D-18-0318.1} \end{APACrefDOI}
\PrintBackRefs{\CurrentBib}

\bibitem [\protect \citeauthoryear {%
Hoffmann%
\ \BBA {} Feingold%
}{%
Hoffmann%
\ \BBA {} Feingold%
}{%
{\protect \APACyear {2023}}%
}]{%
Hoffmann_and_Feingold_2023}
\APACinsertmetastar {%
Hoffmann_and_Feingold_2023}%
\begin{APACrefauthors}%
Hoffmann, F.%
\BCBT {}\ \BBA {} Feingold, G.%
\end{APACrefauthors}%
\unskip\
\newblock
\APACrefYearMonthDay{2023}{}{}.
\newblock
{\BBOQ}\APACrefatitle {A Note on Aerosol Processing by Droplet
  Collision-Coalescence} {A note on aerosol processing by droplet
  collision-coalescence}.{\BBCQ}
\newblock
\APACjournalVolNumPages{Geophys. Res. Lett.}{}{}{}.
\newblock
\begin{APACrefDOI} \doi{10.1029/2023GL103716} \end{APACrefDOI}
\PrintBackRefs{\CurrentBib}

\bibitem [\protect \citeauthoryear {%
Hoose%
\ \BBA {} M\"ohler%
}{%
Hoose%
\ \BBA {} M\"ohler%
}{%
{\protect \APACyear {2012}}%
}]{%
Hoose_and_Moehler_2012}
\APACinsertmetastar {%
Hoose_and_Moehler_2012}%
\begin{APACrefauthors}%
Hoose, C.%
\BCBT {}\ \BBA {} M\"ohler, O.%
\end{APACrefauthors}%
\unskip\
\newblock
\APACrefYearMonthDay{2012}{}{}.
\newblock
{\BBOQ}\APACrefatitle {Heterogeneous ice nucleation on atmospheric aerosols: a
  review of results from laboratory experiments} {Heterogeneous ice nucleation
  on atmospheric aerosols: a review of results from laboratory
  experiments}.{\BBCQ}
\newblock
\APACjournalVolNumPages{Atmos. Chem. Phys.}{12}{}{}.
\newblock
\begin{APACrefDOI} \doi{10.5194/acp-12-9817-2012} \end{APACrefDOI}
\PrintBackRefs{\CurrentBib}

\bibitem [\protect \citeauthoryear {%
Isaac%
\ \BBA {} Douglas%
}{%
Isaac%
\ \BBA {} Douglas%
}{%
{\protect \APACyear {1972}}%
}]{%
Isaac_and_Douglas_1972}
\APACinsertmetastar {%
Isaac_and_Douglas_1972}%
\begin{APACrefauthors}%
Isaac, G.%
\BCBT {}\ \BBA {} Douglas, R.%
\end{APACrefauthors}%
\unskip\
\newblock
\APACrefYearMonthDay{1972}{}{}.
\newblock
{\BBOQ}\APACrefatitle {Another ``Time Lag'' in the Activation of Atmospheric
  Ice Nuclei} {Another ``time lag'' in the activation of atmospheric ice
  nuclei}.{\BBCQ}
\newblock
\APACjournalVolNumPages{J. Appl. Meteorol.}{11}{}{}.
\newblock
\begin{APACrefDOI} \doi{10.1175/1520-0450(1972)011<0490:ALITAO>2.0.CO;2}
  \end{APACrefDOI}
\PrintBackRefs{\CurrentBib}

\bibitem [\protect \citeauthoryear {%
Jaruga%
\ \BBA {} Pawlowska%
}{%
Jaruga%
\ \BBA {} Pawlowska%
}{%
{\protect \APACyear {2018}}%
}]{%
Jaruga_and_Pawlowska_2018}
\APACinsertmetastar {%
Jaruga_and_Pawlowska_2018}%
\begin{APACrefauthors}%
Jaruga, A.%
\BCBT {}\ \BBA {} Pawlowska, H.%
\end{APACrefauthors}%
\unskip\
\newblock
\APACrefYearMonthDay{2018}{}{}.
\newblock
{\BBOQ}\APACrefatitle {libcloudph++ 2.0: aqueous-phase chemistry extension of
  the particle-based cloud microphysics scheme} {libcloudph++ 2.0:
  aqueous-phase chemistry extension of the particle-based cloud microphysics
  scheme}.{\BBCQ}
\newblock
\APACjournalVolNumPages{Geosci. Model Dev.}{11}{}{}.
\newblock
\begin{APACrefDOI} \doi{10.5194/gmd-11-3623-2018} \end{APACrefDOI}
\PrintBackRefs{\CurrentBib}

\bibitem [\protect \citeauthoryear {%
Jensen%
\ \BBA {} Pfister%
}{%
Jensen%
\ \BBA {} Pfister%
}{%
{\protect \APACyear {2004}}%
}]{%
Jensen_and_Pfister_2004}
\APACinsertmetastar {%
Jensen_and_Pfister_2004}%
\begin{APACrefauthors}%
Jensen, E.%
\BCBT {}\ \BBA {} Pfister, L.%
\end{APACrefauthors}%
\unskip\
\newblock
\APACrefYearMonthDay{2004}{}{}.
\newblock
{\BBOQ}\APACrefatitle {Transport and freeze-drying in the tropical tropopause
  layer} {Transport and freeze-drying in the tropical tropopause layer}.{\BBCQ}
\newblock
\APACjournalVolNumPages{J. Geophys. Res. Atmos.}{109}{}{}.
\newblock
\begin{APACrefDOI} \doi{10.1029/2003JD004022} \end{APACrefDOI}
\PrintBackRefs{\CurrentBib}

\bibitem [\protect \citeauthoryear {%
Kanji%
\ \protect \BOthers {.}}{%
Kanji%
\ \protect \BOthers {.}}{%
{\protect \APACyear {2017}}%
}]{%
Kanji_et_al_2017}
\APACinsertmetastar {%
Kanji_et_al_2017}%
\begin{APACrefauthors}%
Kanji, Z.%
, Ladino, L.%
, Wex, H.%
, Boose, Y.%
, Burkert-Kohn, M.%
, Cziczo, D.%
\BCBL {}\ \BBA {} Kr\"amer, M.%
\end{APACrefauthors}%
\unskip\
\newblock
\APACrefYearMonthDay{2017}{}{}.
\newblock
{\BBOQ}\APACrefatitle {Overview of Ice Nucleating Particles} {Overview of ice
  nucleating particles}.{\BBCQ}
\newblock
\APACjournalVolNumPages{Meteorol. Monogr.}{58}{}{}.
\newblock
\begin{APACrefDOI} \doi{10.1175/AMSMONOGRAPHS-D-16-0006.1} \end{APACrefDOI}
\PrintBackRefs{\CurrentBib}

\bibitem [\protect \citeauthoryear {%
K\"archer%
\ \BBA {} Marcolli%
}{%
K\"archer%
\ \BBA {} Marcolli%
}{%
{\protect \APACyear {2021}}%
}]{%
Kaercher_and_Marcolli_2021}
\APACinsertmetastar {%
Kaercher_and_Marcolli_2021}%
\begin{APACrefauthors}%
K\"archer, B.%
\BCBT {}\ \BBA {} Marcolli, C.%
\end{APACrefauthors}%
\unskip\
\newblock
\APACrefYearMonthDay{2021}{}{}.
\newblock
{\BBOQ}\APACrefatitle {Aerosol--cloud interactions: the representation of
  heterogeneous ice activation in cloud models} {Aerosol--cloud interactions:
  the representation of heterogeneous ice activation in cloud models}.{\BBCQ}
\newblock
\APACjournalVolNumPages{Atmos. Chem. Phys.}{21}{}{}.
\newblock
\begin{APACrefDOI} \doi{10.5194/acp-21-15213-2021} \end{APACrefDOI}
\PrintBackRefs{\CurrentBib}

\bibitem [\protect \citeauthoryear {%
Kaufmann%
, Marcolli%
, Luo%
\BCBL {}\ \BBA {} Peter%
}{%
Kaufmann%
\ \protect \BOthers {.}}{%
{\protect \APACyear {2017}}%
}]{%
Kaufmann_et_al_2017}
\APACinsertmetastar {%
Kaufmann_et_al_2017}%
\begin{APACrefauthors}%
Kaufmann, L.%
, Marcolli, C.%
, Luo, B.%
\BCBL {}\ \BBA {} Peter, T.%
\end{APACrefauthors}%
\unskip\
\newblock
\APACrefYearMonthDay{2017}{}{}.
\newblock
{\BBOQ}\APACrefatitle {Refreeze experiments with water droplets containing
  different types of ice nuclei interpreted by classical nucleation theory}
  {Refreeze experiments with water droplets containing different types of ice
  nuclei interpreted by classical nucleation theory}.{\BBCQ}
\newblock
\APACjournalVolNumPages{Atmos. Chem. Phys.}{17}{}{}.
\newblock
\begin{APACrefDOI} \doi{10.5194/acp-17-3525-2017} \end{APACrefDOI}
\PrintBackRefs{\CurrentBib}

\bibitem [\protect \citeauthoryear {%
Kessler%
}{%
Kessler%
}{%
{\protect \APACyear {1969}}%
}]{%
Kessler_1969}
\APACinsertmetastar {%
Kessler_1969}%
\begin{APACrefauthors}%
Kessler, E.%
\end{APACrefauthors}%
\unskip\
\newblock
\APACrefYearMonthDay{1969}{}{}.
\newblock
{\BBOQ}\APACrefatitle {On the Distribution and Continuity of Water Substance in
  Atmospheric Circulations} {On the distribution and continuity of water
  substance in atmospheric circulations}.{\BBCQ}
\newblock
\APACjournalVolNumPages{Meteorol. Monogr.}{10}{}{}.
\newblock
\begin{APACrefDOI} \doi{10.1007/978-1-935704-36-2_1} \end{APACrefDOI}
\PrintBackRefs{\CurrentBib}

\bibitem [\protect \citeauthoryear {%
Khain%
, Ovtchinnikov%
, Pinsky%
, Pokrovsky%
\BCBL {}\ \BBA {} Krugliak%
}{%
Khain%
\ \protect \BOthers {.}}{%
{\protect \APACyear {2000}}%
}]{%
Khain_et_al_2000}
\APACinsertmetastar {%
Khain_et_al_2000}%
\begin{APACrefauthors}%
Khain, A.%
, Ovtchinnikov, M.%
, Pinsky, M.%
, Pokrovsky, A.%
\BCBL {}\ \BBA {} Krugliak, H.%
\end{APACrefauthors}%
\unskip\
\newblock
\APACrefYearMonthDay{2000}{}{}.
\newblock
{\BBOQ}\APACrefatitle {Notes on the state-of-the-art numerical modeling of
  cloud microphysics} {Notes on the state-of-the-art numerical modeling of
  cloud microphysics}.{\BBCQ}
\newblock
\APACjournalVolNumPages{Atmos. Res.}{55}{}{}.
\newblock
\begin{APACrefDOI} \doi{10.1016/S0169-8095(00)00064-8} \end{APACrefDOI}
\PrintBackRefs{\CurrentBib}

\bibitem [\protect \citeauthoryear {%
Kilchhofer%
, Mahrt%
\BCBL {}\ \BBA {} Kanji%
}{%
Kilchhofer%
\ \protect \BOthers {.}}{%
{\protect \APACyear {2021}}%
}]{%
Kilchhofer_et_al_2021}
\APACinsertmetastar {%
Kilchhofer_et_al_2021}%
\begin{APACrefauthors}%
Kilchhofer, K.%
, Mahrt, F.%
\BCBL {}\ \BBA {} Kanji, Z.%
\end{APACrefauthors}%
\unskip\
\newblock
\APACrefYearMonthDay{2021}{}{}.
\newblock
{\BBOQ}\APACrefatitle {The Role of Cloud Processing for the Ice Nucleating
  Ability of Organic Aerosol and Coal Fly Ash Particles} {The role of cloud
  processing for the ice nucleating ability of organic aerosol and coal fly ash
  particles}.{\BBCQ}
\newblock
\APACjournalVolNumPages{J. Geophys. Res. Atmos.}{126}{}{}.
\newblock
\begin{APACrefDOI} \doi{10.1029/2020JD033338} \end{APACrefDOI}
\PrintBackRefs{\CurrentBib}

\bibitem [\protect \citeauthoryear {%
Knopf%
\ \BBA {} Alpert%
}{%
Knopf%
\ \BBA {} Alpert%
}{%
{\protect \APACyear {2013}}%
}]{%
Knopf_and_Alpert_2013}
\APACinsertmetastar {%
Knopf_and_Alpert_2013}%
\begin{APACrefauthors}%
Knopf, D.%
\BCBT {}\ \BBA {} Alpert, P.%
\end{APACrefauthors}%
\unskip\
\newblock
\APACrefYearMonthDay{2013}{}{}.
\newblock
{\BBOQ}\APACrefatitle {A water activity based model of heterogeneous ice
  nucleation kinetics for freezing of water and aqueous solution droplets} {A
  water activity based model of heterogeneous ice nucleation kinetics for
  freezing of water and aqueous solution droplets}.{\BBCQ}
\newblock
\APACjournalVolNumPages{Faraday Discuss.}{165}{}{}.
\newblock
\begin{APACrefDOI} \doi{10.1039/C3FD00035D} \end{APACrefDOI}
\PrintBackRefs{\CurrentBib}

\bibitem [\protect \citeauthoryear {%
Knopf%
\ \BBA {} Alpert%
}{%
Knopf%
\ \BBA {} Alpert%
}{%
{\protect \APACyear {2023}}%
}]{%
Knopf_and_Alpert_2023}
\APACinsertmetastar {%
Knopf_and_Alpert_2023}%
\begin{APACrefauthors}%
Knopf, D.%
\BCBT {}\ \BBA {} Alpert, P.%
\end{APACrefauthors}%
\unskip\
\newblock
\APACrefYearMonthDay{2023}{}{}.
\newblock
{\BBOQ}\APACrefatitle {Atmospheric ice nucleation} {Atmospheric ice
  nucleation}.{\BBCQ}
\newblock
\APACjournalVolNumPages{Nat. Rev. Phys.}{5}{}{203--217}.
\newblock
\begin{APACrefDOI} \doi{10.1038/s42254-023-00570-7} \end{APACrefDOI}
\PrintBackRefs{\CurrentBib}

\bibitem [\protect \citeauthoryear {%
Knopf%
, Alpert%
\BCBL {}\ \BBA {} Wang%
}{%
Knopf%
\ \protect \BOthers {.}}{%
{\protect \APACyear {2018}}%
}]{%
Knopf_et_al_2018}
\APACinsertmetastar {%
Knopf_et_al_2018}%
\begin{APACrefauthors}%
Knopf, D.%
, Alpert, P.%
\BCBL {}\ \BBA {} Wang, B.%
\end{APACrefauthors}%
\unskip\
\newblock
\APACrefYearMonthDay{2018}{}{}.
\newblock
{\BBOQ}\APACrefatitle {The role of organic aerosol in atmospheric ice
  nucleation: A review} {The role of organic aerosol in atmospheric ice
  nucleation: A review}.{\BBCQ}
\newblock
\APACjournalVolNumPages{ACS Earth and Space Chemistry}{2}{3}{168--202}.
\newblock
\begin{APACrefDOI} \doi{10.1021/acsearthspacechem.7b00120} \end{APACrefDOI}
\PrintBackRefs{\CurrentBib}

\bibitem [\protect \citeauthoryear {%
Knopf%
, Alpert%
, Zipori%
, Reicher%
\BCBL {}\ \BBA {} Rudich%
}{%
Knopf%
\ \protect \BOthers {.}}{%
{\protect \APACyear {2020}}%
}]{%
Knopf_et_al_2020}
\APACinsertmetastar {%
Knopf_et_al_2020}%
\begin{APACrefauthors}%
Knopf, D.%
, Alpert, P.%
, Zipori, A.%
, Reicher, N.%
\BCBL {}\ \BBA {} Rudich, Y.%
\end{APACrefauthors}%
\unskip\
\newblock
\APACrefYearMonthDay{2020}{}{}.
\newblock
{\BBOQ}\APACrefatitle {Stochastic nucleation processes and substrate abundance
  explain time-dependent freezing in supercooled droplets} {Stochastic
  nucleation processes and substrate abundance explain time-dependent freezing
  in supercooled droplets}.{\BBCQ}
\newblock
\APACjournalVolNumPages{npj Clim. Atmos. Sci.}{3}{}{}.
\newblock
\begin{APACrefDOI} \doi{10.1038/s41612-020-0106-4} \end{APACrefDOI}
\PrintBackRefs{\CurrentBib}

\bibitem [\protect \citeauthoryear {%
Knopf%
\ \protect \BOthers {.}}{%
Knopf%
\ \protect \BOthers {.}}{%
{\protect \APACyear {2021}}%
}]{%
Knopf_et_al_2021}
\APACinsertmetastar {%
Knopf_et_al_2021}%
\begin{APACrefauthors}%
Knopf, D.%
, Barry, K\BPBI R.%
, Brubaker, T\BPBI A.%
, Jahl, L\BPBI G.%
, K.~A., L\BPBI J.%
, Li, J.%
\BDBL {}Liu, X.%
\end{APACrefauthors}%
\unskip\
\newblock
\APACrefYearMonthDay{2021}{}{}.
\newblock
{\BBOQ}\APACrefatitle {Aerosol-Ice Formation Closure: A Southern Great Plains
  Field Campaign} {Aerosol-ice formation closure: A southern great plains field
  campaign}.{\BBCQ}
\newblock
\APACjournalVolNumPages{Bull. Am. Meteorol. Soc.}{102}{}{}.
\newblock
\begin{APACrefDOI} \doi{10.1175/BAMS-D-20-0151.1} \end{APACrefDOI}
\PrintBackRefs{\CurrentBib}

\bibitem [\protect \citeauthoryear {%
Knopf%
\ \protect \BOthers {.}}{%
Knopf%
\ \protect \BOthers {.}}{%
{\protect \APACyear {2022}}%
}]{%
Knopf_et_al_2022}
\APACinsertmetastar {%
Knopf_et_al_2022}%
\begin{APACrefauthors}%
Knopf, D.%
, Charnawskas, J.%
, Wang, P.%
, Wong, B.%
, Tomlin, J.%
, Jankowski, K.%
\BDBL {}Wang, J.%
\end{APACrefauthors}%
\unskip\
\newblock
\APACrefYearMonthDay{2022}{}{}.
\newblock
{\BBOQ}\APACrefatitle {Micro-spectroscopic and freezing characterization of
  ice-nucleating particles collected in the marine boundary layer in the
  eastern North Atlantic} {Micro-spectroscopic and freezing characterization of
  ice-nucleating particles collected in the marine boundary layer in the
  eastern north atlantic}.{\BBCQ}
\newblock
\APACjournalVolNumPages{Atmos. Chem. Phys.}{22}{}{}.
\newblock
\begin{APACrefDOI} \doi{10.5194/acp-22-5377-2022} \end{APACrefDOI}
\PrintBackRefs{\CurrentBib}

\bibitem [\protect \citeauthoryear {%
Knopf%
, Silber%
, Riemer%
, Fridlind%
\BCBL {}\ \BBA {} Ackerman%
}{%
Knopf%
, Silber%
\BCBL {}\ \protect \BOthers {.}}{%
{\protect \APACyear {2023}}%
}]{%
Knopf_et_al_2023}
\APACinsertmetastar {%
Knopf_et_al_2023}%
\begin{APACrefauthors}%
Knopf, D.%
, Silber, I.%
, Riemer, N.%
, Fridlind, A.%
\BCBL {}\ \BBA {} Ackerman, A.%
\end{APACrefauthors}%
\unskip\
\newblock
\APACrefYearMonthDay{2023}{}{}.
\newblock
{\BBOQ}\APACrefatitle {A {1D} Model for Nucleation of Ice from Aerosol
  Particles: An Application to a Mixed-Phase Arctic Stratus Cloud Layer} {A
  {1D} model for nucleation of ice from aerosol particles: An application to a
  mixed-phase arctic stratus cloud layer}.{\BBCQ}
\newblock
\APACjournalVolNumPages{J. Adv. Model. Earth Syst.}{15}{}{}.
\newblock
\begin{APACrefDOI} \doi{10.1029/2023MS003663} \end{APACrefDOI}
\PrintBackRefs{\CurrentBib}

\bibitem [\protect \citeauthoryear {%
Knopf%
, Wang%
\BCBL {}\ \protect \BOthers {.}}{%
Knopf%
, Wang%
\BCBL {}\ \protect \BOthers {.}}{%
{\protect \APACyear {2023}}%
}]{%
Knopf_et_al_2023_ACP}
\APACinsertmetastar {%
Knopf_et_al_2023_ACP}%
\begin{APACrefauthors}%
Knopf, D.%
, Wang, P.%
, Wong, B.%
, Tomlin, J.%
, Veghte, D.%
, Lata, N.%
\BDBL {}Wang, J.%
\end{APACrefauthors}%
\unskip\
\newblock
\APACrefYearMonthDay{2023}{}{}.
\newblock
{\BBOQ}\APACrefatitle {Physicochemical characterization of free troposphere and
  marine boundary layer ice-nucleating particles collected by aircraft in the
  eastern North Atlantic} {Physicochemical characterization of free troposphere
  and marine boundary layer ice-nucleating particles collected by aircraft in
  the eastern north atlantic}.{\BBCQ}
\newblock
\APACjournalVolNumPages{Atmos. Chem. Phys.}{23}{}{8659--8681}.
\newblock
\begin{APACrefDOI} \doi{10.5194/acp-23-8659-2023} \end{APACrefDOI}
\PrintBackRefs{\CurrentBib}

\bibitem [\protect \citeauthoryear {%
Koop%
}{%
Koop%
}{%
{\protect \APACyear {2002}}%
}]{%
Koop_2002}
\APACinsertmetastar {%
Koop_2002}%
\begin{APACrefauthors}%
Koop, T.%
\end{APACrefauthors}%
\unskip\
\newblock
\APACrefYearMonthDay{2002}{}{}.
\newblock
{\BBOQ}\APACrefatitle {The Water Activity of Aqueous Solutions in Equilibrium
  with Ice} {The water activity of aqueous solutions in equilibrium with
  ice}.{\BBCQ}
\newblock
\APACjournalVolNumPages{Bull. Chem. Soc. Japan}{75}{12}{2587--2588}.
\newblock
\begin{APACrefDOI} \doi{10.1246/bcsj.75.2587} \end{APACrefDOI}
\PrintBackRefs{\CurrentBib}

\bibitem [\protect \citeauthoryear {%
Kotalczyk%
, Devi%
\BCBL {}\ \BBA {} Kruis%
}{%
Kotalczyk%
\ \protect \BOthers {.}}{%
{\protect \APACyear {2017}}%
}]{%
Kotalczyk_et_al_2017}
\APACinsertmetastar {%
Kotalczyk_et_al_2017}%
\begin{APACrefauthors}%
Kotalczyk, G.%
, Devi, J.%
\BCBL {}\ \BBA {} Kruis, F.%
\end{APACrefauthors}%
\unskip\
\newblock
\APACrefYearMonthDay{2017}{}{}.
\newblock
{\BBOQ}\APACrefatitle {A time-driven constant-number {Monte Carlo} method for
  the {GPU}-simulation of particle breakage based on weighted simulation
  particles} {A time-driven constant-number {Monte Carlo} method for the
  {GPU}-simulation of particle breakage based on weighted simulation
  particles}.{\BBCQ}
\newblock
\APACjournalVolNumPages{Powder Tech.}{317}{}{}.
\newblock
\begin{APACrefDOI} \doi{10.1016/j.powtec.2017.05.002} \end{APACrefDOI}
\PrintBackRefs{\CurrentBib}

\bibitem [\protect \citeauthoryear {%
Kubota%
}{%
Kubota%
}{%
{\protect \APACyear {2019}}%
}]{%
Kubota_2019}
\APACinsertmetastar {%
Kubota_2019}%
\begin{APACrefauthors}%
Kubota, N.%
\end{APACrefauthors}%
\unskip\
\newblock
\APACrefYearMonthDay{2019}{}{}.
\newblock
{\BBOQ}\APACrefatitle {Random distribution active site model for ice nucleation
  in water droplets} {Random distribution active site model for ice nucleation
  in water droplets}.{\BBCQ}
\newblock
\APACjournalVolNumPages{Cryst. Eng. Comm.}{21}{}{}.
\newblock
\begin{APACrefDOI} \doi{10.1039/C9CE00246D} \end{APACrefDOI}
\PrintBackRefs{\CurrentBib}

\bibitem [\protect \citeauthoryear {%
Laaksonen%
\ \BBA {} Malila%
}{%
Laaksonen%
\ \BBA {} Malila%
}{%
{\protect \APACyear {2022}}%
}]{%
Laaksonen_and_Malila_2022}
\APACinsertmetastar {%
Laaksonen_and_Malila_2022}%
\begin{APACrefauthors}%
Laaksonen, A.%
\BCBT {}\ \BBA {} Malila, J.%
\end{APACrefauthors}%
\unskip\
\newblock
\APACrefYearMonthDay{2022}{}{}.
\newblock
{\BBOQ}\APACrefatitle {Ice nucleation} {Ice nucleation}.{\BBCQ}
\newblock
\BIn{} \APACrefbtitle {Nucleation of Water} {Nucleation of water}\
  (\BPG~209-248).
\newblock
\APACaddressPublisher{}{Elsevier}.
\newblock
\begin{APACrefDOI} \doi{10.1016/B978-0-12-814321-6.00018-X} \end{APACrefDOI}
\PrintBackRefs{\CurrentBib}

\bibitem [\protect \citeauthoryear {%
Langham%
\ \BBA {} Mason%
}{%
Langham%
\ \BBA {} Mason%
}{%
{\protect \APACyear {1958}}%
}]{%
Langham_and_Mason_1958}
\APACinsertmetastar {%
Langham_and_Mason_1958}%
\begin{APACrefauthors}%
Langham, E.%
\BCBT {}\ \BBA {} Mason, B.%
\end{APACrefauthors}%
\unskip\
\newblock
\APACrefYearMonthDay{1958}{}{}.
\newblock
{\BBOQ}\APACrefatitle {The heterogeneous and homogeneous nucleation of
  supercooled water} {The heterogeneous and homogeneous nucleation of
  supercooled water}.{\BBCQ}
\newblock
\APACjournalVolNumPages{Proc. Royal Soc. A}{}{}{}.
\newblock
\begin{APACrefDOI} \doi{10.1098/rspa.1958.0207} \end{APACrefDOI}
\PrintBackRefs{\CurrentBib}

\bibitem [\protect \citeauthoryear {%
Lata%
\ \protect \BOthers {.}}{%
Lata%
\ \protect \BOthers {.}}{%
{\protect \APACyear {2021}}%
}]{%
Lata_et_al_2021}
\APACinsertmetastar {%
Lata_et_al_2021}%
\begin{APACrefauthors}%
Lata, N.%
, Zhang, B.%
, Schum, S.%
, Mazzoleni, L.%
, Brimberry, R.%
, Marcus, M.%
\BDBL {}China, S.%
\end{APACrefauthors}%
\unskip\
\newblock
\APACrefYearMonthDay{2021}{}{}.
\newblock
{\BBOQ}\APACrefatitle {Aerosol Composition, Mixing State, and Phase State of
  Free Tropospheric Particles and Their Role in Ice Cloud Formation} {Aerosol
  composition, mixing state, and phase state of free tropospheric particles and
  their role in ice cloud formation}.{\BBCQ}
\newblock
\APACjournalVolNumPages{ACS Earth Space Chem.}{5}{}{}.
\newblock
\begin{APACrefDOI} \doi{10.1021/acsearthspacechem.1c00315} \end{APACrefDOI}
\PrintBackRefs{\CurrentBib}

\bibitem [\protect \citeauthoryear {%
Lebo%
\ \BBA {} Morrison%
}{%
Lebo%
\ \BBA {} Morrison%
}{%
{\protect \APACyear {2013}}%
}]{%
Lebo_and_Morrison_2013}
\APACinsertmetastar {%
Lebo_and_Morrison_2013}%
\begin{APACrefauthors}%
Lebo, Z\BPBI J.%
\BCBT {}\ \BBA {} Morrison, H.%
\end{APACrefauthors}%
\unskip\
\newblock
\APACrefYearMonthDay{2013}{}{}.
\newblock
{\BBOQ}\APACrefatitle {A Novel Scheme for Parameterizing Aerosol Processing in
  Warm Clouds} {A novel scheme for parameterizing aerosol processing in warm
  clouds}.{\BBCQ}
\newblock
\APACjournalVolNumPages{J. Atmos. Sci.}{70}{}{}.
\newblock
\begin{APACrefDOI} \doi{10.1175/JAS-D-13-045.1} \end{APACrefDOI}
\PrintBackRefs{\CurrentBib}

\bibitem [\protect \citeauthoryear {%
Lee%
\ \BBA {} Matsoukas%
}{%
Lee%
\ \BBA {} Matsoukas%
}{%
{\protect \APACyear {2000}}%
}]{%
Lee_and_Matsoukas_2000}
\APACinsertmetastar {%
Lee_and_Matsoukas_2000}%
\begin{APACrefauthors}%
Lee, K.%
\BCBT {}\ \BBA {} Matsoukas, T.%
\end{APACrefauthors}%
\unskip\
\newblock
\APACrefYearMonthDay{2000}{}{}.
\newblock
{\BBOQ}\APACrefatitle {Simultaneous coagulation and break-up using constant-{N}
  {Monte Carlo}} {Simultaneous coagulation and break-up using constant-{N}
  {Monte Carlo}}.{\BBCQ}
\newblock
\APACjournalVolNumPages{Powder Tech.}{110}{}{}.
\newblock
\begin{APACrefDOI} \doi{10.1016/S0032-5910(99)00270-3} \end{APACrefDOI}
\PrintBackRefs{\CurrentBib}

\bibitem [\protect \citeauthoryear {%
Leonard%
\ \BBA {} Im%
}{%
Leonard%
\ \BBA {} Im%
}{%
{\protect \APACyear {1999}}%
}]{%
Leonard_and_Im_1999}
\APACinsertmetastar {%
Leonard_and_Im_1999}%
\begin{APACrefauthors}%
Leonard, J.%
\BCBT {}\ \BBA {} Im, J.%
\end{APACrefauthors}%
\unskip\
\newblock
\APACrefYearMonthDay{1999}{}{}.
\newblock
{\BBOQ}\APACrefatitle {Modelling Solid Nucleation and Growth In Supercooled
  Liquid} {Modelling solid nucleation and growth in supercooled liquid}.{\BBCQ}
\newblock
\APACjournalVolNumPages{Mat. Res. Soc. Symp. Proc.}{580}{}{}.
\newblock
\begin{APACrefDOI} \doi{10.1557/proc-580-233} \end{APACrefDOI}
\PrintBackRefs{\CurrentBib}

\bibitem [\protect \citeauthoryear {%
Levine%
}{%
Levine%
}{%
{\protect \APACyear {1950}}%
}]{%
Levine_1950}
\APACinsertmetastar {%
Levine_1950}%
\begin{APACrefauthors}%
Levine, J.%
\end{APACrefauthors}%
\unskip\
\newblock
\APACrefYearMonthDay{1950}{}{}.
\newblock
\APACrefbtitle {Statistical explanation of spontaneous freezing of water
  droplets.} {Statistical explanation of spontaneous freezing of water
  droplets.}
\newblock
\APACaddressPublisher{Washington}{}.
\newblock
\begin{APACrefURL}
  \url{https://web.archive.org/web/*/https://core.ac.uk/download/pdf/42803258.pdf}
  \end{APACrefURL}
\newblock
\APACrefnote{NACA Tech. Note 2234}
\PrintBackRefs{\CurrentBib}

\bibitem [\protect \citeauthoryear {%
Li%
, Mehlig%
, Svensson%
, Brandenburg%
\BCBL {}\ \BBA {} Haugen%
}{%
Li%
\ \protect \BOthers {.}}{%
{\protect \APACyear {2022}}%
}]{%
Li_et_al_2022}
\APACinsertmetastar {%
Li_et_al_2022}%
\begin{APACrefauthors}%
Li, X\BHBI Y.%
, Mehlig, B.%
, Svensson, G.%
, Brandenburg, A.%
\BCBL {}\ \BBA {} Haugen, N\BPBI E\BPBI L.%
\end{APACrefauthors}%
\unskip\
\newblock
\APACrefYearMonthDay{2022}{}{}.
\newblock
{\BBOQ}\APACrefatitle {Collision fluctuations of lucky droplets with
  superdroplets} {Collision fluctuations of lucky droplets with
  superdroplets}.{\BBCQ}
\newblock
\APACjournalVolNumPages{J. Atmos. Sci.}{}{}{}.
\newblock
\begin{APACrefDOI} \doi{10.1175/JAS-D-20-0371.1} \end{APACrefDOI}
\PrintBackRefs{\CurrentBib}

\bibitem [\protect \citeauthoryear {%
Marinescu%
\ \protect \BOthers {.}}{%
Marinescu%
\ \protect \BOthers {.}}{%
{\protect \APACyear {2021}}%
}]{%
Marinescu_et_al_2021}
\APACinsertmetastar {%
Marinescu_et_al_2021}%
\begin{APACrefauthors}%
Marinescu, P\BPBI J.%
, van~den Heever, S\BPBI C.%
, Heikenfeld, M.%
, Barrett, A\BPBI I.%
, Barthlott, C.%
, Hoose, C.%
\BDBL {}Zhang, Y.%
\end{APACrefauthors}%
\unskip\
\newblock
\APACrefYearMonthDay{2021}{}{}.
\newblock
{\BBOQ}\APACrefatitle {Impacts of Varying Concentrations of Cloud Condensation
  Nuclei on Deep Convective Cloud Updrafts—A Multimodel Assessment} {Impacts
  of varying concentrations of cloud condensation nuclei on deep convective
  cloud updrafts—a multimodel assessment}.{\BBCQ}
\newblock
\APACjournalVolNumPages{J. Atmos. Sci.}{78}{}{}.
\newblock
\begin{APACrefDOI} \doi{10.1175/JAS-D-20-0200.1} \end{APACrefDOI}
\PrintBackRefs{\CurrentBib}

\bibitem [\protect \citeauthoryear {%
Marshall%
}{%
Marshall%
}{%
{\protect \APACyear {1961}}%
}]{%
Marshall_1961}
\APACinsertmetastar {%
Marshall_1961}%
\begin{APACrefauthors}%
Marshall, J.%
\end{APACrefauthors}%
\unskip\
\newblock
\APACrefYearMonthDay{1961}{}{}.
\newblock
{\BBOQ}\APACrefatitle {Heterogeneous nucleations is a stochastic process}
  {Heterogeneous nucleations is a stochastic process}.{\BBCQ}
\newblock
\APACjournalVolNumPages{Nubila: rivista di fisica delle nubi}{4}{}{}.
\newblock
\begin{APACrefURL}
  \url{https://web.archive.org/web/*/https://cma.entecra.it/Astro2_sito/doc/Nubila_1_1961.pdf}
  \end{APACrefURL}
\PrintBackRefs{\CurrentBib}

\bibitem [\protect \citeauthoryear {%
Matsushima%
, Nishizawa%
\BCBL {}\ \BBA {} Shima%
}{%
Matsushima%
\ \protect \BOthers {.}}{%
{\protect \APACyear {2023}}%
}]{%
Matsushima_et_al_2023}
\APACinsertmetastar {%
Matsushima_et_al_2023}%
\begin{APACrefauthors}%
Matsushima, T.%
, Nishizawa, S.%
\BCBL {}\ \BBA {} Shima, S.%
\end{APACrefauthors}%
\unskip\
\newblock
\APACrefYearMonthDay{2023}{}{}.
\newblock
{\BBOQ}\APACrefatitle {Overcoming computational challenges to realize meter- to
  submeter-scale resolution in cloud simulations using the super-droplet
  method} {Overcoming computational challenges to realize meter- to
  submeter-scale resolution in cloud simulations using the super-droplet
  method}.{\BBCQ}
\newblock
\APACjournalVolNumPages{Geosci. Model Dev.}{16}{}{}.
\newblock
\begin{APACrefDOI} \doi{10.5194/gmd-16-6211-2023} \end{APACrefDOI}
\PrintBackRefs{\CurrentBib}

\bibitem [\protect \citeauthoryear {%
Maxey%
\ \BBA {} Corrsin%
}{%
Maxey%
\ \BBA {} Corrsin%
}{%
{\protect \APACyear {1986}}%
}]{%
Maxey_and_Corrsin_1986}
\APACinsertmetastar {%
Maxey_and_Corrsin_1986}%
\begin{APACrefauthors}%
Maxey, M.%
\BCBT {}\ \BBA {} Corrsin, S.%
\end{APACrefauthors}%
\unskip\
\newblock
\APACrefYearMonthDay{1986}{}{}.
\newblock
{\BBOQ}\APACrefatitle {Gravitational Settling of Aerosol Particles in Randomly
  Oriented Cellular Flow Fields} {Gravitational settling of aerosol particles
  in randomly oriented cellular flow fields}.{\BBCQ}
\newblock
\APACjournalVolNumPages{J. Atmos. Sci.}{43}{}{}.
\newblock
\begin{APACrefDOI} \doi{10.1175/1520-0469(1986)043<1112:GSOAPI>2.0.CO;2}
  \end{APACrefDOI}
\PrintBackRefs{\CurrentBib}

\bibitem [\protect \citeauthoryear {%
Michel%
}{%
Michel%
}{%
{\protect \APACyear {1967}}%
}]{%
Michel_1967}
\APACinsertmetastar {%
Michel_1967}%
\begin{APACrefauthors}%
Michel, B.%
\end{APACrefauthors}%
\unskip\
\newblock
\APACrefYearMonthDay{1967}{}{}.
\newblock
{\BBOQ}\APACrefatitle {From the Nucleation of Ice Crystals in Clouds to the
  Formation of Frazil Ice in Rivers} {From the nucleation of ice crystals in
  clouds to the formation of frazil ice in rivers}.{\BBCQ}
\newblock
\BIn{} \APACrefbtitle {International Conference on Low Temperature Science,
  1966, Sapporo, Japan.} {International conference on low temperature science,
  1966, sapporo, japan.}
\newblock
\begin{APACrefURL}
  \url{https://web.archive.org/web/*/https://eprints.lib.hokudai.ac.jp/dspace/bitstream/2115/20291/1/1_p129-136.pdf}
  \end{APACrefURL}
\PrintBackRefs{\CurrentBib}

\bibitem [\protect \citeauthoryear {%
Morrison%
\ \protect \BOthers {.}}{%
Morrison%
\ \protect \BOthers {.}}{%
{\protect \APACyear {2012}}%
}]{%
Morrison_et_al_2012}
\APACinsertmetastar {%
Morrison_et_al_2012}%
\begin{APACrefauthors}%
Morrison, H.%
, de Boer, G.%
, Feingold, G.%
, Harrington, J.%
, Shupe, M.%
\BCBL {}\ \BBA {} Sulia, K.%
\end{APACrefauthors}%
\unskip\
\newblock
\APACrefYearMonthDay{2012}{}{}.
\newblock
{\BBOQ}\APACrefatitle {Resilience of persistent Arctic mixed-phase clouds}
  {Resilience of persistent arctic mixed-phase clouds}.{\BBCQ}
\newblock
\APACjournalVolNumPages{Nat. Geosci.}{5}{}{}.
\newblock
\begin{APACrefDOI} \doi{10.1038/ngeo1332} \end{APACrefDOI}
\PrintBackRefs{\CurrentBib}

\bibitem [\protect \citeauthoryear {%
Morrison%
\ \BBA {} Grabowski%
}{%
Morrison%
\ \BBA {} Grabowski%
}{%
{\protect \APACyear {2007}}%
}]{%
Morrison_and_Grabowski_2007}
\APACinsertmetastar {%
Morrison_and_Grabowski_2007}%
\begin{APACrefauthors}%
Morrison, H.%
\BCBT {}\ \BBA {} Grabowski, W.%
\end{APACrefauthors}%
\unskip\
\newblock
\APACrefYearMonthDay{2007}{}{}.
\newblock
{\BBOQ}\APACrefatitle {Comparison of Bulk and Bin Warm-Rain Microphysics Models
  Using a Kinematic Framework} {Comparison of bulk and bin warm-rain
  microphysics models using a kinematic framework}.{\BBCQ}
\newblock
\APACjournalVolNumPages{J. Atmos. Sci.}{64}{}{}.
\newblock
\begin{APACrefDOI} \doi{10.1175/JAS3980} \end{APACrefDOI}
\PrintBackRefs{\CurrentBib}

\bibitem [\protect \citeauthoryear {%
Morrison%
\ \protect \BOthers {.}}{%
Morrison%
\ \protect \BOthers {.}}{%
{\protect \APACyear {2020}}%
}]{%
Morrison_et_al_2020}
\APACinsertmetastar {%
Morrison_et_al_2020}%
\begin{APACrefauthors}%
Morrison, H.%
, van Lier-Walqui, M.%
, Fridlind, A.%
, Grabowski, W.%
, Harrington, J.%
, Hoose, C.%
\BDBL {}Xue, L.%
\end{APACrefauthors}%
\unskip\
\newblock
\APACrefYearMonthDay{2020}{}{}.
\newblock
{\BBOQ}\APACrefatitle {Confronting the Challenge of Modeling Cloud and
  Precipitation Microphysics} {Confronting the challenge of modeling cloud and
  precipitation microphysics}.{\BBCQ}
\newblock
\APACjournalVolNumPages{J. Adv. Model. Earth Syst.}{12}{}{}.
\newblock
\begin{APACrefDOI} \doi{10.1029/2019MS001689} \end{APACrefDOI}
\PrintBackRefs{\CurrentBib}

\bibitem [\protect \citeauthoryear {%
Morrison%
\ \protect \BOthers {.}}{%
Morrison%
\ \protect \BOthers {.}}{%
{\protect \APACyear {2011}}%
}]{%
Morrison_et_al_2011}
\APACinsertmetastar {%
Morrison_et_al_2011}%
\begin{APACrefauthors}%
Morrison, H.%
, Zuidema, P.%
, Ackerman, A.%
, Avramov, A.%
, de Boer, G.%
, Fan, J.%
\BDBL {}Shipway, B.%
\end{APACrefauthors}%
\unskip\
\newblock
\APACrefYearMonthDay{2011}{}{}.
\newblock
{\BBOQ}\APACrefatitle {Intercomparison of cloud model simulations of {A}rctic
  mixed-phase boundary layer clouds observed during {SHEBA/FIRE-ACE}}
  {Intercomparison of cloud model simulations of {A}rctic mixed-phase boundary
  layer clouds observed during {SHEBA/FIRE-ACE}}.{\BBCQ}
\newblock
\APACjournalVolNumPages{J. Adv. Model. Earth Syst.}{3}{}{}.
\newblock
\begin{APACrefDOI} \doi{10.1029/2011MS000066} \end{APACrefDOI}
\PrintBackRefs{\CurrentBib}

\bibitem [\protect \citeauthoryear {%
Mossop%
}{%
Mossop%
}{%
{\protect \APACyear {1955}}%
}]{%
Mossop_1955}
\APACinsertmetastar {%
Mossop_1955}%
\begin{APACrefauthors}%
Mossop, S.%
\end{APACrefauthors}%
\unskip\
\newblock
\APACrefYearMonthDay{1955}{}{}.
\newblock
{\BBOQ}\APACrefatitle {The Freezing of Supercooled Water} {The freezing of
  supercooled water}.{\BBCQ}
\newblock
\APACjournalVolNumPages{Proc. Phys. Soc. B}{68}{4}{}.
\newblock
\begin{APACrefDOI} \doi{10.1088/0370-1301/68/4/301} \end{APACrefDOI}
\PrintBackRefs{\CurrentBib}

\bibitem [\protect \citeauthoryear {%
Muhlbauer%
\ \protect \BOthers {.}}{%
Muhlbauer%
\ \protect \BOthers {.}}{%
{\protect \APACyear {2013}}%
}]{%
Muhlbauer_et_al_2013}
\APACinsertmetastar {%
Muhlbauer_et_al_2013}%
\begin{APACrefauthors}%
Muhlbauer, A.%
, Grabowski, W.%
, Malinowski, S\BPBI P.%
, Ackerman, T\BPBI P.%
, Bryan, G\BPBI H.%
, Lebo, Z\BPBI J.%
\BDBL {}Thompson, G.%
\end{APACrefauthors}%
\unskip\
\newblock
\APACrefYearMonthDay{2013}{}{}.
\newblock
{\BBOQ}\APACrefatitle {Reexamination of the State of the Art of Cloud Modeling
  Shows Real Improvements} {Reexamination of the state of the art of cloud
  modeling shows real improvements}.{\BBCQ}
\newblock
\APACjournalVolNumPages{Bull. Amer. Meteor. Soc.}{94}{}{}.
\newblock
\begin{APACrefDOI} \doi{10.1175/BAMS-D-12-00188.1} \end{APACrefDOI}
\PrintBackRefs{\CurrentBib}

\bibitem [\protect \citeauthoryear {%
Murray%
, Broadley%
, Wilson%
, Atkinson%
\BCBL {}\ \BBA {} Wills%
}{%
Murray%
\ \protect \BOthers {.}}{%
{\protect \APACyear {2011}}%
}]{%
Murray_et_al_2011}
\APACinsertmetastar {%
Murray_et_al_2011}%
\begin{APACrefauthors}%
Murray, B.%
, Broadley, S.%
, Wilson, T.%
, Atkinson, J.%
\BCBL {}\ \BBA {} Wills, R.%
\end{APACrefauthors}%
\unskip\
\newblock
\APACrefYearMonthDay{2011}{}{}.
\newblock
{\BBOQ}\APACrefatitle {Heterogeneous freezing of water droplets containing
  kaolinite particles} {Heterogeneous freezing of water droplets containing
  kaolinite particles}.{\BBCQ}
\newblock
\APACjournalVolNumPages{Atmos. Chem. Phys.}{11}{}{}.
\newblock
\begin{APACrefDOI} \doi{10.5194/acp-11-4191-2011} \end{APACrefDOI}
\PrintBackRefs{\CurrentBib}

\bibitem [\protect \citeauthoryear {%
Niedermeier%
\ \protect \BOthers {.}}{%
Niedermeier%
\ \protect \BOthers {.}}{%
{\protect \APACyear {2015}}%
}]{%
Niedermeier_et_al_2015}
\APACinsertmetastar {%
Niedermeier_et_al_2015}%
\begin{APACrefauthors}%
Niedermeier, D.%
, Augustin-Bauditz, S.%
, Hartmann, S.%
, Wex, H.%
, Ignatius, K.%
\BCBL {}\ \BBA {} Stratmann, F.%
\end{APACrefauthors}%
\unskip\
\newblock
\APACrefYearMonthDay{2015}{}{}.
\newblock
{\BBOQ}\APACrefatitle {Can we define an asymptotic value for the ice active
  surface site density for heterogeneous ice nucleation?} {Can we define an
  asymptotic value for the ice active surface site density for heterogeneous
  ice nucleation?}{\BBCQ}
\newblock
\APACjournalVolNumPages{J. Geophys. Res. Atmos.}{120}{}{}.
\newblock
\begin{APACrefDOI} \doi{10.1002/2014JD022814} \end{APACrefDOI}
\PrintBackRefs{\CurrentBib}

\bibitem [\protect \citeauthoryear {%
Niemand%
\ \protect \BOthers {.}}{%
Niemand%
\ \protect \BOthers {.}}{%
{\protect \APACyear {2012}}%
}]{%
Niemand_et_al_2012}
\APACinsertmetastar {%
Niemand_et_al_2012}%
\begin{APACrefauthors}%
Niemand, M.%
, Möhler, O.%
, Vogel, B.%
, Vogel, H.%
, Hoose, C.%
, Connolly, P.%
\BDBL {}Leisner, T.%
\end{APACrefauthors}%
\unskip\
\newblock
\APACrefYearMonthDay{2012}{}{}.
\newblock
{\BBOQ}\APACrefatitle {A particle-surface-area-based parameterization of
  immersion freezing on desert dust particles} {A particle-surface-area-based
  parameterization of immersion freezing on desert dust particles}.{\BBCQ}
\newblock
\APACjournalVolNumPages{J. Atmos. Sci.}{69}{}{}.
\newblock
\begin{APACrefDOI} \doi{10.1175/JAS-D-11-0249.1} \end{APACrefDOI}
\PrintBackRefs{\CurrentBib}

\bibitem [\protect \citeauthoryear {%
Nordam%
, Kristiansen%
, Nepstad%
, van Sebille%
\BCBL {}\ \BBA {} Booth%
}{%
Nordam%
\ \protect \BOthers {.}}{%
{\protect \APACyear {2023}}%
}]{%
Nordam_et_al_2023}
\APACinsertmetastar {%
Nordam_et_al_2023}%
\begin{APACrefauthors}%
Nordam, T.%
, Kristiansen, R.%
, Nepstad, R.%
, van Sebille, E.%
\BCBL {}\ \BBA {} Booth, A.%
\end{APACrefauthors}%
\unskip\
\newblock
\APACrefYearMonthDay{2023}{}{}.
\newblock
{\BBOQ}\APACrefatitle {A comparison of {E}ulerian and {L}agrangian methods for
  vertical particle transport in the water column} {A comparison of {E}ulerian
  and {L}agrangian methods for vertical particle transport in the water
  column}.{\BBCQ}
\newblock
\APACjournalVolNumPages{Geosci. Model Dev.}{16}{}{}.
\newblock
\begin{APACrefDOI} \doi{10.5194/gmd-16-5339-2023} \end{APACrefDOI}
\PrintBackRefs{\CurrentBib}

\bibitem [\protect \citeauthoryear {%
Paoli%
, Hélie%
\BCBL {}\ \BBA {} Poinstot%
}{%
Paoli%
\ \protect \BOthers {.}}{%
{\protect \APACyear {2004}}%
}]{%
Paoli_et_al_2004}
\APACinsertmetastar {%
Paoli_et_al_2004}%
\begin{APACrefauthors}%
Paoli, R.%
, Hélie, J.%
\BCBL {}\ \BBA {} Poinstot, T.%
\end{APACrefauthors}%
\unskip\
\newblock
\APACrefYearMonthDay{2004}{}{}.
\newblock
{\BBOQ}\APACrefatitle {Contrail formation in aircraft wakes} {Contrail
  formation in aircraft wakes}.{\BBCQ}
\newblock
\APACjournalVolNumPages{J. Fluid. Mech.}{502}{}{}.
\newblock
\begin{APACrefDOI} \doi{10.1017/S0022112003007808} \end{APACrefDOI}
\PrintBackRefs{\CurrentBib}

\bibitem [\protect \citeauthoryear {%
Petters%
\ \BBA {} Kreidenweis%
}{%
Petters%
\ \BBA {} Kreidenweis%
}{%
{\protect \APACyear {2007}}%
}]{%
Petters_and_Kreidenweis_2007}
\APACinsertmetastar {%
Petters_and_Kreidenweis_2007}%
\begin{APACrefauthors}%
Petters, M.%
\BCBT {}\ \BBA {} Kreidenweis, S.%
\end{APACrefauthors}%
\unskip\
\newblock
\APACrefYearMonthDay{2007}{}{}.
\newblock
{\BBOQ}\APACrefatitle {A single parameter representation of hygroscopic growth
  and cloud condensation nucleus activity} {A single parameter representation
  of hygroscopic growth and cloud condensation nucleus activity}.{\BBCQ}
\newblock
\APACjournalVolNumPages{Atmos. Chem. Phys.}{7}{}{}.
\newblock
\begin{APACrefDOI} \doi{10.5194/acp-7-1961-2007} \end{APACrefDOI}
\PrintBackRefs{\CurrentBib}

\bibitem [\protect \citeauthoryear {%
Pincus%
\ \BBA {} Chepfer%
}{%
Pincus%
\ \BBA {} Chepfer%
}{%
{\protect \APACyear {2020}}%
}]{%
Pincus_and_Chepfer_2020}
\APACinsertmetastar {%
Pincus_and_Chepfer_2020}%
\begin{APACrefauthors}%
Pincus, R.%
\BCBT {}\ \BBA {} Chepfer, H.%
\end{APACrefauthors}%
\unskip\
\newblock
\APACrefYearMonthDay{2020}{}{}.
\newblock
{\BBOQ}\APACrefatitle {Clouds as Light} {Clouds as light}.{\BBCQ}
\newblock
\BIn{} A.~Siebesma, S.~Bony, C.~Jakob\BCBL {}\ \BBA {} B.~Stevens\ (\BEDS),
  \APACrefbtitle {Clouds and Climate: Climate Science's Greatest Challenge}
  {Clouds and climate: Climate science's greatest challenge}\ (\BPGS\ 99--122).
\newblock
\begin{APACrefDOI} \doi{10.1017/9781107447738.005} \end{APACrefDOI}
\PrintBackRefs{\CurrentBib}

\bibitem [\protect \citeauthoryear {%
Pruppacher%
\ \BBA {} Klett%
}{%
Pruppacher%
\ \BBA {} Klett%
}{%
{\protect \APACyear {2010}}%
}]{%
Pruppacher_and_Klett_2010}
\APACinsertmetastar {%
Pruppacher_and_Klett_2010}%
\begin{APACrefauthors}%
Pruppacher, H.%
\BCBT {}\ \BBA {} Klett, J.%
\end{APACrefauthors}%
\unskip\
\newblock
\APACrefYearMonthDay{2010}{}{}.
\newblock
{\BBOQ}\APACrefatitle {Heterogeneous Nucleation} {Heterogeneous
  nucleation}.{\BBCQ}
\newblock
\BIn{} \APACrefbtitle {Microphysics of Clouds and Precipitation.} {Microphysics
  of clouds and precipitation.}
\newblock
\APACaddressPublisher{}{Springer}.
\newblock
\begin{APACrefDOI} \doi{10.1007/978-0-306-48100-0_9} \end{APACrefDOI}
\PrintBackRefs{\CurrentBib}

\bibitem [\protect \citeauthoryear {%
Rasinski%
, Pawlowska%
\BCBL {}\ \BBA {} Grabowski%
}{%
Rasinski%
\ \protect \BOthers {.}}{%
{\protect \APACyear {2011}}%
}]{%
Rasinski_et_al_2011}
\APACinsertmetastar {%
Rasinski_et_al_2011}%
\begin{APACrefauthors}%
Rasinski, P.%
, Pawlowska, H.%
\BCBL {}\ \BBA {} Grabowski, W.%
\end{APACrefauthors}%
\unskip\
\newblock
\APACrefYearMonthDay{2011}{}{}.
\newblock
{\BBOQ}\APACrefatitle {Observations and kinematic modeling of drizzling marine
  stratocumulus} {Observations and kinematic modeling of drizzling marine
  stratocumulus}.{\BBCQ}
\newblock
\APACjournalVolNumPages{Atmos. Res.}{102}{}{}.
\newblock
\begin{APACrefDOI} \doi{10.1016/j.atmosres.2011.06.020} \end{APACrefDOI}
\PrintBackRefs{\CurrentBib}

\bibitem [\protect \citeauthoryear {%
Reisner%
, Rasmussen%
\BCBL {}\ \BBA {} Bruintjes%
}{%
Reisner%
\ \protect \BOthers {.}}{%
{\protect \APACyear {1998}}%
}]{%
Reisner_et_al_1998}
\APACinsertmetastar {%
Reisner_et_al_1998}%
\begin{APACrefauthors}%
Reisner, J.%
, Rasmussen, R.%
\BCBL {}\ \BBA {} Bruintjes, R.%
\end{APACrefauthors}%
\unskip\
\newblock
\APACrefYearMonthDay{1998}{}{}.
\newblock
{\BBOQ}\APACrefatitle {Explicit forecasting of supercooled liquid water in
  winter storms using the {MM5} mesoscale model} {Explicit forecasting of
  supercooled liquid water in winter storms using the {MM5} mesoscale
  model}.{\BBCQ}
\newblock
\APACjournalVolNumPages{Q. J. Royal Meteorol. Soc.}{124}{548}{1071--1107}.
\newblock
\begin{APACrefDOI} \doi{10.1002/qj.49712454804} \end{APACrefDOI}
\PrintBackRefs{\CurrentBib}

\bibitem [\protect \citeauthoryear {%
Richter%
, MacMillan%
\BCBL {}\ \BBA {} Wainwright%
}{%
Richter%
\ \protect \BOthers {.}}{%
{\protect \APACyear {2021}}%
}]{%
Richter_et_al_2021}
\APACinsertmetastar {%
Richter_et_al_2021}%
\begin{APACrefauthors}%
Richter, D.%
, MacMillan, T.%
\BCBL {}\ \BBA {} Wainwright, C.%
\end{APACrefauthors}%
\unskip\
\newblock
\APACrefYearMonthDay{2021}{}{}.
\newblock
{\BBOQ}\APACrefatitle {A {L}agrangian Cloud Model for the Study of Marine Fog}
  {A {L}agrangian cloud model for the study of marine fog}.{\BBCQ}
\newblock
\APACjournalVolNumPages{Boundary-Layer Meteorol.}{181}{}{}.
\newblock
\begin{APACrefDOI} \doi{10.1007/s10546-020-00595-w} \end{APACrefDOI}
\PrintBackRefs{\CurrentBib}

\bibitem [\protect \citeauthoryear {%
Riemer%
, Ault%
, West%
, Craig%
\BCBL {}\ \BBA {} Curtis%
}{%
Riemer%
\ \protect \BOthers {.}}{%
{\protect \APACyear {2019}}%
}]{%
Riemer_et_al_2019}
\APACinsertmetastar {%
Riemer_et_al_2019}%
\begin{APACrefauthors}%
Riemer, N.%
, Ault, A.%
, West, M.%
, Craig, R.%
\BCBL {}\ \BBA {} Curtis, J.%
\end{APACrefauthors}%
\unskip\
\newblock
\APACrefYearMonthDay{2019}{}{}.
\newblock
{\BBOQ}\APACrefatitle {Aerosol Mixing State: Measurements, Modeling, and
  Impacts} {Aerosol mixing state: Measurements, modeling, and impacts}.{\BBCQ}
\newblock
\APACjournalVolNumPages{Rev. Geophys.}{57}{}{}.
\newblock
\begin{APACrefDOI} \doi{10.1029/2018RG000615} \end{APACrefDOI}
\PrintBackRefs{\CurrentBib}

\bibitem [\protect \citeauthoryear {%
Rigg%
, Alpert%
\BCBL {}\ \BBA {} Knopf%
}{%
Rigg%
\ \protect \BOthers {.}}{%
{\protect \APACyear {2013}}%
}]{%
Rigg_et_al_2013}
\APACinsertmetastar {%
Rigg_et_al_2013}%
\begin{APACrefauthors}%
Rigg, Y.%
, Alpert, P.%
\BCBL {}\ \BBA {} Knopf, D.%
\end{APACrefauthors}%
\unskip\
\newblock
\APACrefYearMonthDay{2013}{}{}.
\newblock
{\BBOQ}\APACrefatitle {Immersion freezing of water and aqueous ammonium sulfate
  droplets initiated by humic-like substances as a function of water activity}
  {Immersion freezing of water and aqueous ammonium sulfate droplets initiated
  by humic-like substances as a function of water activity}.{\BBCQ}
\newblock
\APACjournalVolNumPages{Atmos. Chem. Phys.}{13}{}{}.
\newblock
\begin{APACrefDOI} \doi{10.5194/acp-13-6603-2013} \end{APACrefDOI}
\PrintBackRefs{\CurrentBib}

\bibitem [\protect \citeauthoryear {%
Schmeller%
\ \BBA {} Geresdi%
}{%
Schmeller%
\ \BBA {} Geresdi%
}{%
{\protect \APACyear {2019}}%
}]{%
Schmeller_and_Geresdi_2019}
\APACinsertmetastar {%
Schmeller_and_Geresdi_2019}%
\begin{APACrefauthors}%
Schmeller, G.%
\BCBT {}\ \BBA {} Geresdi, I.%
\end{APACrefauthors}%
\unskip\
\newblock
\APACrefYearMonthDay{2019}{}{}.
\newblock
{\BBOQ}\APACrefatitle {Study of interaction between cloud microphysics and
  chemistry using coupled bin microphysics and bin aqueous chemistry scheme}
  {Study of interaction between cloud microphysics and chemistry using coupled
  bin microphysics and bin aqueous chemistry scheme}.{\BBCQ}
\newblock
\APACjournalVolNumPages{Atmos. Env.}{198}{}{}.
\newblock
\begin{APACrefDOI} \doi{10.1016/j.atmosenv.2018.10.064} \end{APACrefDOI}
\PrintBackRefs{\CurrentBib}

\bibitem [\protect \citeauthoryear {%
Sch\"olzel%
\ \BBA {} Friederichs%
}{%
Sch\"olzel%
\ \BBA {} Friederichs%
}{%
{\protect \APACyear {2008}}%
}]{%
Schoelzel_and_Friederichs_2008}
\APACinsertmetastar {%
Schoelzel_and_Friederichs_2008}%
\begin{APACrefauthors}%
Sch\"olzel, C.%
\BCBT {}\ \BBA {} Friederichs, P.%
\end{APACrefauthors}%
\unskip\
\newblock
\APACrefYearMonthDay{2008}{}{}.
\newblock
{\BBOQ}\APACrefatitle {Multivariate non-normally distributed random variables
  in climate research – introduction to the copula approach} {Multivariate
  non-normally distributed random variables in climate research –
  introduction to the copula approach}.{\BBCQ}
\newblock
\APACjournalVolNumPages{Nonlin. Processes Geophys.}{15}{}{761--772}.
\newblock
\begin{APACrefDOI} \doi{10.5194/npg-15-761-2008} \end{APACrefDOI}
\PrintBackRefs{\CurrentBib}

\bibitem [\protect \citeauthoryear {%
Seifert%
\ \BBA {} Beheng%
}{%
Seifert%
\ \BBA {} Beheng%
}{%
{\protect \APACyear {2006}}%
}]{%
Seifert_and_Beheng_2006}
\APACinsertmetastar {%
Seifert_and_Beheng_2006}%
\begin{APACrefauthors}%
Seifert, A.%
\BCBT {}\ \BBA {} Beheng, K.%
\end{APACrefauthors}%
\unskip\
\newblock
\APACrefYearMonthDay{2006}{}{}.
\newblock
{\BBOQ}\APACrefatitle {A two-moment cloud microphysics parameterization for
  mixed-phase clouds. Part 1: Model description} {A two-moment cloud
  microphysics parameterization for mixed-phase clouds. part 1: Model
  description}.{\BBCQ}
\newblock
\APACjournalVolNumPages{Meteorol. Atmos. Phys.}{92}{}{}.
\newblock
\begin{APACrefDOI} \doi{10.1007/s00703-005-0112-4} \end{APACrefDOI}
\PrintBackRefs{\CurrentBib}

\bibitem [\protect \citeauthoryear {%
Seifert%
, Leinonen%
, Siewert%
\BCBL {}\ \BBA {} Kneifel%
}{%
Seifert%
\ \protect \BOthers {.}}{%
{\protect \APACyear {2019}}%
}]{%
Seifert_et_al_2019}
\APACinsertmetastar {%
Seifert_et_al_2019}%
\begin{APACrefauthors}%
Seifert, A.%
, Leinonen, J.%
, Siewert, C.%
\BCBL {}\ \BBA {} Kneifel, S.%
\end{APACrefauthors}%
\unskip\
\newblock
\APACrefYearMonthDay{2019}{}{}.
\newblock
{\BBOQ}\APACrefatitle {The Geometry of Rimed Aggregate Snowflakes: A Modeling
  Study} {The geometry of rimed aggregate snowflakes: A modeling study}.{\BBCQ}
\newblock
\APACjournalVolNumPages{J. Adv. Model. Earth Syst.}{11}{}{}.
\newblock
\begin{APACrefDOI} \doi{10.1029/2018MS001519} \end{APACrefDOI}
\PrintBackRefs{\CurrentBib}

\bibitem [\protect \citeauthoryear {%
Shima%
, Sato%
, Hashimoto%
\BCBL {}\ \BBA {} Misumi%
}{%
Shima%
\ \protect \BOthers {.}}{%
{\protect \APACyear {2020}}%
}]{%
Shima_et_al_2020}
\APACinsertmetastar {%
Shima_et_al_2020}%
\begin{APACrefauthors}%
Shima, S.%
, Sato, Y.%
, Hashimoto, A.%
\BCBL {}\ \BBA {} Misumi, R.%
\end{APACrefauthors}%
\unskip\
\newblock
\APACrefYearMonthDay{2020}{}{}.
\newblock
{\BBOQ}\APACrefatitle {Predicting the morphology of ice particles in deep
  convection using the super-droplet method: development and evaluation of
  {SCALE-SDM} 0.2.5-2.2.0, -2.2.1, and -2.2.2} {Predicting the morphology of
  ice particles in deep convection using the super-droplet method: development
  and evaluation of {SCALE-SDM} 0.2.5-2.2.0, -2.2.1, and -2.2.2}.{\BBCQ}
\newblock
\APACjournalVolNumPages{Geosci. Model Dev.}{13}{}{}.
\newblock
\begin{APACrefDOI} \doi{10.5194/gmd-13-4107-2020} \end{APACrefDOI}
\PrintBackRefs{\CurrentBib}

\bibitem [\protect \citeauthoryear {%
Shirgaonkar%
\ \BBA {} Lele%
}{%
Shirgaonkar%
\ \BBA {} Lele%
}{%
{\protect \APACyear {2012}}%
}]{%
Shirgaonkar_and_Lele_2012}
\APACinsertmetastar {%
Shirgaonkar_and_Lele_2012}%
\begin{APACrefauthors}%
Shirgaonkar, A.%
\BCBT {}\ \BBA {} Lele, S.%
\end{APACrefauthors}%
\unskip\
\newblock
\APACrefYearMonthDay{2012}{}{}.
\newblock
{\BBOQ}\APACrefatitle {Large Eddy Simulation of Early Stage Contrails: Effect
  of Atmospheric Properties} {Large eddy simulation of early stage contrails:
  Effect of atmospheric properties}.{\BBCQ}
\newblock
\BIn{} \APACrefbtitle {44th AIAA Aerospace Sciences Meeting and Exhibit.} {44th
  aiaa aerospace sciences meeting and exhibit.}
\newblock
\begin{APACrefDOI} \doi{10.2514/6.2006-1414} \end{APACrefDOI}
\PrintBackRefs{\CurrentBib}

\bibitem [\protect \citeauthoryear {%
Silber%
}{%
Silber%
}{%
{\protect \APACyear {2023}}%
}]{%
Silber_2023}
\APACinsertmetastar {%
Silber_2023}%
\begin{APACrefauthors}%
Silber, I.%
\end{APACrefauthors}%
\unskip\
\newblock
\APACrefYearMonthDay{2023}{}{}.
\newblock
{\BBOQ}\APACrefatitle {Arctic Cloud-Base Ice Precipitation Properties Retrieved
  Using Bayesian Inference} {Arctic cloud-base ice precipitation properties
  retrieved using bayesian inference}.{\BBCQ}
\newblock
\APACjournalVolNumPages{J. Geophys. Res. Atmos.}{128}{}{}.
\newblock
\begin{APACrefDOI} \doi{10.1029/2022JD038202} \end{APACrefDOI}
\PrintBackRefs{\CurrentBib}

\bibitem [\protect \citeauthoryear {%
Silber%
, Fridlind%
, Verlinde%
, Russell%
\BCBL {}\ \BBA {} Ackerman%
}{%
Silber%
\ \protect \BOthers {.}}{%
{\protect \APACyear {2020}}%
}]{%
Silber_et_al_2020}
\APACinsertmetastar {%
Silber_et_al_2020}%
\begin{APACrefauthors}%
Silber, I.%
, Fridlind, A.%
, Verlinde, J.%
, Russell, L.%
\BCBL {}\ \BBA {} Ackerman, A.%
\end{APACrefauthors}%
\unskip\
\newblock
\APACrefYearMonthDay{2020}{}{}.
\newblock
{\BBOQ}\APACrefatitle {Nonturbulent liquid‐bearing polar clouds: Observed
  frequency of occurrence and simulated sensitivity to gravity waves}
  {Nonturbulent liquid‐bearing polar clouds: Observed frequency of occurrence
  and simulated sensitivity to gravity waves}.{\BBCQ}
\newblock
\APACjournalVolNumPages{Geophys. Res. Lett.}{47}{}{}.
\newblock
\begin{APACrefDOI} \doi{10.1029/2020GL087099} \end{APACrefDOI}
\PrintBackRefs{\CurrentBib}

\bibitem [\protect \citeauthoryear {%
Slawinska%
, Grabowski%
\BCBL {}\ \BBA {} Morrison%
}{%
Slawinska%
\ \protect \BOthers {.}}{%
{\protect \APACyear {2009}}%
}]{%
Slawinska_et_al_2009}
\APACinsertmetastar {%
Slawinska_et_al_2009}%
\begin{APACrefauthors}%
Slawinska, J.%
, Grabowski, W.%
\BCBL {}\ \BBA {} Morrison, H.%
\end{APACrefauthors}%
\unskip\
\newblock
\APACrefYearMonthDay{2009}{}{}.
\newblock
{\BBOQ}\APACrefatitle {The impact of atmospheric aerosols on precipitation from
  deep organized convection: A prescribed-flow model study using double-moment
  bulk microphysics} {The impact of atmospheric aerosols on precipitation from
  deep organized convection: A prescribed-flow model study using double-moment
  bulk microphysics}.{\BBCQ}
\newblock
\APACjournalVolNumPages{Q. J. Royal Meteorol. Soc.}{135}{}{}.
\newblock
\begin{APACrefDOI} \doi{10.1002/qj.450} \end{APACrefDOI}
\PrintBackRefs{\CurrentBib}

\bibitem [\protect \citeauthoryear {%
Smolarkiewicz%
}{%
Smolarkiewicz%
}{%
{\protect \APACyear {2006}}%
}]{%
Smolarkiewicz_2006}
\APACinsertmetastar {%
Smolarkiewicz_2006}%
\begin{APACrefauthors}%
Smolarkiewicz, P.%
\end{APACrefauthors}%
\unskip\
\newblock
\APACrefYearMonthDay{2006}{}{}.
\newblock
{\BBOQ}\APACrefatitle {Multidimensional positive definite advection transport
  algorithm: an overview} {Multidimensional positive definite advection
  transport algorithm: an overview}.{\BBCQ}
\newblock
\APACjournalVolNumPages{Int. J. Numer. Meth. Fluids}{50}{}{}.
\newblock
\begin{APACrefDOI} \doi{10.1002/fld.1071} \end{APACrefDOI}
\PrintBackRefs{\CurrentBib}

\bibitem [\protect \citeauthoryear {%
S\"olch%
\ \BBA {} K\"archer%
}{%
S\"olch%
\ \BBA {} K\"archer%
}{%
{\protect \APACyear {2012}}%
}]{%
Soelch_and_Kaercher_2012}
\APACinsertmetastar {%
Soelch_and_Kaercher_2012}%
\begin{APACrefauthors}%
S\"olch, I.%
\BCBT {}\ \BBA {} K\"archer, B.%
\end{APACrefauthors}%
\unskip\
\newblock
\APACrefYearMonthDay{2012}{}{}.
\newblock
{\BBOQ}\APACrefatitle {A large-eddy model for cirrus clouds with explicit
  aerosol and ice microphysics and {L}agrangian ice particle tracking} {A
  large-eddy model for cirrus clouds with explicit aerosol and ice microphysics
  and {L}agrangian ice particle tracking}.{\BBCQ}
\newblock
\APACjournalVolNumPages{Q. J. Royal Meteorol. Soc.}{136}{}{}.
\newblock
\begin{APACrefDOI} \doi{10.1002/qj.689} \end{APACrefDOI}
\PrintBackRefs{\CurrentBib}

\bibitem [\protect \citeauthoryear {%
Solomon%
, Feingold%
\BCBL {}\ \BBA {} Shupe%
}{%
Solomon%
\ \protect \BOthers {.}}{%
{\protect \APACyear {2015}}%
}]{%
Solomon_et_al_2015}
\APACinsertmetastar {%
Solomon_et_al_2015}%
\begin{APACrefauthors}%
Solomon, A.%
, Feingold, G.%
\BCBL {}\ \BBA {} Shupe, M.%
\end{APACrefauthors}%
\unskip\
\newblock
\APACrefYearMonthDay{2015}{}{}.
\newblock
{\BBOQ}\APACrefatitle {The role of ice nuclei recycling in the maintenance of
  cloud ice in Arctic mixed-phase stratocumulus} {The role of ice nuclei
  recycling in the maintenance of cloud ice in arctic mixed-phase
  stratocumulus}.{\BBCQ}
\newblock
\APACjournalVolNumPages{Atmos. Chem. Phys.}{15}{}{}.
\newblock
\begin{APACrefDOI} \doi{10.5194/acp-15-10631-2015} \end{APACrefDOI}
\PrintBackRefs{\CurrentBib}

\bibitem [\protect \citeauthoryear {%
Stansbury%
}{%
Stansbury%
}{%
{\protect \APACyear {1961}}%
}]{%
Stansbury_1961}
\APACinsertmetastar {%
Stansbury_1961}%
\begin{APACrefauthors}%
Stansbury, E.%
\end{APACrefauthors}%
\unskip\
\newblock
\APACrefYearMonthDay{1961}{}{}.
\newblock
\APACrefbtitle {Stochastic freezing} {Stochastic freezing}\
  \APACbVolEdTR{}{\BTR{}\ \BNUM\ MW-35}.
\newblock
\APACaddressInstitution{Montreal}{McGill Univ.}
\PrintBackRefs{\CurrentBib}

\bibitem [\protect \citeauthoryear {%
Stevens%
\ \protect \BOthers {.}}{%
Stevens%
\ \protect \BOthers {.}}{%
{\protect \APACyear {2018}}%
}]{%
Stevens_et_al_2018}
\APACinsertmetastar {%
Stevens_et_al_2018}%
\begin{APACrefauthors}%
Stevens, R.%
, Loewe, K.%
, Dearden, C.%
, Dimitrelos, A.%
, Possner, A.%
, Eirund, G.%
\BDBL {}Field, P\BPBI R.%
\end{APACrefauthors}%
\unskip\
\newblock
\APACrefYearMonthDay{2018}{}{}.
\newblock
{\BBOQ}\APACrefatitle {A model intercomparison of {CCN}-limited tenuous clouds
  in the high {Arctic}} {A model intercomparison of {CCN}-limited tenuous
  clouds in the high {Arctic}}.{\BBCQ}
\newblock
\APACjournalVolNumPages{Atmos. Chem. Phys.}{18}{}{}.
\newblock
\begin{APACrefDOI} \doi{10.5194/acp-18-11041-2018} \end{APACrefDOI}
\PrintBackRefs{\CurrentBib}

\bibitem [\protect \citeauthoryear {%
Sulia%
, Harrington%
\BCBL {}\ \BBA {} Morrison%
}{%
Sulia%
\ \protect \BOthers {.}}{%
{\protect \APACyear {2013}}%
}]{%
Sulia_et_al_2013}
\APACinsertmetastar {%
Sulia_et_al_2013}%
\begin{APACrefauthors}%
Sulia, K\BPBI J.%
, Harrington, J\BPBI Y.%
\BCBL {}\ \BBA {} Morrison, H.%
\end{APACrefauthors}%
\unskip\
\newblock
\APACrefYearMonthDay{2013}{}{}.
\newblock
{\BBOQ}\APACrefatitle {A Method for Adaptive Habit Prediction in Bulk
  Microphysical Models. Part III: Applications and Studies within a
  Two-Dimensional Kinematic Model} {A method for adaptive habit prediction in
  bulk microphysical models. part iii: Applications and studies within a
  two-dimensional kinematic model}.{\BBCQ}
\newblock
\APACjournalVolNumPages{J. Atmos. Sci.}{70}{}{}.
\newblock
\begin{APACrefDOI} \doi{10.1175/JAS-D-12-0316.1} \end{APACrefDOI}
\PrintBackRefs{\CurrentBib}

\bibitem [\protect \citeauthoryear {%
Sullivan%
\ \protect \BOthers {.}}{%
Sullivan%
\ \protect \BOthers {.}}{%
{\protect \APACyear {2018}}%
}]{%
Sullivan_et_al_2018}
\APACinsertmetastar {%
Sullivan_et_al_2018}%
\begin{APACrefauthors}%
Sullivan, S.%
, Barthlott, C.%
, Crosier, J.%
, Zhukov, I.%
, Nenes, A.%
\BCBL {}\ \BBA {} Hoose, C.%
\end{APACrefauthors}%
\unskip\
\newblock
\APACrefYearMonthDay{2018}{}{}.
\newblock
{\BBOQ}\APACrefatitle {The effect of secondary ice production parameterization
  on the simulation of a cold frontal rainband} {The effect of secondary ice
  production parameterization on the simulation of a cold frontal
  rainband}.{\BBCQ}
\newblock
\APACjournalVolNumPages{Atmos. Chem. Phys.}{18}{}{}.
\newblock
\begin{APACrefDOI} \doi{10.5194/acp-18-16461-2018} \end{APACrefDOI}
\PrintBackRefs{\CurrentBib}

\bibitem [\protect \citeauthoryear {%
Szabó-Takács%
}{%
Szabó-Takács%
}{%
{\protect \APACyear {2011}}%
}]{%
Szabo_2011}
\APACinsertmetastar {%
Szabo_2011}%
\begin{APACrefauthors}%
Szabó-Takács, B.%
\end{APACrefauthors}%
\unskip\
\newblock
\APACrefYearMonthDay{2011}{}{}.
\newblock
{\BBOQ}\APACrefatitle {Numerical simulation of the cycle of aerosol particles
  in stratocumulus clouds with a two-dimensional kinematic model} {Numerical
  simulation of the cycle of aerosol particles in stratocumulus clouds with a
  two-dimensional kinematic model}.{\BBCQ}
\newblock
\APACjournalVolNumPages{Q. J. Hungarian Meteorol. Service}{115}{3}{147--165}.
\newblock
\begin{APACrefURL}
  \url{https://web.archive.org/web/*/https://www.met.hu/downloads.php?fn=/metadmin/newspaper/2012/04/115-3-2-szabo.pdf}
  \end{APACrefURL}
\PrintBackRefs{\CurrentBib}

\bibitem [\protect \citeauthoryear {%
Szumowski%
, Grabowski%
\BCBL {}\ \BBA {} III%
}{%
Szumowski%
\ \protect \BOthers {.}}{%
{\protect \APACyear {1998}}%
}]{%
Szumowski_et_al_1998}
\APACinsertmetastar {%
Szumowski_et_al_1998}%
\begin{APACrefauthors}%
Szumowski, M.%
, Grabowski, W.%
\BCBL {}\ \BBA {} III, O\BPBI H.%
\end{APACrefauthors}%
\unskip\
\newblock
\APACrefYearMonthDay{1998}{}{}.
\newblock
{\BBOQ}\APACrefatitle {Simple two-dimensional kinematic framework designed to
  test warm rain microphysical models} {Simple two-dimensional kinematic
  framework designed to test warm rain microphysical models}.{\BBCQ}
\newblock
\APACjournalVolNumPages{Atmos. Res.}{45}{}{}.
\newblock
\begin{APACrefDOI} \doi{10.1016/S0169-8095(97)00082-3} \end{APACrefDOI}
\PrintBackRefs{\CurrentBib}

\bibitem [\protect \citeauthoryear {%
Unterstrasser%
, Hoffmann%
\BCBL {}\ \BBA {} Lerch%
}{%
Unterstrasser%
\ \protect \BOthers {.}}{%
{\protect \APACyear {2017}}%
}]{%
Unterstrasser_et_al_2017}
\APACinsertmetastar {%
Unterstrasser_et_al_2017}%
\begin{APACrefauthors}%
Unterstrasser, S.%
, Hoffmann, F.%
\BCBL {}\ \BBA {} Lerch, M.%
\end{APACrefauthors}%
\unskip\
\newblock
\APACrefYearMonthDay{2017}{}{}.
\newblock
{\BBOQ}\APACrefatitle {Collection/aggregation algorithms in {L}agrangian cloud
  microphysical models: rigorous evaluation in box model simulations}
  {Collection/aggregation algorithms in {L}agrangian cloud microphysical
  models: rigorous evaluation in box model simulations}.{\BBCQ}
\newblock
\APACjournalVolNumPages{Geosci. Model Dev.}{10}{}{}.
\newblock
\begin{APACrefDOI} \doi{10.5194/gmd-10-1521-2017} \end{APACrefDOI}
\PrintBackRefs{\CurrentBib}

\bibitem [\protect \citeauthoryear {%
Vali%
}{%
Vali%
}{%
{\protect \APACyear {1971}}%
}]{%
Vali_1971}
\APACinsertmetastar {%
Vali_1971}%
\begin{APACrefauthors}%
Vali, G.%
\end{APACrefauthors}%
\unskip\
\newblock
\APACrefYearMonthDay{1971}{}{}.
\newblock
{\BBOQ}\APACrefatitle {Quantitative Evaluation of Experimental Results an the
  Heterogeneous Freezing Nucleation of Supercooled Liquids} {Quantitative
  evaluation of experimental results an the heterogeneous freezing nucleation
  of supercooled liquids}.{\BBCQ}
\newblock
\APACjournalVolNumPages{J. Atmos. Sci.}{28}{}{}.
\newblock
\begin{APACrefDOI} \doi{10.1175/1520-0469(1971)028<0402:QEOERA>2.0.CO;2}
  \end{APACrefDOI}
\PrintBackRefs{\CurrentBib}

\bibitem [\protect \citeauthoryear {%
Vali%
}{%
Vali%
}{%
{\protect \APACyear {1994}}%
}]{%
Vali_1994}
\APACinsertmetastar {%
Vali_1994}%
\begin{APACrefauthors}%
Vali, G.%
\end{APACrefauthors}%
\unskip\
\newblock
\APACrefYearMonthDay{1994}{}{}.
\newblock
{\BBOQ}\APACrefatitle {Freezing Rate Due to Heterogeneous Nucleation} {Freezing
  rate due to heterogeneous nucleation}.{\BBCQ}
\newblock
\APACjournalVolNumPages{J. Atmos. Sci.}{51}{}{}.
\newblock
\begin{APACrefDOI} \doi{10.1175/1520-0469(1994)051<1843:FRDTHN>2.0.CO;2}
  \end{APACrefDOI}
\PrintBackRefs{\CurrentBib}

\bibitem [\protect \citeauthoryear {%
Vali%
}{%
Vali%
}{%
{\protect \APACyear {2014}}%
}]{%
Vali_2014}
\APACinsertmetastar {%
Vali_2014}%
\begin{APACrefauthors}%
Vali, G.%
\end{APACrefauthors}%
\unskip\
\newblock
\APACrefYearMonthDay{2014}{}{}.
\newblock
{\BBOQ}\APACrefatitle {Interpretation of freezing nucleation experiments:
  singular and stochastic; sites and surfaces} {Interpretation of freezing
  nucleation experiments: singular and stochastic; sites and surfaces}.{\BBCQ}
\newblock
\APACjournalVolNumPages{Atmos. Chem. Phys.}{14}{}{}.
\newblock
\begin{APACrefDOI} \doi{10.5194/acp-14-5271-2014} \end{APACrefDOI}
\PrintBackRefs{\CurrentBib}

\bibitem [\protect \citeauthoryear {%
Vali%
}{%
Vali%
}{%
{\protect \APACyear {2019}}%
}]{%
Vali_2019}
\APACinsertmetastar {%
Vali_2019}%
\begin{APACrefauthors}%
Vali, G.%
\end{APACrefauthors}%
\unskip\
\newblock
\APACrefYearMonthDay{2019}{}{}.
\newblock
{\BBOQ}\APACrefatitle {Revisiting the differential freezing nucleus spectra
  derived from drop-freezing experiments: methods of calculation, applications,
  and confidence limits} {Revisiting the differential freezing nucleus spectra
  derived from drop-freezing experiments: methods of calculation, applications,
  and confidence limits}.{\BBCQ}
\newblock
\APACjournalVolNumPages{Atmos. Meas. Tech.}{12}{}{}.
\newblock
\begin{APACrefDOI} \doi{10.5194/amt-12-1219-2019} \end{APACrefDOI}
\PrintBackRefs{\CurrentBib}

\bibitem [\protect \citeauthoryear {%
Vali%
, DeMott%
, M\"ohler%
\BCBL {}\ \BBA {} Whale%
}{%
Vali%
\ \protect \BOthers {.}}{%
{\protect \APACyear {2015}}%
}]{%
Vali_et_al_2015}
\APACinsertmetastar {%
Vali_et_al_2015}%
\begin{APACrefauthors}%
Vali, G.%
, DeMott, P.%
, M\"ohler, O.%
\BCBL {}\ \BBA {} Whale, T.%
\end{APACrefauthors}%
\unskip\
\newblock
\APACrefYearMonthDay{2015}{}{}.
\newblock
{\BBOQ}\APACrefatitle {Technical Note: A proposal for ice nucleation
  terminology} {Technical note: A proposal for ice nucleation
  terminology}.{\BBCQ}
\newblock
\APACjournalVolNumPages{Atmos. Chem. Phys.}{15}{}{}.
\newblock
\begin{APACrefDOI} \doi{10.5194/acp-15-10263-2015} \end{APACrefDOI}
\PrintBackRefs{\CurrentBib}

\bibitem [\protect \citeauthoryear {%
Vali%
\ \BBA {} Snider%
}{%
Vali%
\ \BBA {} Snider%
}{%
{\protect \APACyear {2015}}%
}]{%
Vali_and_Snider_2015}
\APACinsertmetastar {%
Vali_and_Snider_2015}%
\begin{APACrefauthors}%
Vali, G.%
\BCBT {}\ \BBA {} Snider, J.%
\end{APACrefauthors}%
\unskip\
\newblock
\APACrefYearMonthDay{2015}{}{}.
\newblock
{\BBOQ}\APACrefatitle {Time-dependent freezing rate parcel model}
  {Time-dependent freezing rate parcel model}.{\BBCQ}
\newblock
\APACjournalVolNumPages{Atmos. Chem. Phys.}{15}{4}{2071--2079}.
\newblock
\begin{APACrefDOI} \doi{10.5194/acp-15-2071-2015} \end{APACrefDOI}
\PrintBackRefs{\CurrentBib}

\bibitem [\protect \citeauthoryear {%
Vali%
\ \BBA {} Stansbury%
}{%
Vali%
\ \BBA {} Stansbury%
}{%
{\protect \APACyear {1966}}%
}]{%
Vali_and_Stansbury_1966}
\APACinsertmetastar {%
Vali_and_Stansbury_1966}%
\begin{APACrefauthors}%
Vali, G.%
\BCBT {}\ \BBA {} Stansbury, E.%
\end{APACrefauthors}%
\unskip\
\newblock
\APACrefYearMonthDay{1966}{}{}.
\newblock
{\BBOQ}\APACrefatitle {Time-dependent characteristics of the heterogeneous
  nucleation of ice} {Time-dependent characteristics of the heterogeneous
  nucleation of ice}.{\BBCQ}
\newblock
\APACjournalVolNumPages{Can. J. Phys.}{44}{}{}.
\newblock
\begin{APACrefDOI} \doi{10.1139/p66-044} \end{APACrefDOI}
\PrintBackRefs{\CurrentBib}

\bibitem [\protect \citeauthoryear {%
Vonnegut%
}{%
Vonnegut%
}{%
{\protect \APACyear {1948}}%
}]{%
Vonnegut_1948}
\APACinsertmetastar {%
Vonnegut_1948}%
\begin{APACrefauthors}%
Vonnegut, B.%
\end{APACrefauthors}%
\unskip\
\newblock
\APACrefYearMonthDay{1948}{}{}.
\newblock
{\BBOQ}\APACrefatitle {Variation with temperature of the nucleation rate of
  supercooled liquid tin and water drops} {Variation with temperature of the
  nucleation rate of supercooled liquid tin and water drops}.{\BBCQ}
\newblock
\APACjournalVolNumPages{J. Colloid Sci.}{3}{}{}.
\newblock
\begin{APACrefDOI} \doi{10.1016/S0095-8522(48)90049-X} \end{APACrefDOI}
\PrintBackRefs{\CurrentBib}

\bibitem [\protect \citeauthoryear {%
Vonnegut%
}{%
Vonnegut%
}{%
{\protect \APACyear {1949}}%
}]{%
Vonnegut_1949}
\APACinsertmetastar {%
Vonnegut_1949}%
\begin{APACrefauthors}%
Vonnegut, B.%
\end{APACrefauthors}%
\unskip\
\newblock
\APACrefYearMonthDay{1949}{}{}.
\newblock
{\BBOQ}\APACrefatitle {Nucleation of Supercooled Water Clouds by Silver Iodide
  Smokes} {Nucleation of supercooled water clouds by silver iodide
  smokes}.{\BBCQ}
\newblock
\APACjournalVolNumPages{Chem. Rev.}{44}{}{}.
\newblock
\begin{APACrefDOI} \doi{10.1021/cr60138a003} \end{APACrefDOI}
\PrintBackRefs{\CurrentBib}

\bibitem [\protect \citeauthoryear {%
Welß%
, von Terzi%
, Kneifel%
, Seifert%
\BCBL {}\ \BBA {} Siewert%
}{%
Welß%
\ \protect \BOthers {.}}{%
{\protect \APACyear {2022}}%
}]{%
Welss_et_al_2022}
\APACinsertmetastar {%
Welss_et_al_2022}%
\begin{APACrefauthors}%
Welß, J\BHBI N.%
, von Terzi, L.%
, Kneifel, S.%
, Seifert, A.%
\BCBL {}\ \BBA {} Siewert, C.%
\end{APACrefauthors}%
\unskip\
\newblock
\APACrefYearMonthDay{2022}{}{}.
\newblock
{\BBOQ}\APACrefatitle {Exploring the origin of increasing ice particle number
  in the dendritic growth zone combining polarimetric radar observations and
  novel {L}agrangian particle modeling} {Exploring the origin of increasing ice
  particle number in the dendritic growth zone combining polarimetric radar
  observations and novel {L}agrangian particle modeling}.{\BBCQ}
\newblock
\BIn{} \APACrefbtitle {EGU22, the 24th EGU General Assembly, held 23-27 May,
  2022 in Vienna, Austria and Online.} {Egu22, the 24th egu general assembly,
  held 23-27 may, 2022 in vienna, austria and online.}
\newblock
\begin{APACrefDOI} \doi{10.5194/egusphere-egu22-5159} \end{APACrefDOI}
\PrintBackRefs{\CurrentBib}

\bibitem [\protect \citeauthoryear {%
Wisner%
, Orville%
\BCBL {}\ \BBA {} Myers%
}{%
Wisner%
\ \protect \BOthers {.}}{%
{\protect \APACyear {1972}}%
}]{%
Wisner_et_al_1972}
\APACinsertmetastar {%
Wisner_et_al_1972}%
\begin{APACrefauthors}%
Wisner, C.%
, Orville, H.%
\BCBL {}\ \BBA {} Myers, C.%
\end{APACrefauthors}%
\unskip\
\newblock
\APACrefYearMonthDay{1972}{}{}.
\newblock
{\BBOQ}\APACrefatitle {A Numerical Model of a Hail-Bearing Cloud} {A numerical
  model of a hail-bearing cloud}.{\BBCQ}
\newblock
\APACjournalVolNumPages{J. Atmos. Sci.}{29}{}{}.
\newblock
\begin{APACrefDOI} \doi{10.1175/1520-0469(1972)029<1160:ANMOAH>2.0.CO;2}
  \end{APACrefDOI}
\PrintBackRefs{\CurrentBib}

\bibitem [\protect \citeauthoryear {%
Wood%
, Leon%
, Lebsock%
, Snider%
\BCBL {}\ \BBA {} Clarke%
}{%
Wood%
\ \protect \BOthers {.}}{%
{\protect \APACyear {2012}}%
}]{%
Wood_et_al_2012}
\APACinsertmetastar {%
Wood_et_al_2012}%
\begin{APACrefauthors}%
Wood, R.%
, Leon, D.%
, Lebsock, M.%
, Snider, J.%
\BCBL {}\ \BBA {} Clarke, A.%
\end{APACrefauthors}%
\unskip\
\newblock
\APACrefYearMonthDay{2012}{}{}.
\newblock
{\BBOQ}\APACrefatitle {Precipitation driving of droplet concentration
  variability in marine low clouds} {Precipitation driving of droplet
  concentration variability in marine low clouds}.{\BBCQ}
\newblock
\APACjournalVolNumPages{J. Geophys. Res. Atmos.}{117}{}{}.
\newblock
\begin{APACrefDOI} \doi{10.1029/2012JD018305} \end{APACrefDOI}
\PrintBackRefs{\CurrentBib}

\bibitem [\protect \citeauthoryear {%
Wright%
\ \BBA {} Petters%
}{%
Wright%
\ \BBA {} Petters%
}{%
{\protect \APACyear {2013}}%
}]{%
Wright_and_Petters_2013}
\APACinsertmetastar {%
Wright_and_Petters_2013}%
\begin{APACrefauthors}%
Wright, T.%
\BCBT {}\ \BBA {} Petters, M.%
\end{APACrefauthors}%
\unskip\
\newblock
\APACrefYearMonthDay{2013}{}{}.
\newblock
{\BBOQ}\APACrefatitle {The role of time in heterogeneous freezing nucleation}
  {The role of time in heterogeneous freezing nucleation}.{\BBCQ}
\newblock
\APACjournalVolNumPages{J. Geophys. Res. Atmos.}{118}{9}{3731--3743}.
\newblock
\begin{APACrefDOI} \doi{10.1002/jgrd.50365} \end{APACrefDOI}
\PrintBackRefs{\CurrentBib}

\bibitem [\protect \citeauthoryear {%
Wright%
, Petters%
, Hader%
, Morton%
\BCBL {}\ \BBA {} Holder%
}{%
Wright%
\ \protect \BOthers {.}}{%
{\protect \APACyear {2013}}%
}]{%
Wright_et_al_2013}
\APACinsertmetastar {%
Wright_et_al_2013}%
\begin{APACrefauthors}%
Wright, T.%
, Petters, M.%
, Hader, J.%
, Morton, T.%
\BCBL {}\ \BBA {} Holder, A.%
\end{APACrefauthors}%
\unskip\
\newblock
\APACrefYearMonthDay{2013}{}{}.
\newblock
{\BBOQ}\APACrefatitle {Minimal cooling rate dependence of ice nuclei activity
  in the immersion mode} {Minimal cooling rate dependence of ice nuclei
  activity in the immersion mode}.{\BBCQ}
\newblock
\APACjournalVolNumPages{J. Geophys. Res. Atmos.}{118}{18}{}.
\newblock
\begin{APACrefDOI} \doi{10.1002/jgrd.50810} \end{APACrefDOI}
\PrintBackRefs{\CurrentBib}

\bibitem [\protect \citeauthoryear {%
Yang%
, Ovchinnikov%
\BCBL {}\ \BBA {} Shaw%
}{%
Yang%
\ \protect \BOthers {.}}{%
{\protect \APACyear {2015}}%
}]{%
Yang_et_al_2015}
\APACinsertmetastar {%
Yang_et_al_2015}%
\begin{APACrefauthors}%
Yang, F.%
, Ovchinnikov, M.%
\BCBL {}\ \BBA {} Shaw, R.%
\end{APACrefauthors}%
\unskip\
\newblock
\APACrefYearMonthDay{2015}{}{}.
\newblock
{\BBOQ}\APACrefatitle {Long-lifetime ice particles in mixed-phase stratiform
  clouds: Quasi-steady and recycled growth} {Long-lifetime ice particles in
  mixed-phase stratiform clouds: Quasi-steady and recycled growth}.{\BBCQ}
\newblock
\APACjournalVolNumPages{J. Geophys. Res. Atmos.}{120}{}{}.
\newblock
\begin{APACrefDOI} \doi{10.1002/2015JD023679} \end{APACrefDOI}
\PrintBackRefs{\CurrentBib}

\bibitem [\protect \citeauthoryear {%
Yao%
, Dawson%
, Dabdub%
\BCBL {}\ \BBA {} Riemer%
}{%
Yao%
\ \protect \BOthers {.}}{%
{\protect \APACyear {2021}}%
}]{%
Yao_et_al_2021}
\APACinsertmetastar {%
Yao_et_al_2021}%
\begin{APACrefauthors}%
Yao, Y.%
, Dawson, M.%
, Dabdub, D.%
\BCBL {}\ \BBA {} Riemer, N.%
\end{APACrefauthors}%
\unskip\
\newblock
\APACrefYearMonthDay{2021}{}{}.
\newblock
{\BBOQ}\APACrefatitle {Evaluating the Impacts of Cloud Processing on
  Resuspended Aerosol Particles After Cloud Evaporation Using a
  Particle-Resolved Model} {Evaluating the impacts of cloud processing on
  resuspended aerosol particles after cloud evaporation using a
  particle-resolved model}.{\BBCQ}
\newblock
\APACjournalVolNumPages{J. Geophys. Res. Atmos.}{126}{}{}.
\newblock
\begin{APACrefDOI} \doi{10.1029/2021JD034992} \end{APACrefDOI}
\PrintBackRefs{\CurrentBib}

\end{thebibliography}
\end{document}